\documentclass[12pt]{article}
\usepackage{amssymb}
\usepackage{amsmath}
\usepackage{amsthm}
\usepackage{color}
\usepackage{comment}
\usepackage{enumitem}		
\usepackage{geometry}		
\usepackage{graphicx}
\usepackage{authblk}
\usepackage{amssymb}
\usepackage{listings}
\usepackage{xcolor}
\usepackage{caption}
\usepackage{color}
\usepackage{algorithm}
\usepackage{algpseudocode}

\usepackage{subcaption}

\makeatletter
	\@addtoreset{equation}{section}
\makeatother

\let\originalleft\left
\let\originalright\right
\renewcommand{\left}{\mathopen{}\mathclose\bgroup\originalleft}
\renewcommand{\right}{\aftergroup\egroup\originalright}

\newlist{romanlist}{enumerate}{3}
\setlist[romanlist]{label=\roman*),ref=(\roman*)}

\begin{document}

\newcommand{\cF}{\mathcal{F}}
\newcommand{\cP}{\mathcal{P}}
\newcommand{\cR}{\mathcal{R}}
\newcommand{\cS}{\mathcal{S}}
\newcommand{\cT}{\mathcal{T}}
\newcommand{\cW}{\mathcal{W}}
\newcommand{\ee}{\varepsilon}
\newcommand{\rD}{{\rm D}}
\newcommand{\re}{{\rm e}}

\newtheorem{theorem}{Theorem}[section]
\newtheorem{corollary}[theorem]{Corollary}
\newtheorem{lemma}[theorem]{Lemma}
\newtheorem{proposition}[theorem]{Proposition}

\theoremstyle{definition}
\newtheorem{definition}{Definition}[section]


\title{
Dynamical properties of a small heterogeneous chain network of neurons in discrete time.
}

\author[1,*]{I.~Ghosh}
\author[2]{A.S.~Nair}
\author[3,4]{H.O.~Fatoyinbo}
\author[2]{S.S.~Muni}

\affil[1]{School of Mathematical and Computational Sciences\\
Massey University\\
Colombo Road, Palmerston North, 4410\\
New Zealand}

\affil[2]{School of Digital Sciences\\
Digital University Kerala\\ 
Technocity campus, Mangalapuram, 695317\\
Kerala, India}

\affil[3]{Department of Mathematical Sciences\\ School of Engineering, Computer and Mathematical Sciences\\ Auckland University of Technology, Auckland 1142\\ New Zealand}

\affil[4]{EpiCentre, School of Veterinary Science\\
Massey University\\
Colombo Road, Palmerston North, 4410\\
New Zealand}

\maketitle


\begin{abstract}

We propose a novel nonlinear bidirectionally coupled heterogeneous chain network whose dynamics evolve in discrete time. The backbone of the model is a pair of popular map-based neuron models, the Chialvo and the Rulkov maps. This model is assumed to proximate the intricate dynamical properties of neurons in the widely complex nervous system. The model is first realized via various nonlinear analysis techniques: fixed point analysis, phase portraits, Jacobian matrix, noninvertibility criterion, and bifurcation diagrams. We observe the coexistence of chaotic and period-$4$ attractors. Various codimension-$1$ and -$2$ patterns for example saddle-node, period-doubling, Neimark-Sacker, double Neimark-Sacker, flip- and fold-Neimark Sacker, and $1:1$ and $1:2$ resonance are also explored. Furthermore, the study employs two synchronization measures to quantify how the oscillators in the network behave in tandem with each other over a long number of iterations. Finally, a time series analysis of the model is performed to investigate its complexity in terms of sample entropy.

\end{abstract}

\section{Introduction}

The fundamental units of the nervous system constitute these special cells called {\em neurons} and {\em neurodynamics} is the study of the dynamical properties of these cells. It is an evolving field of mathematical biology majorly dealing with quantifying the complexity of the nervous system in terms of mathematical tools like ordinary differential equations, partial differential equations, delay differential equations, maps, cellular automata, etc. The nervous system is the center for controlling all major activities in the human body, enabling communication between different parts of the body primarily through the transmission of electrochemical signals~\cite{Lo08} among the neurons. This complex functioning of signal processing and information transfer demands subtle nonlinear dynamical techniques~\cite{Iz07, GeKi14, HeYa21} applied on a single neuron and a network of neurons for investigating various phenomena such as bifurcation analysis~\cite{HamSA02}, identifying the transition between different states like stable fixed points, limit cycles, and chaotic attractors. The complexity of the network dynamics can be reduced using nonlinear dynamics by preserving the essential features of the neural interactions. Some popular neurodynamical models studied extensively by the research community are the Hindmarsh-Rose, model~\cite{HiRo84}, the Hodgkin-Huxley model~\cite{HoHu52}, the Morris–Lecar model~\cite{MoLe81}, and the FitzHugh-Nagumo Model~\cite{Fi61, Sh14}. The Hindmarsh-Rose neuron model studies the spiking-bursting behavior of neurons with different variables such as membrane potential, ion channels, and adaptation current~\cite{HiRo84}. The Hodgkin-Huxley Model describes the generation of action potential in neurons and describes the neuronal membrane potential dynamics based on the dynamics of voltage-gated ion channels~\cite{HoHu52}. The Morris–Lecar model is a biological neuron model that describes the dynamics of action potentials in neurons. This model simplifies the Hodgkin-Huxley model by reducing the number of state variables and parameters while retaining essential features of neuronal excitability~\cite{MoLe81, HFetal20}. FitzHugh-Nagumo model is another two-dimensional model built on simplifying the Hodgkin-Huxley nonlinear model that contains two coupled nonlinear ordinary differential equations describing the evolution of a neuron membrane voltage and representing the recovery action~\cite{Sh14}. 

The majority of studies in neurodynamics focus on continuous-time systems, leaving discrete systems relatively understudied~\cite{Da12}. However, discrete models play a significant role in understanding neural dynamics. Notable examples include neuron models such as the Chialvo model~\cite{Ch95}, the Rulkov~\cite{Ru01, Ru02} model, and the Nekorkin model~\cite{NeVd07}. Furthermore, discrete neuron maps have been also generated from their continuous counterparts using Euler discretization techniques, for example, the discrete Izhikevich model~\cite{MuRa22} and the discrete Hindmarsh Rose model~\cite{Mu23}. In discrete models, each neuron is represented by a finite number of states, with transmissions between states governed by specific rules. The Chialvo and the Rulkov models are prominently studied neuron models esteemed for their capacity to accurately replicate diverse spatiotemporal patterns discerned in neural activity.


Discrete models offer valuable insights into how the firing patterns of neurons depend on factors such as the refractory period, firing threshold, and network architecture, enabling efficient prediction of chaotic time series and demodulation of FSK (Frequency-Shift Keying) signals~\cite{HaKa90}. Moreover, these models are computationally efficient, making them particularly useful for large-scale simulations and analyses. In our current investigation, we specifically focus on a heterogeneous discrete model to explore the dynamics of coupled neurons. This choice allows us to delve into the intricate interactions and behaviors within such networks, leveraging the advantages offered by discrete modeling approaches.

Millions of neurons are interconnected in the nervous system in a complex network structure. Recent studies indicate that neurons can establish, modify, and adjust their connections and signaling properties, allowing for adaptation to spatial and temporal patterns of neural signals~\cite{LaSe03}. This remarkable adaptability enables neuron networks to autonomously organize, calibrate, and retain information based on experience. Despite the extensive research focused on unraveling the intricate structure of these networks, delving into simple networks remains imperative. It offers foundational insights into comprehending brain function, modeling complex neural systems, deciphering neurological disorders, fostering technological innovation, and facilitating educational advancements in neuroscience. Some of the simple networks studied include the ring network~\cite{OmLa22, TsDr22, KhPa19, LoDe20}, star network~\cite{ChLi21, YaMa22, RaPy21, KiKu09}, ring-star network~\cite{MuPr20, MuFa22, GhMu23, MuRa22}, lattice network~\cite{VaWa05, PaMo14, ReSu23}, and multiplex network~\cite{LeYa21, WeWu15}. In ring structures, each neuron within the network is exclusively connected solely to its two nearest neighbors. The hippocampus, cerebellum, and neocortex are regions of the brain known for their complex processing capabilities and involvement in various cognitive functions. The presence of ring architectures has been identified in these brain regions, as well as in domains beyond neuroscience, such as chemistry and electrical engineering~\cite{KaSi12}. The star configuration comprises a single central node, often referred to as the hub, to which all other nodes within the network are connected. Notably, in this configuration, nodes are linked exclusively to the central hub and are not interconnected with one another.

There is an enormous need to study complex systems in terms of minimal mathematical models. The nervous system being one of the most important studied complex systems, requires mathematical modelers to study it in terms of a reduced model of coupled oscillators forming a small network that can be treated as a unit of a macroscopic ensemble of neurons and the nervous system as a whole. Most of the studies focus on the dynamical analysis of a single neuron to an ensemble of neurons consisting of at least a hundred units. Note that for a bigger ensemble of neurons, the analytical studies of the dynamics of the system become quite intractable. Thus studying a ``small network'' is indispensable. First of all, because it acts as a bridge between a single oscillator and a network of coupled oscillators where there exists at least a hundred, and secondly because the mathematical analysis of the smallest possible oscillator network is still tractable. {\em Small networks} are reduced models consisting of three to ten coupled oscillators giving rise to collective dynamical properties. They can be thought of as network units that repeat themselves to form a more complex topological structure. Our focus is on a heterogeneous system of oscillators representing the nervous system.

Heterogeneous neuron networks have been recently modeled and studied in various forms. One important work is by Shen {\em et al.}~\cite{HetN1} where the authors coupled a Fithugh-Nagumo neuron with a Hindmarsh-Rose neuron and studied their dynamical properties. Njitacke {\em et al.}~\cite{HetN4} coupled a two-dimensional Hindmarsh-Rose neuron to a three-dimensional version through a multistable memristive connection. The same authors also did a similar study with Hindmarsh-Rose neurons but with a gap-junction~\cite{HetN2} to induce heterogeneity in the neuron system. A variation of energy influx is also able to incorporate heterogeneity in the neuron network over time as shown by Yang {\em et al.}~\cite{HetN5}. Bradley {\em et al.}~\cite{HetN8} studied a weakly coupled network of Wang-Buzskai and Hodgekin-Huxley neurons. Xie {\em et al.}~\cite{HetN3} in their work showed that even continuous energy accumulation in neurons can incorporate heterogeneity via shape deformation under external stimulation. Thus heterogeneity in neuron networks is an invigorating phenomenon and requires the attention of mathematical modelers and neuroscientists. We recently studied a heterogeneous ring-star network of Chialvo neurons where the heterogeneities were realized with the introduction of additive noise to the central-peripheral and the peripheral-peripheral nodes with atleast a hundred nodes in the network, see Ghosh {\em et al.}~\cite{GhMu23}. However, because of the complexity and the large number of nodes in that system, not many rooms were left to perform analytical calculations. To address this issue, this paper tries to study the smallest possible chain network of oscillators where heterogeneity is inculcated. In this case, the heterogeneity was included not only in the coupling strengths but also in the type of oscillators. This system is of course complicated than a single oscillator model, nonetheless is analytically tractable. Thus, we perform both algebraic calculations supported by numerics to make the study even more robust. The goal in mind was to delve into the intricate dynamical properties of a small network modeling the dynamics of the nervous system. By investigating the behavior of even the simplest network configurations, researchers can gain insights into phenomena such as synchronization, pattern formation, and information transfer, which are essential for understanding more complex neural networks and neuron functions as a whole. Furthermore, researchers could potentially predict and control the behavior of neurons, with implications for neural engineering and the development of neural prosthetics. Overall, simulating small networks of coupled neurons offers a powerful approach to advancing our understanding of neural function and dysfunction in both engineering and biological research contexts. heterogeneous neuron networks hold quite a potential to be applied in a wide array of fields like robust learning (See Perez-Nieves {\em et al.}~\cite{HetN6}), reliable neuronal systems (See Lengler{\em et al.}~\cite{LeJu13}), and image encrypting procedures (See Yunliang {\em et al.}~\cite{HetN7}).

The motivation for taking a small heterogeneous chain network made of three oscillators whose dynamics are governed by neuron maps is rooted in the fact that there are three types of neurons based on their functionalities: motor neurons, sensory neurons, and interneurons~\cite{Fu00}. Motor neurons act as transmission media for synaptic impulses from the central nervous system to the organs and tissues, whereas sensory neurons act as transmission media from the organs and tissues to the central nervous system. The interneurons, however, act as a bridge between the sensory and the motor neurons. Both motor and sensory neurons have similar functionalities and thus could correspond to the two end oscillators in our model represented by the Chialvo map. Whereas, an interneuron could be modeled by the Rulkov map which acts as a bridge between the two Chialvo neurons, representing a unit of the much more complex nervous system. The interlinks between the three types of neurons further correspond to the simplest type of couplings shown in our model. A chain network made of three neurons can be regarded as a special case of a star network. Moreover, studying the dynamics of small networks of interconnected neurons can provide insights into the functioning of larger brain circuits. A topologically similar tri-oscillator chain network model, but in continuous time, has been studied by Njitacke {\em et al.}~\cite{Nj22}. In a very recent study led by Cao {\em et al.}~\cite{CaWa24}, the authors have considered a Chialvo neuron coupled with a Rulkov neuron through a memristence and have studied the corresponding dynamical properties. Although the base models of Cao {\em et al.} are similar to this work, it is to be noted that the major difference in this work from Cao's work is the topology of the network. First of all, we do not consider any memristence in this system, and secondly, our system has two Chialvo neurons on the edge with a Rulkov neuron at the center making a good unit candidate for a bigger ensemble of a star-network model. The motivation behind taking this model was to imitate the real-world functionalities of the three types of neurons mentioned above.

Once we have our model, it is imperative to dive right into unfolding its dynamical properties. The model is complex but can be to some extent analytically tractable. The first step is showcasing a collection of typical phase portraits of the system which exhibits chaotic attractors as expected. We also perform fixed point analysis, establish the Jacobian matrix of the model at the fixed point, explore the properties of the eigenvalues, study the concept of noninvertibility, and finally traverse the field of bifurcation analysis using sophisticated tools like \textsc{MatContM}~\cite{MeGo17}.

As mentioned before, the advantage of a small network is it not only serves its purpose of being studied as a dynamical unit but also provides a model with which we can traverse a batch of spatiotemporal patterns that arise due to the collective behaviors of the oscillators that make up the model. Concerning this, we study the synchronization behavior of the neurons involved in this network model and try to unfold what kind of collective behavior it portrays in general. Synchronization is one of the sophisticated phenomena that drives neural communication. Synchronization refers to how two neurons arrange themselves to form a functionally specialized ensemble. To study synchronization, we employ two measures called the {\em cross-correlation coefficient}~\cite{SeVa16, SeVa18} and the {\em Kuramoto order parameter}~\cite{Ku84, St00, BiGo20}. The motivation behind taking two measures is to corroborate the respective results with each other. The first is computed as a displacement between two oscillators involved in the network and the second is computed as the phase of an oscillator in the network. The concept of cross-correlation coefficient has been widely used to quantify synchronization in various network topologies~\cite{SeVa16, SeVa18, ShMu21, ShBu22, RySc23, VaSt16, NjTa23, GuXi23b}. Similarly, the Kuramoto order parameter has also had a prolific application in unfolding the same~\cite{Pr23, AnPr22, ScTi17}. Furthermore, we require a separate measure to quantify the abundant complexity of a neuron network. This can be realized in terms of information production, i.e, the {\em entropy} of the system. Shannon introduced the concept of entropy in the context of information theory~\cite{Sh01}, which since then has been successfully employed in the field of neuroscience~\cite{FaDe23, TiLa18, HaRo22, ZbRa21, ViWi11, ItHa11}. The entropy measure that we follow in this paper is by Richmond {\em et al.}~\cite{RiMo00}, called the {\em sample entropy}. The authors devised the sample entropy to quantify the complexity in physiological time series data and thus serve as a perfect candidate in this paper quantifying the complexity in neuron-based dynamical systems. This is by far the most popular entropy measure besides {\em approximate entropy} put forward by Pincus~\cite{Pi91}.

The goals of this paper are cataloged herewith:
\begin{enumerate}
    \item Introduce a novel heterogeneous small network of neurons coupled bi-directionally through linear coupling strengths, where each of these neurons is an oscillator whose dynamical properties are governed by popular discrete-time neuron maps,
    \item report the dynamical properties of this network unit through phase portraits, fixed point analysis, and noninvertibility criterion,
    \item analytically and numerically study various bifurcation patterns (codimension-$1$, and -$2$) of the network, especially using \textsc{MatContM}, 
    \item investigate the synchronization behavior in the small network using two quantitative measures called the cross-correlation coefficient and the Kuramoto order parameter,
    \item statistically explore how complex the small network is in terms of information processing within the network via a time series analysis measure called the sample entropy, and
    \item develop an overall grasp of a heterogeneous network that can eventually act as the unit of a large ensemble of neurons connected in a complicated topology.
\end{enumerate}

We organize the paper as follows: In \S\ref{sec:2DMaps}, we review the Chialvo and the Rulkov neuron maps which act as the building blocks of our heterogeneous tri-oscillator chain network. In \S\ref{sec:model} we put forward the novel heterogeneous network model constituting a nonlinear system of six coupled equations. We comment on the topology of the network and give an overview of what a typical phase portrait of this system looks like. In \S\ref{sec:fixedPoints}, we analyze the fixed points of this system, build the Jacobian of the system, and explore its eigenvalues at the fixed points, giving a birds-eye view of the dynamical properties of this network. In \S\ref{sec:noninv} we study where this discrete-time system based on neuron maps is noninvertible. Next, we delve into the innate bifurcation properties of the network in \S\ref{sec:bif1} and \S\ref{sec:bif2}. In \S\ref{sec:sync} we set up the synchronization measures for our network in terms of the cross-correlation coefficient and the Kuramoto order parameter and finally in \S\ref{sec:sampleEntropy} we perform a time series analysis of our model in terms of sample entropy to quantify the model's complexity. Concluding remarks and future directions are provided in \S\ref{sec:conclusions}. Note that all the numerical simulations have been performed in \texttt{Python 3.9.7} with extensive usage of \texttt{numpy}, \texttt{pandas}, and \texttt{matplotlib}, except for the codimension-$1$ and -$2$ bifurcation patterns, where it is \texttt{MATLAB} and \textsc{MatContM} which have been employed.

\section{Two-dimensional neuron maps}
\label{sec:2DMaps}
In this section, we review the dynamical structure of the maps that act as the building blocks of the tri-oscillator chain. Each of these oscillators is a two-dimensional iterated map used for modeling the dynamics of a neuron governed by either the {\em Chialvo} or the {\em Rulkov} map. For a detailed in-depth review of these topics please refer to Ibarz {\em et al.}~\cite{IbCa11}. The Chialvo map~\cite{Ch95} is given by
\begin{align}
\label{eq:Chialvo1}
  x(n+1) &= x(n)^2e^{(y(n) - x(n))}+k_0, \\
\label{eq:Chialvo2}
  y(n+1) &= ay(n) - bx(n) + c,
\end{align}
where the dynamical variables $x$ and $y$ represent the activation and the recovery variables respectively for the action potential. Chialvo map has four control parameters $a$, $b$, $c$, and $k_0$, where the first two represent the time constant of recovery and the activation dependence of recovery. These are kept at less than $1$. The latter two represent the offset and the time-dependent additive perturbation respectively. This map represents a model showcasing excitable dynamics where $y$ produces not slow but fast recovery. As mentioned in~\cite{IbCa11}, the model exhibits an ensemble of important dynamics, for example, subthreshold oscillations, bistability, and chaotic orbits among many others. Thus it serves as a good candidate for map-based neuron models. This model has attracted a wide array of recent works from the research community, see~\cite{MuFa22, XuHu23, GaRo23, RoPo20, GhMu23}. \par
The Rulkov map, specifically the chaotic family of the map~\cite{Ru01}, is the main focus of this paper. This map is given by the following pair of equations
\begin{align}
\label{eq:Rulkov1}
u(n+1) &= \frac{\alpha}{1+u(n)^2}+v(n), \\
\label{eq:Rulkov2}
v(n+1) &= v(n) - \mu(u(n) - \gamma),
\end{align}
where $u$ is again the activation variable whereas $v$ represents the slow variable of the model because $0<\mu<<1$. The parameter $\alpha$ induces nonlinearity in the model, whereas $\gamma$ is a DC modulator. There are two other versions of the map corresponding to non-chaotic behavior: the {\em non-chaotic} family~\cite{Ru02}, and the {\em supercritical} family~\cite{ShRu04}. For a detailed review of these please refer to Ibarz {\em et al.}~\cite{IbCa11}. The chaotic family does not constitute a piecewise behavior on $u$ and $v$, whereas the non-chaotic family does. In~\eqref{eq:Rulkov1}--\eqref{eq:Rulkov2} it can be seen that the map is unimodal. There exists a pair of fixed points, one stable and the other unstable, which disappears through saddle-node bifurcation on variation of parameters. Also, the spikes exhibit chaotic orbits, which can be made evident from the phase-plane plot of~\eqref{eq:Rulkov1}--\eqref{eq:Rulkov2}. This is left as an exercise for the reader. Like the Chialvo map, the Rulkov model has attracted quite a lot of attention from the research community recently, see~\cite{WaCa15,LoCo23, LiBa21, BiGu22, BaLi23}.\par
In the next section, we put forward a novel map-based small network model made up of the above chaotic maps. This generates a linearly coupled six-dimensional dynamical system in discrete time which can be reflected as a unit of a larger ensemble of neurons modeling the nervous system. 

\section{Network Model}
\label{sec:model}
\begin{figure}
\centering
\includegraphics[width=1\linewidth]{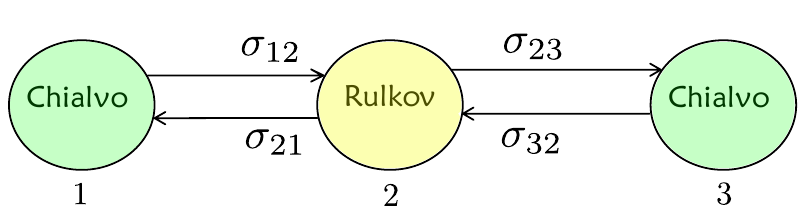}
\caption{A heterogeneous network of a tri-oscillator chain composed of end nodes (Chialvo neuron map) and central node (Rulkov neuron map). Bidirectional coupling strengths between node $1$ and $2$ are denoted by $\sigma_{12}, \sigma_{21}$. Similarly, bidirectional coupling strengths between node $2$ and $3$ are denoted by $\sigma_{23}, \sigma_{32}$. }\label{fig:triNode}
\end{figure}

We have considered a small heterogeneous network of a tri-oscillator chain where the dynamics of the end nodes are represented by the Chialvo map and that of the central node by the Rulkov map. Thus, the system is a {\em Chialvo-Rulkov-Chialvo} chain (See Fig.~\ref{fig:triNode}), giving rise to a nonlinear system of six coupled equations given by
\begin{align}
\label{eq:model1}
    x_1(n+1)  &= x_1(n)^2e^{(y_1(n) - x_1(n))}+k_0+\sigma_{12}(x_2(n) - x_1(n)), \\
\label{eq:model2}
    y_1(n+1)  &= ay_1(n) - bx_1(n) + c, \\
\label{eq:model3}
    x_2(n+1) &= \frac{\alpha}{1+x_2(n)^2} + y_2(n) + \sigma_{21}(x_1(n) - x_2(n)) + \sigma_{23}(x_3(n) - x_2(n)),\\
\label{eq:model4}
    y_2(n+1) &= y_2(n) - \mu(x_2(n) - \gamma), \\
\label{eq:model5}
    x_3(n+1) &= x_3(n)^2e^{(y_3(n) - x_3(n))}+k_0+\sigma_{32}(x_2(n) - x_3(n)), \\
\label{eq:model6}
    y_3(n+1) &= ay_3(n) - bx_3(n) + c.
\end{align}

Let the dynamical variable set for~\eqref{eq:model1}--\eqref{eq:model6} is given by
\begin{align}
    X(n) = \left\{x_1(n), y_1(n), x_2(n), y_2(n), x_3(n), y_3(n)\right\}.
\end{align}
The indices $1$ and $3$ represent the oscillators following the Chialvo map and the index $2$ represents an oscillator following the Rulkov map. The linear coupling strengths between the oscillators are represented by $\sigma_{ij}$. Here $\sigma_{ij} \ne \sigma_{ji}$, that is the coupling between two oscillators is bidirectional. Note that the system is asymmetric because it changes its form under the transformation $X \to -X$. The dynamics of the map are realized on the oscillators, however, the coupling links are static. Before delving into the deeper dynamics of the network, it would be interesting to look into the topological properties in simple mathematical terms. One such tool is the {\em adjacency matrix}. The adjacency matrix of this small network given in Fig.~\ref{fig:triNode} is given by
\begin{align}
\mathcal{A} = \begin{bmatrix}
    0 & \sigma_{12} & 0\\
    \sigma_{21} & 0 & \sigma_{23}\\
    0 & \sigma_{32} & 0
\end{bmatrix}.
\end{align}
with the spectrum of the graph given as 
$$
\Lambda = {\rm \texttt{sort}}(0, \sqrt{\sigma_{12}\sigma_{21} + \sigma_{23}\sigma_{32}}, -\sqrt{\sigma_{12}\sigma_{21} + \sigma_{23}\sigma_{32}}),
$$
where the function \texttt{sort} sorts an array in ascending order. Here $\Lambda$ is the set of eigenvalues of the adjacency matrix arranged in an ascending order. The first and the third oscillators (following the dynamics set by the Chialvo map) have degrees $\sigma_{12}+\sigma_{21}$ and $\sigma_{23}+\sigma_{32}$ respectively, whereas the degree of the second (middle) oscillator is given by $\sigma_{12}+\sigma_{21}+\sigma_{23}+\sigma_{32}$. The network is topologically invariant when $\sigma_{12}=\sigma_{21}=\sigma_{32}=\sigma_{23}=\sigma$.

\begin{figure}[]
\begin{subfigure}{.45\linewidth}
\centering
\includegraphics[scale=0.4]{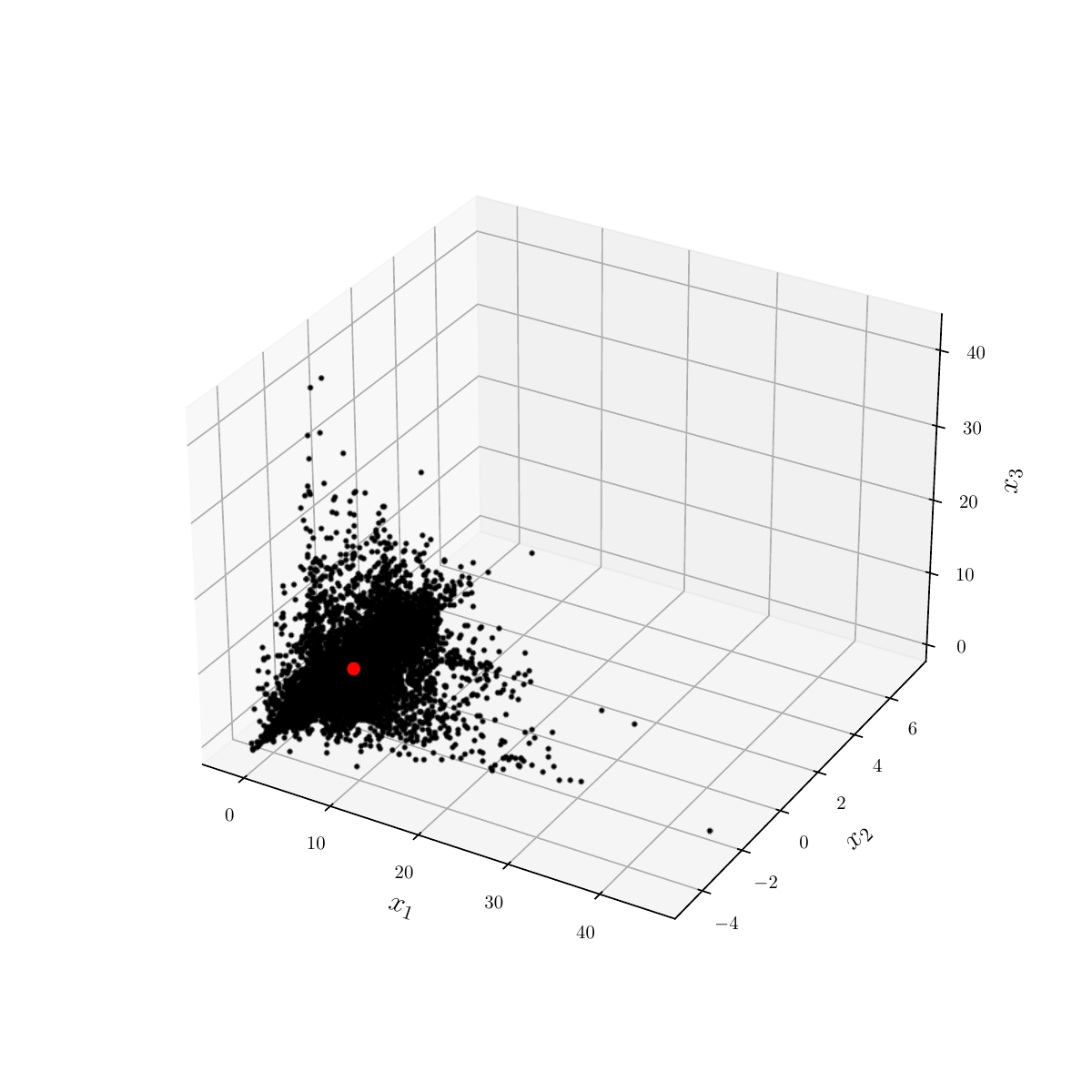}
\caption{$\sigma_{12}=-0.095$}
\end{subfigure}%
\begin{subfigure}{.45\linewidth}
\centering
\includegraphics[scale=0.4]{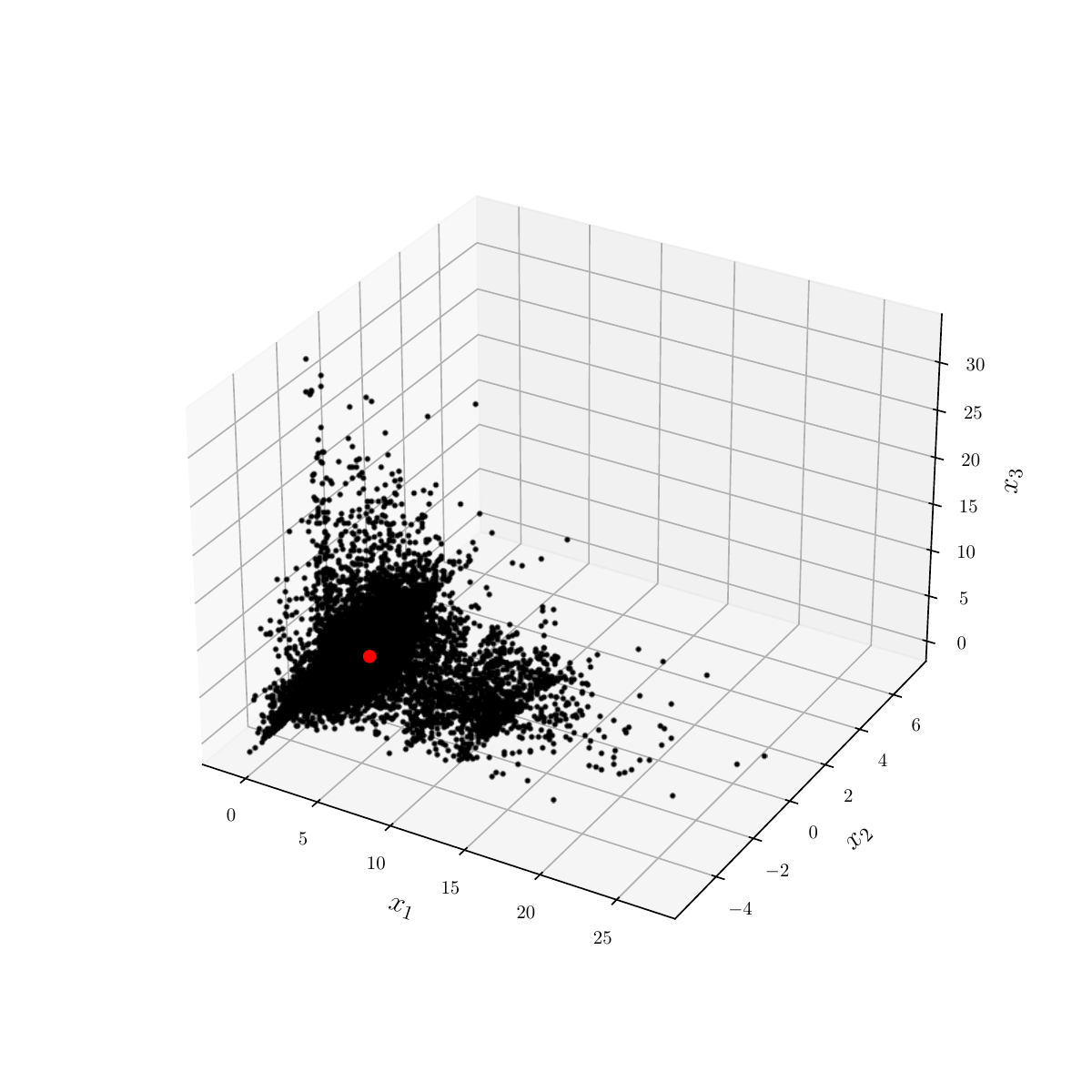}
\caption{$\sigma_{12}=0$}
\end{subfigure}\\[1ex]
\begin{subfigure}{1\linewidth}
\centering
\includegraphics[scale=0.4]{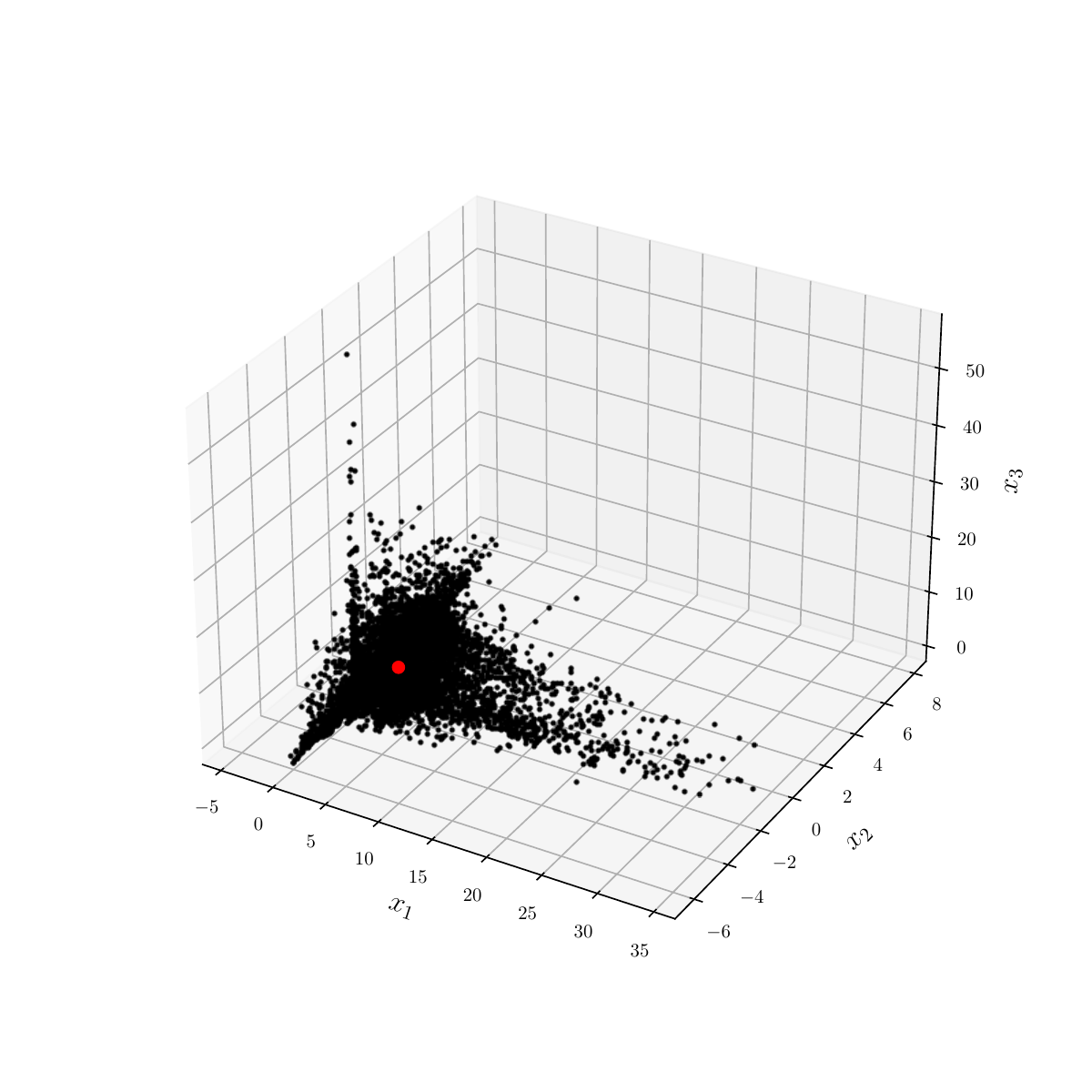}
\caption{$\sigma_{12}=0.085$}
\end{subfigure}
\caption{Phase portraits of~\eqref{eq:model1}--\eqref{eq:model6} (last $60000$ points of the simulation are plotted) with changing $\sigma_{12}$. Other coupling strengths are $\sigma_{21}=0.2$, $\sigma_{23}=0.085$, and $\sigma_{32}=-0.08$. Local parameters are $a=0.6$, $b = 0.6$, $c = 0.89$, $k_0=-1$,  $\alpha = 5$, $\mu = 0.0001$, and $\gamma = -0.5$. The red dot in each panel represents the point $(x_1^*, x_2^*, x_3^*)$ reduced from $X^*$.}
\label{fig:pp}
\end{figure}

Typical phase portraits of the system~\eqref{eq:model1}--\eqref{eq:model6} showing chaotic attractors under the varying coupling strength $\sigma_{12}$ are illustrated in Fig.~\ref{fig:pp}. The local parameters of the oscillators are fixed as $a=0.6$, $b = 0.6$, $c = 0.89$, $k_0=-1$,  $\alpha = 5$, $\mu = 0.0001$, and $\gamma = -0.5$. The other coupling strengths are set as $\sigma_{21}=0.2$, $\sigma_{23}=0.085$, and $\sigma_{32}=-0.08$. The simulation is run for $80000$ iterates out of which the last $60000$ are shown to make sure no transients creep in. All the dynamical variables are randomly initialized from the uniform distribution $[0.2, 0.3]$. We are going to maintain these initial conditions for all our numerical simulations throughout the paper until specified otherwise.

\section{Fixed point analysis of the network}
\label{sec:fixedPoints}
Analyzing the fixed points of a system (continuous or discrete time) is the first step toward unfolding its complex dynamics. The fixed point of a map-based model $f(\mathbf{x})$ is the point $\mathbf{x}^*\in \mathbb{R}^N$, such that $f(\mathbf{x}^*)=\mathbf{x}^*$. Let the fixed point of the system~\eqref{eq:model1}--\eqref{eq:model6} be given by 
\begin{align}
\label{eq:fp}
X^* = \left(x_1^*, y_1^*, x_2^*, y_2^*, x_3^*, y_3^* \right).
\end{align} 
To compute $X^*$, the following list of equations needs to be solved:
\begin{align}
\label{eq:model_1}
    x_1^*  &= {x_1^*}^2e^{(y_1^* - x_1^*)}+k_0+\sigma_{12}(x_2^* - x_1^*), \\
\label{eq:model_2}
    y_1^*  &= ay_1^* - bx_1^* + c, \\
\label{eq:model_3}
    x_2^* &= \frac{\alpha}{1+{x_2^*}^2} + y_2 + \sigma_{21}(x_1^* - x_2^*) + \sigma_{23}(x_3^* - x_2^*), \\
\label{eq:model_4}
    y_2^* &= y_2^* - \mu(x_2^* - \gamma), \\
\label{eq:model_5}
    x_3^* &= {x_3^2}^*e^{(y_3^* - x_3^*)}+k_0+\sigma_{32}(x_2^* - x_3^*), \\
\label{eq:model_6}
    y_3^* &= ay_3^* - bx_3^*+ c.
\end{align}
A step towards gaining this is to try eliminating the $y_i^*$ terms such that we end up with two transcendental equations involving $x_1^*$ and $x_3^*$. For $x_2^*$, we can directly see that
\begin{align}
\label{eq:x2_star}
x_2^* = \gamma,
\end{align}
from~\eqref{eq:model_4}, which we substitute in~\eqref{eq:model_5} to have
\begin{align}
\label{eq:y3_star}
    y_3^* = x_3^* + \ln{\left\{(\sigma_{32}+1)x_3^*-k_0-\gamma \sigma_{32}\right\}} - 2\ln{(x_3^*)},
\end{align}
requiring us to solve $x_3^*$. Thus, we substitute the above two equations in~\eqref{eq:model_6} to get the following transcendental equation of $x_3^*$,
\begin{align}
\label{eq:x3_star}
(1-a+b)x_3^* +(1-a)\ln{\left\{(\sigma_{32}+1)x_3^*-k_0-\gamma \sigma_{32}\right\}} + 2(a-1)\ln{(x_3^*)} - c = 0.
\end{align}
Solving~\eqref{eq:x3_star} numerically will give us the corresponding value of $y_3^*$. Next, we substitute~\eqref{eq:x2_star} in~\eqref{eq:model_1} to have
\begin{align}
\label{eq:y1_star}
    y_1^* = x_1^* + \ln{\left\{(\sigma_{12}+1)x_1^*-k_0-\gamma \sigma_{12}\right\}} - 2\ln{(x_1^*)},
\end{align}
requiring us to solve $x_1^*$. Again, similar substitutions in~\eqref{eq:model_2} give us
\begin{align}
\label{eq:x1_star}
(1-a+b)x_1^* +(1-a)\ln{\left\{(\sigma_{12}+1)x_1^*-k_0-\gamma \sigma_{12}\right\}} + 2(a-1)\ln{(x_1^*)} - c = 0.
\end{align}
Solving~\eqref{eq:x1_star} numerically will give us the corresponding value of $y_1^*$. Finally, we substitute the values of $x_1^*$, $x_2^*$, and $x_3^*$ in~\eqref{eq:model_3} to have
\begin{align}
\label{eq:y2_star}
y_2^* = \gamma - \frac{\alpha}{1+\gamma^2} - \sigma_{21}(x_1^* - \gamma) - \sigma_{23}(x_3^* - \gamma).
\end{align}
We need to numerically compute $x_1^*$ and $x_3^*$ using any standard computational solvers, for example, \texttt{fsolve()} function from the \texttt{scipy.optimize} module. Once this is achieved, we substitute the values of $x_i^*$'s in the $y_i^*$'s to eventually evaluate $X^*$.

For $X^*$, the dynamics of the perturbation vector $\delta X = X - X^*$ from $X^*$ is given by
\begin{align}
    \begin{bmatrix}
        x_1(n+1) \\
        y_1(n+1) \\
        x_2(n+1) \\
        y_2(n+1) \\
        x_3(n+1) \\
        y_3(n+1) 
    \end{bmatrix} = 
    \mathcal{J}.\begin{bmatrix}
        x_1(n) \\
        y_1(n) \\
        x_2(n) \\
        y_2(n) \\
        x_3(n) \\
        y_3(n),
    \end{bmatrix}
\end{align}
where $X_i = \left(x_1, y_1, x_2, y_2, x_3, y_3 \right)$ and $\mathcal{J}$ is the Jacobian of the system,
\begin{align}
\mathcal{J} = \begin{bmatrix}
    x_1(2-x_1)e^{y_1-x_1}-\sigma_{12} & x_1^2e^{y_1-x_1} & \sigma_{12}& 0& 0& 0 \\
    -b &a &0 &0 &0 &0 \\
    -\sigma_{21}& 0& -\frac{2\alpha x_2}{(1+x_2^2)^2}+\sigma_{21}-\sigma_{23} & 1& \sigma_{23}& 0 \\
    0& 0& -\mu& 1& 0& 0\\
    0& 0& \sigma_{32}& 0 & x_3(2-x_3)e^{y_3-x_3}-\sigma_{32}& x_3^2e^{y_3-x_3}\\
    0& 0& 0& 0& -b& a
\end{bmatrix}.
\end{align}
The linear stability analysis of the fixed point depends on the absolute values of the eigenvalues of $\mathcal{J}$. The eigenvalues $\lambda_i$, $i=1, \ldots, 6$ can be evaluated from $\mathcal{J}$ at the fixed point $X^*$ by solving the equation on the sixth order polynomial $P_6(\lambda)=0$, where
\begin{align}
\label{eq:poly}
P_6(\lambda)=a_0\lambda^6+a_1\lambda^5+a_2\lambda^4+a_3\lambda^3+a_4\lambda^2+a_5\lambda+a_6.
\end{align}
Again, solving this can be achieved by using any type of available standard computational software. Note that in~\eqref{eq:poly} we have
\begin{align}
\label{eq:a0}
a_0 &= 1, \\
\label{eq:a1}
a_1 &= J_3+(\sigma_{12}-D_1-a)J_4, \\
\label{eq:a2}
a_2 &= J_2+(\sigma_{12}-D_1-a)J_3 + (D_1a-\sigma_{12}a+D_2b)J_4+L_4, \\
\label{eq:a3}
a_3 &= J_1+(\sigma_{12}-D_1-a)J_2 + (D_1a-\sigma_{12}a+D_2b)J_3+L_3,\\
\label{eq:a4}
a_4 &= J_0+(\sigma_{12}-D_1-a)J_1 + (D_1a-\sigma_{12}a+D_2b)J_2+L_2,\\
\label{eq:a5}
a_5 &= (\sigma_{12}-D_1-a)J_0 + (D_1a-\sigma_{12}a+D_2b)J_1+L_1,\\
\label{eq:a6}
a_6 &=(D_1a-\sigma_{12}a+D_2b)J_0+L_0.
\end{align}
We observe a clear pattern in the equations of $a_0$--$a_6$. Furthermore, the terms $J_i$ are defined as
\begin{align}
\label{eq:J0}
J_0 &= (\sigma_{21}-\sigma_{23}+D_3+\mu)(D_4a-\sigma_{32}a+D_5b)-\sigma_{23}\sigma_{32}b, \\
\label{eq:J1}
J_1 &= (\sigma_{21}-\sigma_{23}+D_3+\mu)(\sigma_{23}+\sigma_{32}-\sigma_{21}-D_3-D_4-a-1) + \sigma_{23}\sigma_{32}b,\\
\label{eq:J2}
J_2 &= (D_4 - \sigma_{32})a +D_5b +(\sigma_{23}-\sigma_{21}-D_3-1)(\sigma_{32}-a-D_4) + \sigma_{21}-\sigma_{23}+D_3+\mu, \\
\label{eq:J3}
J_3 &= \sigma_{23}+\sigma_{32}-\sigma_{21}-D_4-D_3-a-1, \\
\label{eq:J4}
J_4 &= 1,
\end{align}
and the terms $L_i$ are defined as
\begin{align}
\label{eq:L0}
L_0 &= a(D_4a-\sigma_{32}a+D_5b), \\
\label{eq:L1}
L_1 &= a(\sigma_{32}-a-D_4) - (a+1)(D_4a - \sigma_{32}a +D_5b), \\
\label{eq:L2}
L_2 &=  a(D_4-\sigma_{32}+1) - (a+1)(\sigma_{32}-a-D_4) +D_5b, \\
\label{eq:L3}
L_3 &= \sigma_{32}-2a-D_4-1, \\
\label{eq:L4}
L_4 &= 1.
\end{align}
Note that the terms $D_i$ are defined in terms of $x_j^*$ as,
\begin{align}
\label{eq:Ds}
   D_1 &= x_1^*(2-x_1^*)e^{y_1^*-x_1^*}, \\
D_2 &= {x_1^*}^2e^{y_1^*-x_1^*},\\
D_3 &= -\frac{2\alpha x_2^*}{(1+{x_2^*}^2)^2},\\
D_4 &= x_3^*(2-x_3^*)e^{y_3^*-x_3^*},\\
D_5 &= {x_3^*}^2e^{y_3^*-x_3^*}. 
\end{align}

\begin{figure}
\centering
\includegraphics[width=0.8\linewidth]{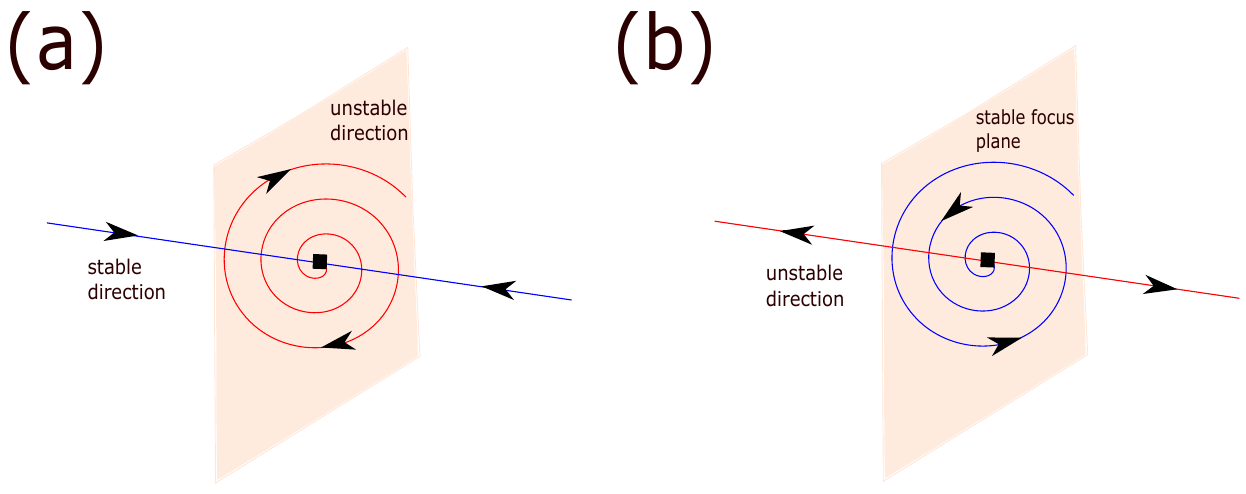}
\caption{Schematic representation of the dynamics at the saddle fixed point (denoted by black square). The red curves denote the unstable manifolds and the blue curves denote the stable manifolds. In (a), we observe a saddle-focus dynamics and in (b), we observe a stable focus dynamics.}\label{fig:SaddStab_Dir}
\end{figure}

To check the stability of $X^*$ we have to monitor the signs of the eigenvalues $\lambda_i$, $i=1, \ldots, 6$. If all $|\lambda_i|<1$, then $X^*$ is locally {\em stable}, if all $|\lambda_i|>1$, then $X^*$ is locally {\em unstable}, otherwise the fixed point is a $k-$saddle where $k$ is the number of eigenvalues whose absolute value is $>1$.\par
\begin{figure}[h]
\begin{tabular}{cc}
  \includegraphics[scale=0.5]{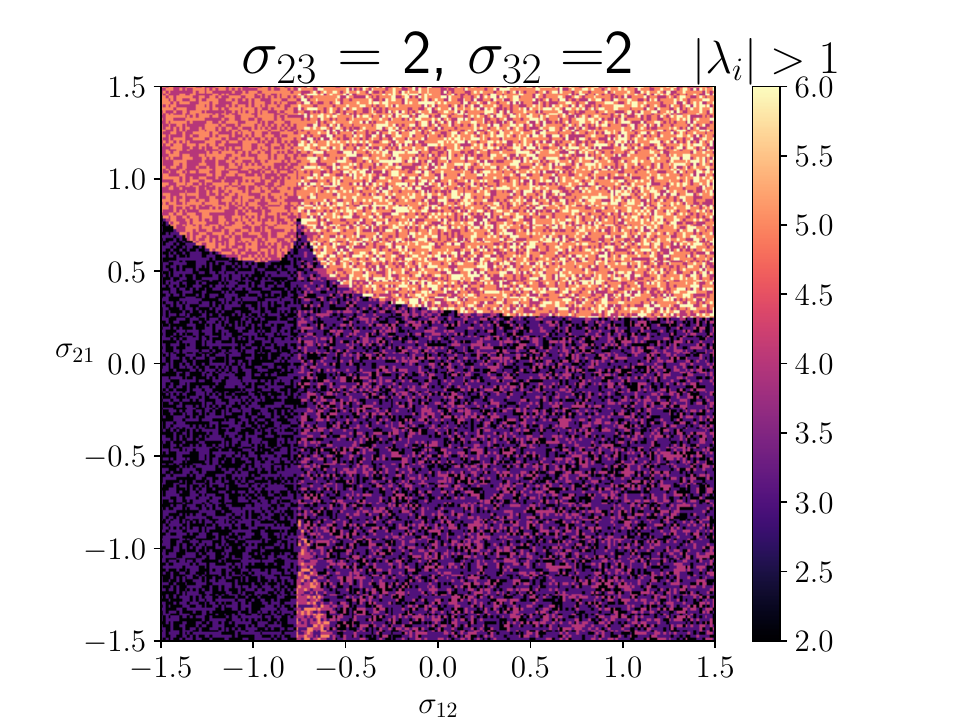} &   \includegraphics[scale=0.5]{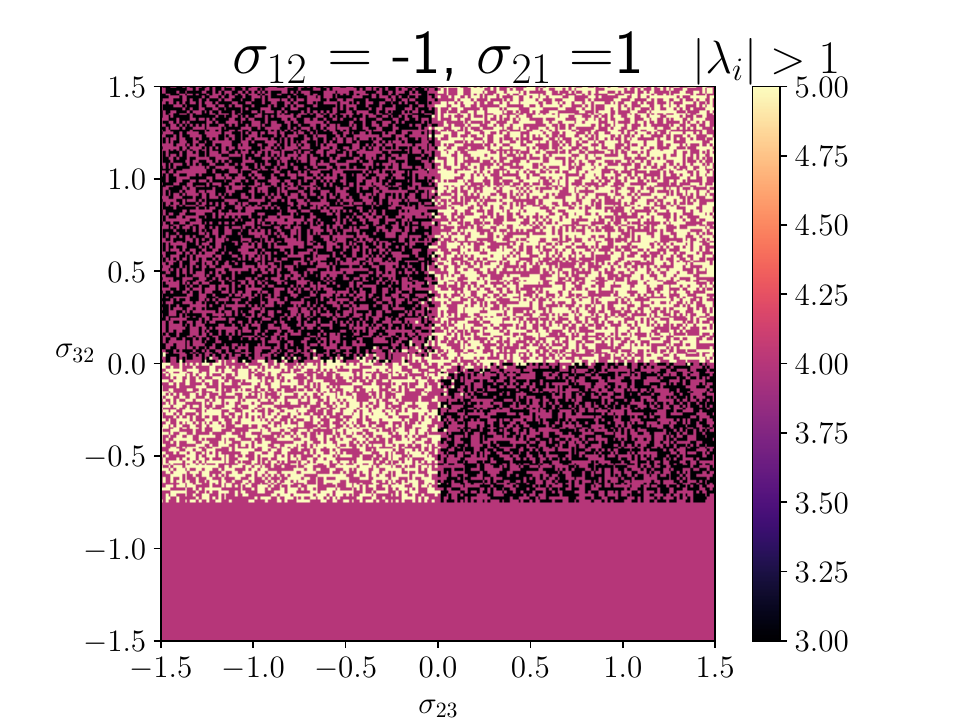} \\
(a) $\sigma_{12}$ vs $\sigma_{21}$ & (b) $\sigma_{23}$ vs $\sigma_{32}$ \\[3pt]
\end{tabular}
\caption{
Two-dimensional color-coded stability region plots. Both panels show $k-$saddle type and unstable fixed points. Panel (a) shows a $(\sigma_{12}, \sigma_{21})$ plane with $\sigma_{23}=\sigma_{32}=2$ and $\mu=0.01$ and panel (b) shows a $(\sigma_{23}, \sigma_{32})$ plane with $\sigma_{12}=-1$, $\sigma_{21}=1$, and $\mu=0.0001$. Other parameters are kept as $a=0.6$, $b=0.6$, $c=0.89$, $k_0=-1$, $\alpha=5$, $\mu=0.01$, and $\gamma=-0.5$. For the saddle type fixed points, $k$ varies from $2$ to $5$. Interestingly no stable fixed points were found in the reported region of $\sigma_{ij}$ values.
}
\label{fig:stab}
\end{figure}

For example, in Table~\ref{tab:stab} we list the set of six eigenvalues $\lambda_i$ on the parameter grid $(\sigma_{12}, \sigma_{21})$ keeping fixed $\sigma_{23}=\sigma_{32}=2$ and $a=0.6, b=0.6, c=0.89, k_0=-1, \alpha=5, \mu=0.01$ and $\gamma=-0.5$. The table lists fixed points of $k-$saddle types for $k=2,\ldots, 5$ and an unstable fixed point. Note that Fig.~\ref{fig:stab} is a two-parameter bifurcation diagram where we color code the grid according to the type of fixed point generated by the system at that specific parameter combination. Fig.~\ref{fig:stab}-(a) shows the color-coded plot on the $(\sigma_{12}, \sigma_{21})$ plane for $\sigma_{23}=2, \sigma_{32}=2$ and $a=0.6$, $b=0.6$, $c=0.89$, $k_0=-1$, $\alpha=5$, $\mu=0.01$, and $\gamma=-0.5$. Note that there exist distinct bifurcation boundaries where the type of the saddle fixed point changes from having $k=2,3,4$ to having $k=4,5,6$ or $k=2,3$ or $k=4,5$. Fig.~\ref{fig:stab}-(a) shows a similar plot but on the parameter plane $(\sigma_{23}, \sigma_{32})$ with $\sigma_{12}=-1, \sigma_{21}=1$ and the other local parameters kept the same except $\mu$ set to $0.0001$. We observe a region where the fixed point is a $4$-saddle and has a boundary with two types of regions where the $k$ values are either $3,4$ or $4,5$. The initial condition for all the dynamical variables is sampled randomly from the uniform distribution $(0.2, 0.3)$.

\begin{table}
    \centering
    \begin{tabular}{|c|c|c|c|c|}
    \hline
         $(\sigma_{12}, \sigma_{21})$& $\lambda_i$ & $\left|\lambda_i \right|$ & $\left|\lambda_i \right|>1$ & Type \\
         \hline \hline
         $(-1, -1.5)$ & $6.91009635$ & $6.91009635$ & T & $2-$saddle \\
         & $2.73275416$ & $2.73275416$ & T & \\
         & $-0.3225025$ & $0.3225025$ & F&\\
         & $0.77122194+0.13862423i$ & $0.78358149$ & F& \\
         & $0.77122194-0.13862423i$ & $0.78358149$ & F& \\
         & $0.99997521$ & $0.99997521$ & F& \\
         \hline
         
         $(-1, -1)$ & $6.94598203$ & $6.94598203$ & T & $3-$saddle \\
         & $2.83166417$ & $2.83166417$ & T & \\
         & $0.13290233$ & $0.13290233$ & F&\\
         & $0.72064294+0.15653426i$ & $0.73744778$ & F& \\
         & $0.72064294-0.15653426i$ & $0.73744778$ & F& \\
         & $1.00001025$ & $1.00001025$ & T& \\
         \hline
         
         $(0.2, 0.75)$ & $7.16390981$ & $7.16390981$ & T & $4-$saddle \\
         & $3.33704297$ & $3.33704297$ & T & \\
         & $1.09923821+0.49181717i$ & $1.20424614$ & T&\\
         & $1.09923821-0.49181717i$ & $1.20424614$ & T& \\
         & $0.99980818$ & $0.99980818$ & F& \\
         & $0.99993496$ & $0.99993496$ & F& \\
         \hline

         $(0.3, 0.9)$ & $7.15754433$ & $7.15754433$ & T & $5-$saddle \\
         & $3.46098435$ & $3.46098435$ & T & \\
         & $1.18869024+0.45586832i$ & $1.27310659$ & T&\\
         & $1.18869024-0.45586832i$ & $1.27310659$ & T& \\
         & $0.99861359$ & $0.99861359$ & F& \\
         & $1.00003121$ & $1.00003121$ & T& \\
         \hline

         $(0.84, 0.5)$ & $7.09290783$ & $7.09290783$ & T & unstable \\
         & $4.33547605$ & $4.33547605$ & T & \\
         & $1.0154801 +0.4854296i$ & $1.12554064$ & T&\\
         & $1.0154801 -0.4854296i$ & $1.12554064$ & T& \\
         & $1.00191419$ & $1.00191419$ & T& \\
         & $1.00004641$ & $1.00004641$ & T& \\
         \hline
    \end{tabular}
    \caption{Eigenvalues associated with the stability analysis of the fixed points at the parameter point $(\sigma_{12}, \sigma_{21})$ with the fixed coupling strengths $\sigma_{23}=2$, and $\sigma_{32}=2$. The other parameters of the system are set as $a=0.6$, $b=0.6$, $c=0.89$, $k_0 = -1$, $\alpha = 5$, $\mu=0.01$, and$\gamma=-0.5$.}
    \label{tab:stab}
\end{table}
From Table \ref{tab:stab}, we can understand the dimensions of the stable and unstable manifolds around the saddle fixed point. Even though this fixed point exists in a six-dimensional space, it's important to consider simpler one-dimensional manifolds and spiral motions to grasp the overall dynamics near it. For instance, when $(\sigma_{12}, \sigma_{21}) = (0.3, 0.9)$, the fixed point has a one-dimensional manifold spiraling expanding outward. This spiral motion occurs within a two-dimensional plane within the six-dimensional space. Additionally, there's a one-dimensional stable manifold perpendicular to this plane, as shown in Fig. \ref{fig:SaddStab_Dir} (a). Similarly, in Fig. \ref{fig:SaddStab_Dir} (b), there's a one-dimensional stable manifold spirally contracting towards the fixed point, with a perpendicular one-dimensional unstable manifold.

\begin{figure}
\centering
\includegraphics[width=0.6\linewidth]{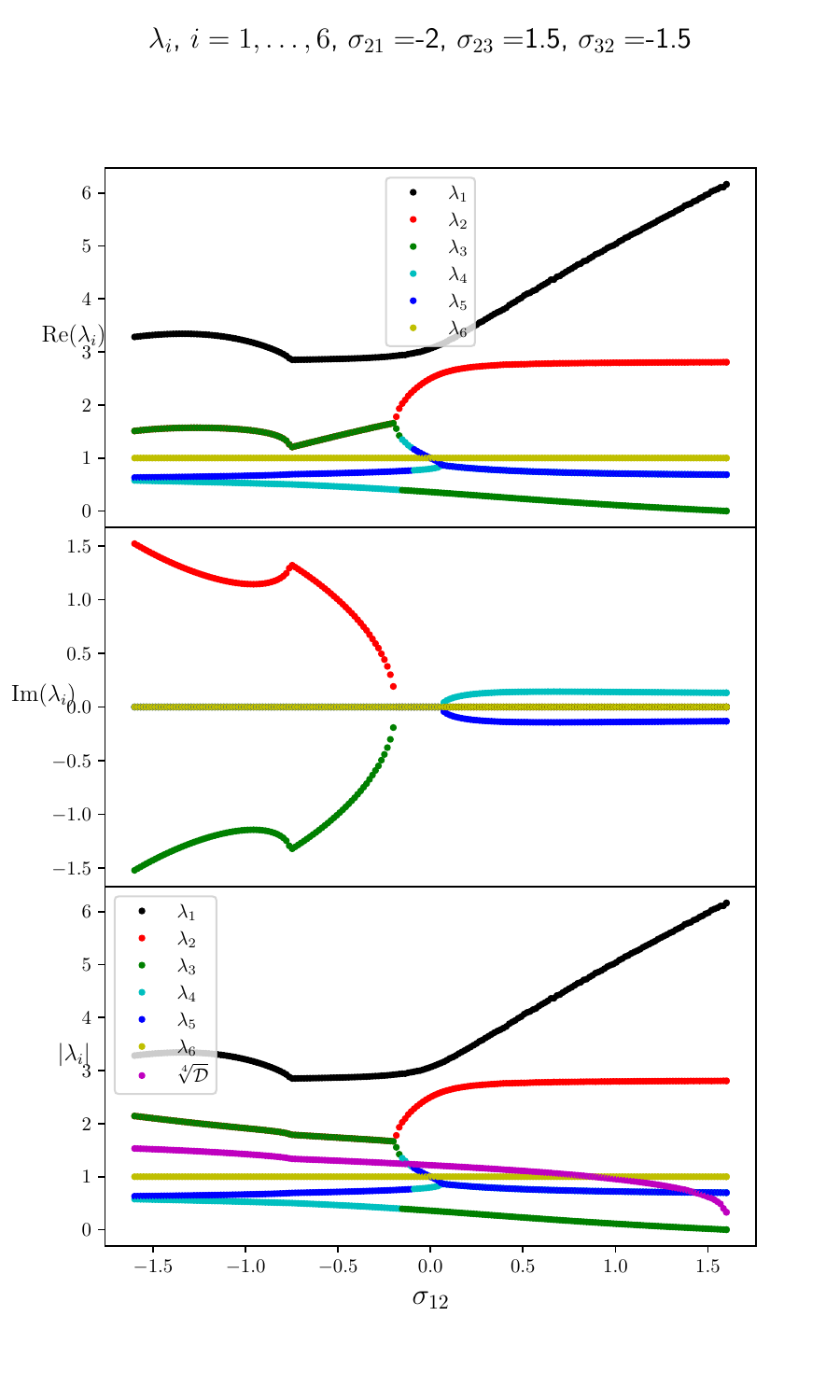}
\caption{Solutions of $P_6(\lambda)$~\eqref{eq:poly} with varying $\sigma_{12}$. The local parameters of the oscillators are $a=0.6$, $b = 0.6$, $c = 0.89$, $k_0=-1$, $\alpha = 5$, $\mu = 0.0001$, and $\gamma = -0.5$. The other coupling strengths are $\sigma_{21}=-2$, $\sigma_{23}=1.5$, and $\sigma_{32}=-1.5$. The six eigenvalues are colored according to the legends mentioned in the plot. The three subplots show the real parts, the imaginary parts, and the absolute values of the eigenvalues. $\lambda_1$ is always real, whereas $\lambda_i$ for $i=2, \ldots, 5$ can be both real and complex-valued. Finally, $\lambda_6$ is always real and $\approx 1$. }\label{fig:eig}
\end{figure}

Note that keeping track of the eigenvalues of the characteristic equation $P_6(\lambda)$ gives us an analytical intuition on different bifurcation patterns that might arise in the dynamics of~\eqref{eq:model1}--\eqref{eq:model6}, see Fig.~\ref{fig:eig}. Here we have plotted the behavior of $\lambda_i$ with the variation of $\sigma_{12}$. The local parameters of the oscillators are $a=0.6$, $b = 0.6$, $c = 0.89$, $k_0=-1$, $\alpha = 5$, $\mu = 0.0001$, and $\gamma = -0.5$. The other coupling strengths are $\sigma_{21}=-2$, $\sigma_{23}=1.5$, and $\sigma_{32}=-1.5$. We have colored the points according to the index of the eigenvalues. Some important observations are noted here. We see that $\lambda_1$ is always real, whereas $\lambda_i$ for $i=2, \ldots, 5$ can be both real and complex-valued. Finally, $\lambda_6$ is always real and $\approx 1$. As soon as the absolute value of one of the eigenvalues crosses $1$, the system will give rise to either a saddle-node (fold) bifurcation or a period-doubling (flip) bifurcation. When a complex eigenvalue has modulus $1$, the system gives rise to a Neimark-Sacker bifurcation. More on these are detailed in \S\ref{sec:bif2}.

\section{Noninvertibility criterion}
\label{sec:noninv}
\begin{figure}[h]
\begin{tabular}{cc}
  \includegraphics[scale=0.5]{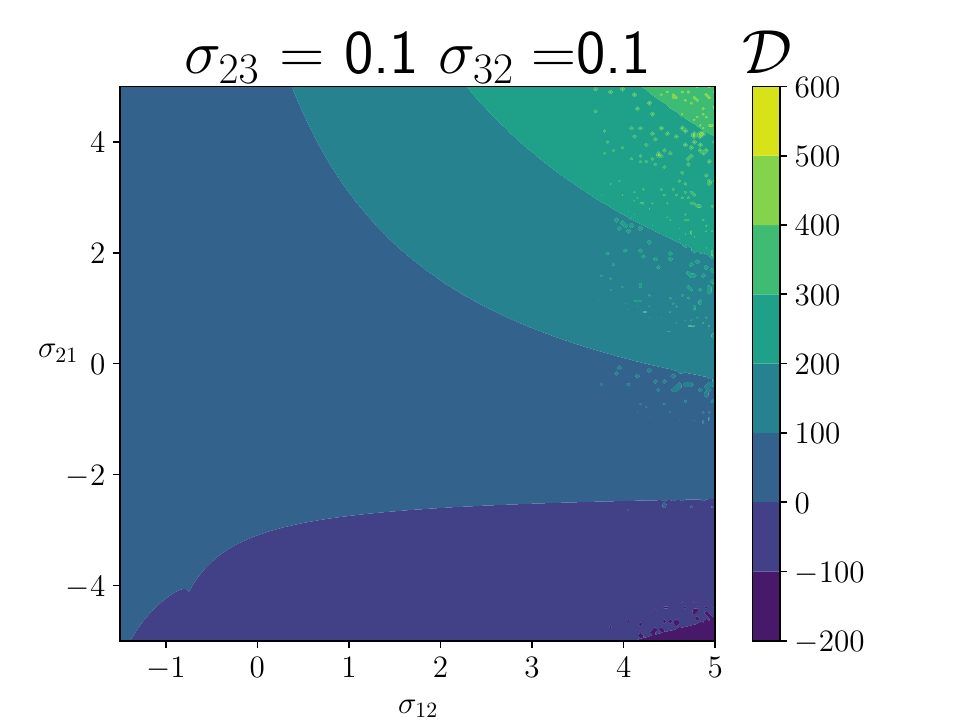} &   \includegraphics[scale=0.5]{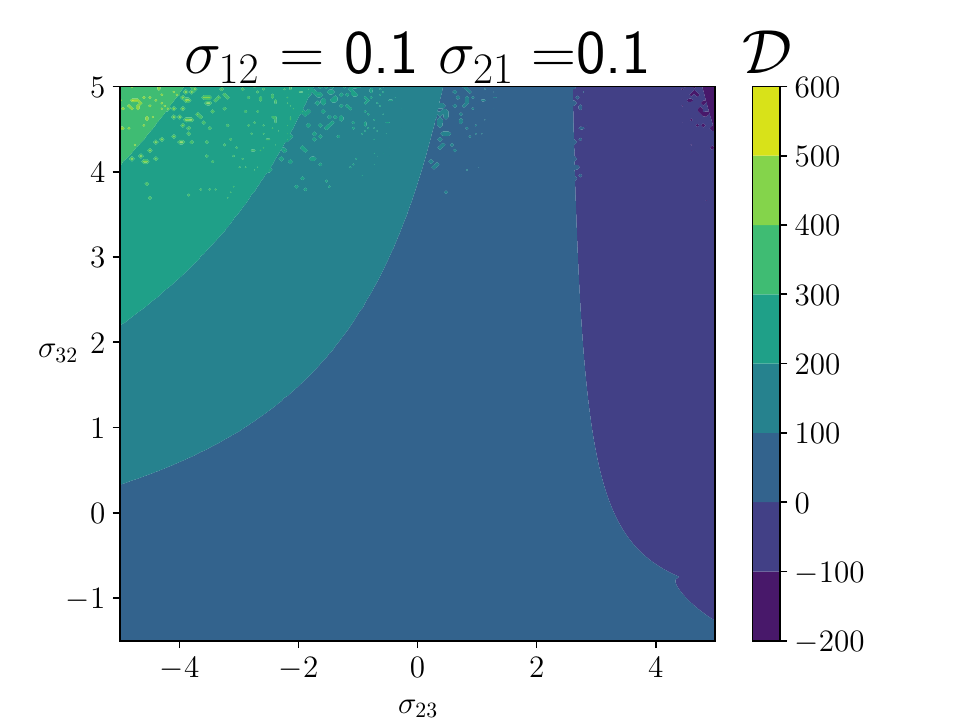} \\
(a) $\sigma_{12}$ vs $\sigma_{21}$ & (b) $\sigma_{23}$ vs $\sigma_{32}$ \\[3pt]
\end{tabular}
\caption{
Two-dimensional color-coded plots for the determinant $\mathcal{D}$ of the Jacobian $\mathcal{J}$ at $X^*$. Panel (a) shows a $(\sigma_{12}, \sigma_{21})$ plane with $\sigma_{23}=\sigma_{32}=0.1$ and panel (b) shows a $(\sigma_{23}, \sigma_{32})$ plane with $\sigma_{12}=\sigma_{21}=0.1$. The local parameters for both the panels are set to be $a=0.6, b=0.6, c=0.89, k_0=-1, \alpha=5, \mu=0.0001$, and $\gamma=-0.5$. The system is noninvertible at $\mathcal{D}=0$.
}
\label{fig:noninv}
\end{figure}

Establishing the {\em noninvertibility} criterion of a map provides useful intricate details about its core dynamics. Noninvertibility in map-based systems has been extensively studied by Mira {\em et al.}\cite{MiGa96, MiCa96}. This property tells us about the stretching and folding behaviors of the concerned map. Depending on whether the map is one-, two-, or three-dimensional, there exists a critical line, curve, or surface that separates the phase plane into domains consisting of distinct preimages. The noninvertibility of the Chialvo neuron model with and without the application of memristive electromagnetic flux has been touched upon by Muni {\em et al.}\cite{MuFa22}. Note that in that work the system was either two- or three-dimensional. In this paper, the system is six-dimensional, and hence the analysis becomes exorbitantly complex. Thus, we just come up with the critical curve ${\rm C}_{-1}$ where the preimages merge and is where the determinant $\mathcal{D}$ of the Jacobian matrix is $0$. This is given by
\begin{align}
\label{eq:LC}
{\rm C}_{-1}= \left\{X(n) \middle|\ \mathcal{D} = 0  \right\},
\end{align}
where,
\begin{align}
\label{eq:detJ}
\mathcal{D} = {\rm det}(\mathcal{J})&= \left[(H_1a +H_2b)(H_3+\sigma_{21}-\sigma_{23}+\mu) - \sigma_{12}a(H_3 - \sigma_{23}+\mu) \right] \nonumber \\
&\times \left[(H_4-\sigma_{32})a +H_5b \right] - \sigma_{23}\sigma_{32}b\left[(H_1-\sigma_{12})a + H_2b \right],
\end{align}
where
\begin{align}
\label{eq:Hs}
   H_1 &= x_1(2-x_1)e^{y_1-x_1}, \\
H_2 &= {x_1}^2e^{y_1-x_1},\\
H_3 &= -\frac{2\alpha x_2}{(1+x_2^2)^2},\\
H_4 &= x_3(2-x_3)e^{y_3-x_3},\\
H_5 &= {x_3}^2e^{y_3-x_3}. 
\end{align}
A pair of two-parameter contour plots illustrating the value of $\mathcal{D}$ at the fixed point $X^*$ of the system is shown in Fig.~\ref{fig:noninv}. Panel (a) shows the $(\sigma_{12}, \sigma_{21})$ plane, and panel (b) shows the $(\sigma_{23}, \sigma_{32})$ plane. The coupling strength in panel (a) is set to be $\sigma_{23}=0.1, \sigma_{32}=0.1$, whereas in (b) is set to be $\sigma_{12}=0.1, \sigma_{21}=0.1$. The local parameters for both the panels are set to be $a=0.6, b=0.6, c=0.89, k_0=-1, \alpha=5, \mu=0.0001$, and $\gamma=-0.5$. The system is invertible and orientation-reversing when $\mathcal{D}<0$ and is invertible and orientation-preserving when $\mathcal{D}>0$.

\section{Bifurcation structure of dynamical variables and coexistence}
\label{sec:bif1}
\begin{figure}
\begin{subfigure}{.5\linewidth}
\centering
\includegraphics[scale=0.45]{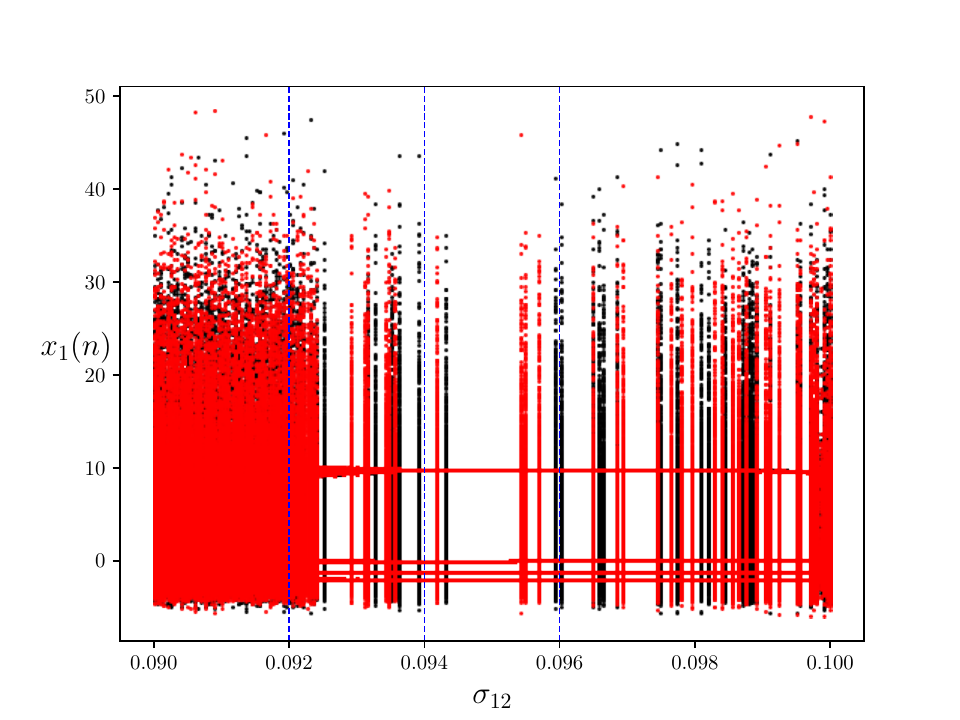}
\caption{$x_1(n)$}
\end{subfigure}%
\begin{subfigure}{.5\linewidth}
\centering
\includegraphics[scale=0.45]{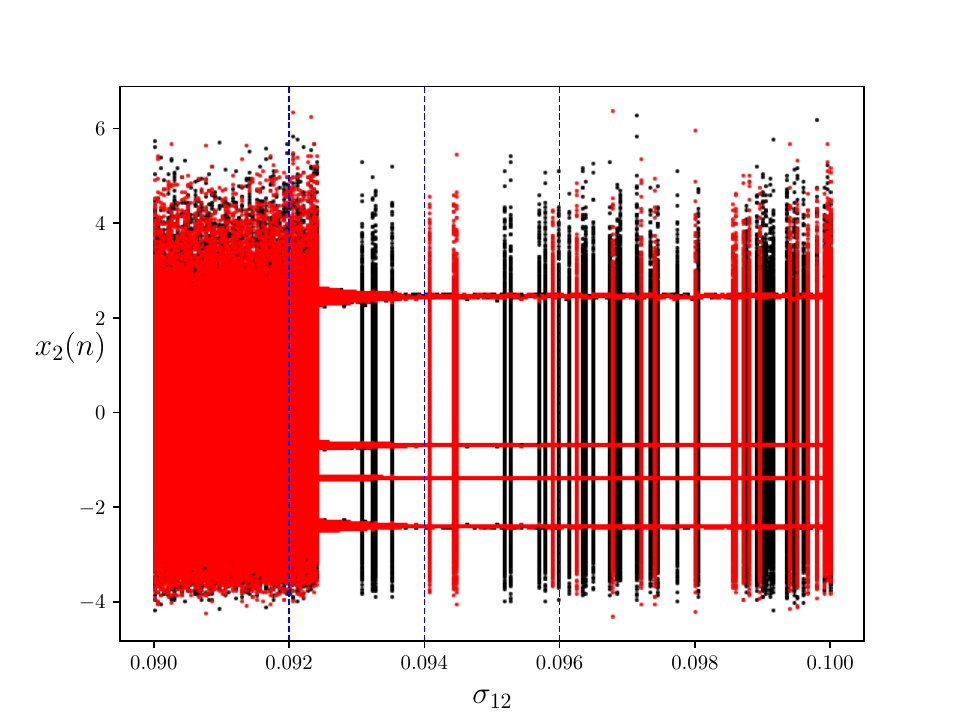}
\caption{$x_2(n)$}
\end{subfigure}\\[1ex]
\begin{subfigure}{1\linewidth}
\centering
\includegraphics[scale=0.45]{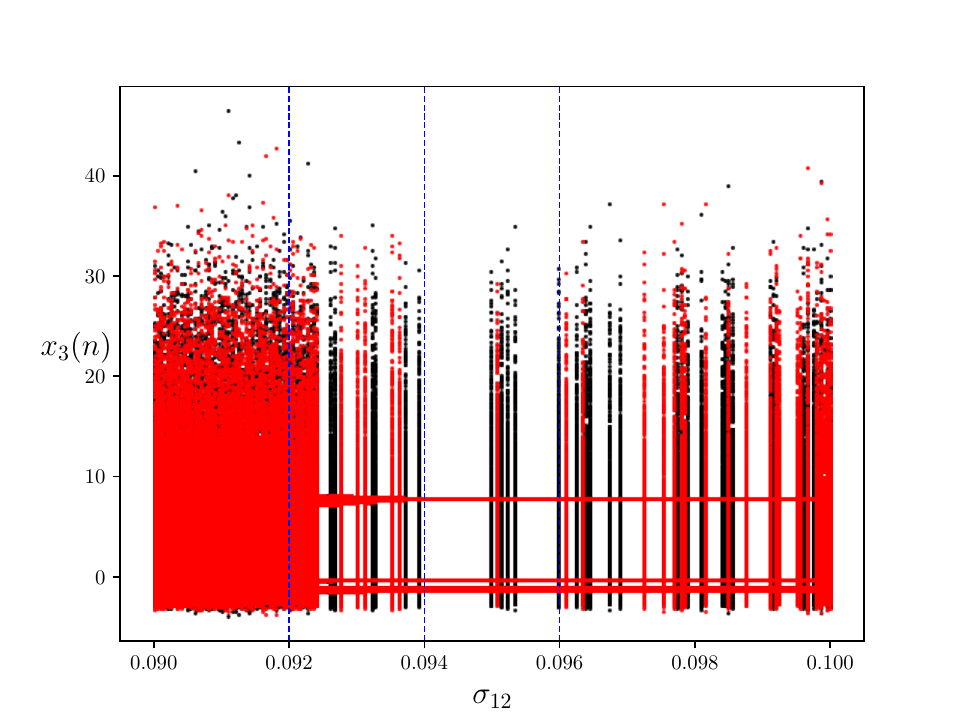}
\caption{$x_3(n)$}
\end{subfigure}
\caption{Bifurcation structures of action potentials computed via a forward (red dots) and backward (black dots) continuation, with the variation of the bifurcation parameter $\sigma_{12}$. The other coupling strengths are $\sigma_{21}=0.1$, $\sigma_{23}=0.05$, and $\sigma_{32}=0.06$. The local parameters for the oscillators are set as $a=0.6$, $b=0.6$, $c=0.89$, $k_0=-1$, $\alpha =5$, $\mu=0.0001$, and $\gamma=-0.5$. Coexistence of chaotic and period-$4$ attractors are observed as exhibited by the presence of hysteresis loops. }
\label{fig:bif_dyn}
\end{figure}

\begin{figure}
\begin{tabular}{cc}
  \includegraphics[scale=0.4]{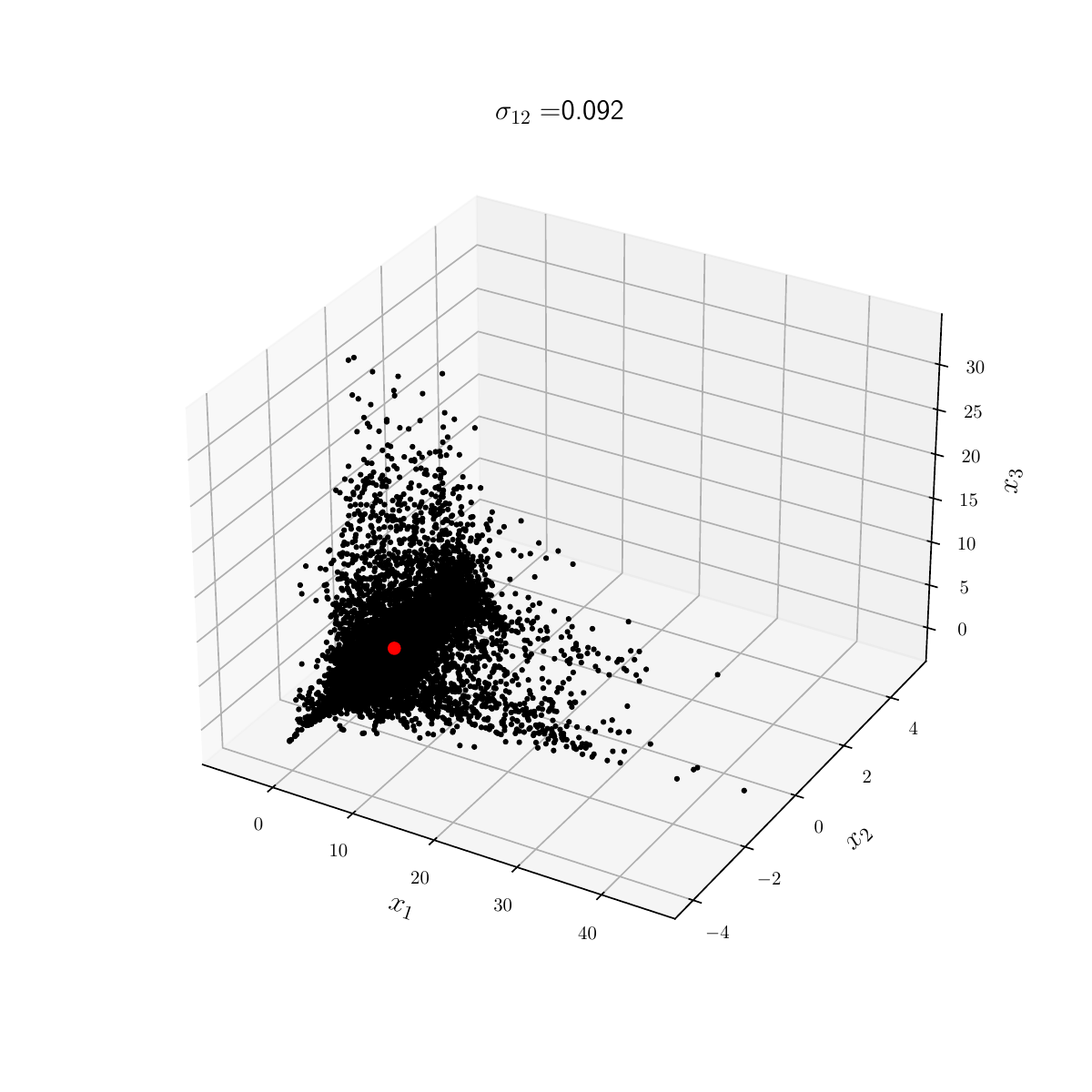} &   \includegraphics[scale=0.4]{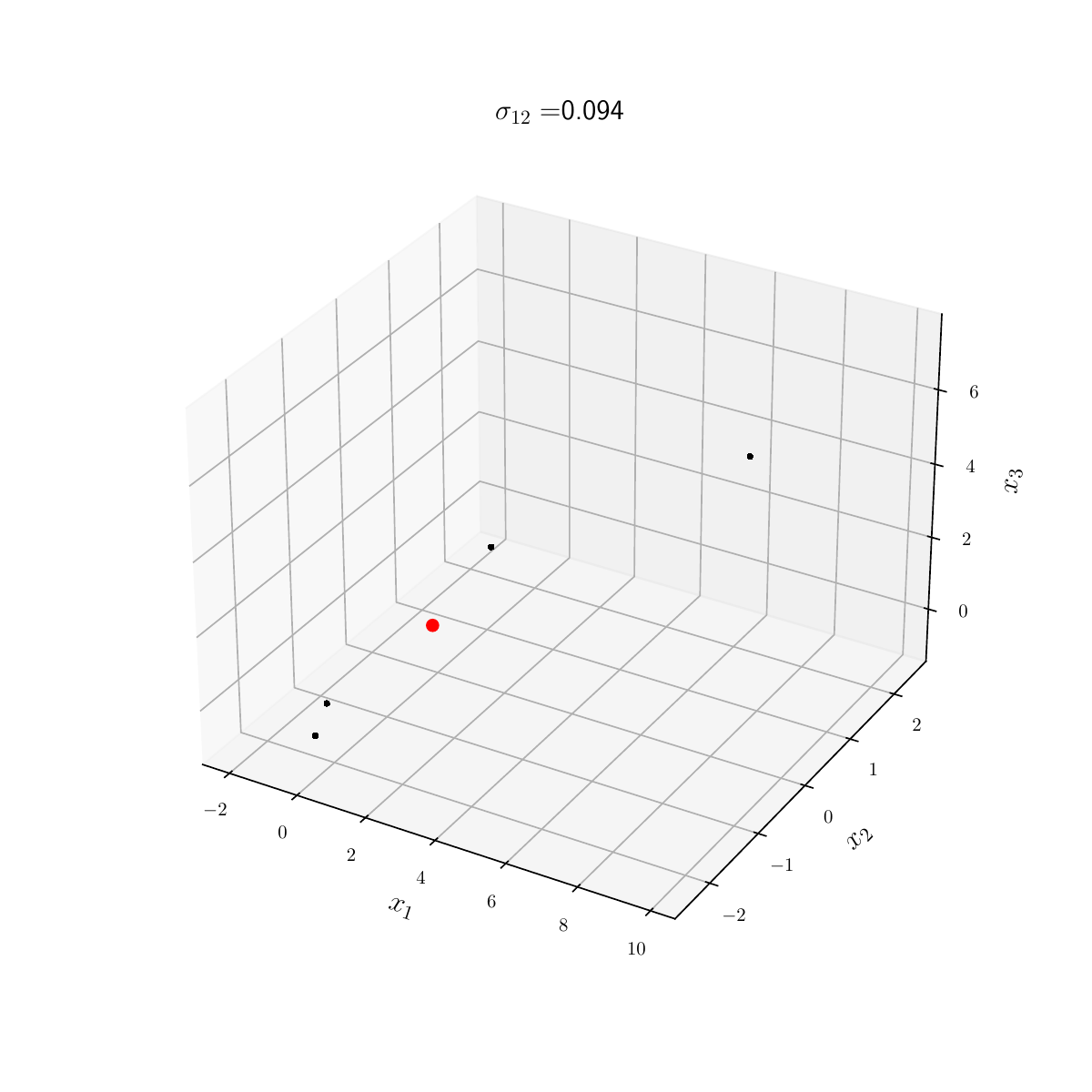} \\
(a) $\sigma_{12} =0.092$ & (b) $\sigma_{12} =0.094$  \\[3pt]
\includegraphics[scale=0.4]{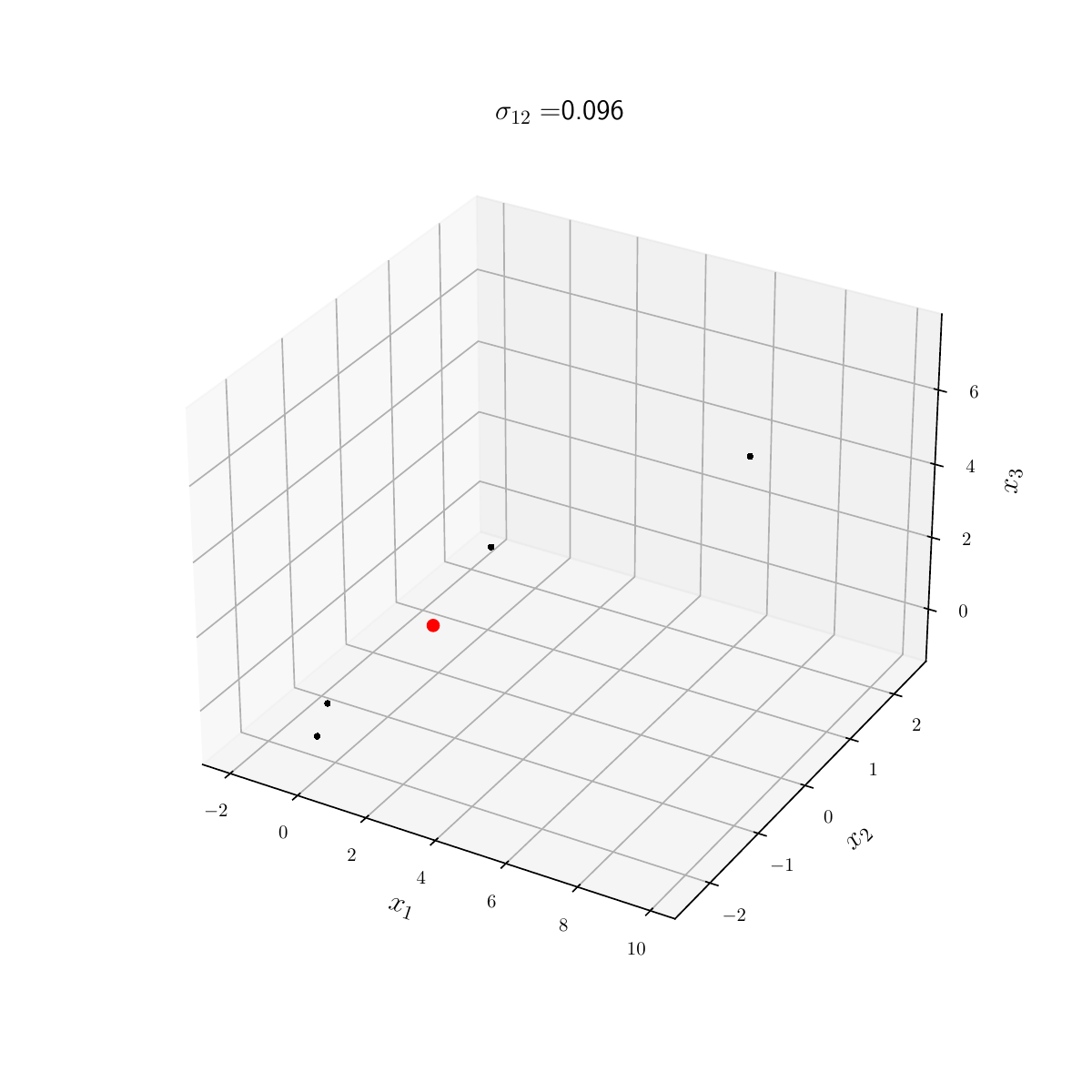} &   \includegraphics[scale=0.4]{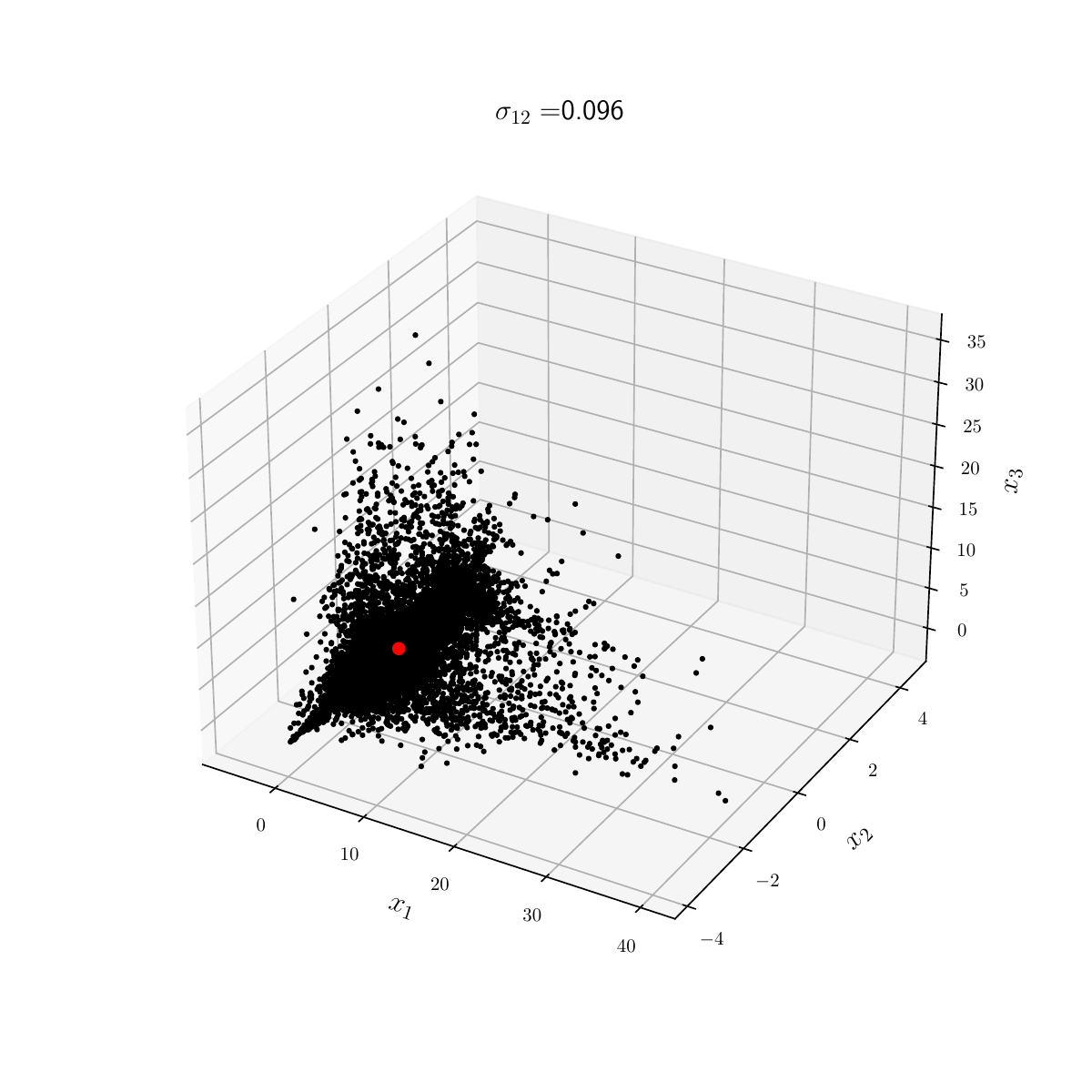} \\
(c) $\sigma_{12} =0.096$  & (d) $\sigma_{12} =0.096$ \\[3pt]
\end{tabular}
\caption{Phase portraits of~\eqref{eq:model1}--\eqref{eq:model6} with changing $\sigma_{12}$. Other coupling strengths are $\sigma_{21}=0.1$, $\sigma_{23}=0.05$, and $\sigma_{32}=0.06$. The red dot in each panel represents the point $(x_1^*, x_2^*, x_3^*)$ reduced from $X^*$. Phase portraits are drawn according to the $\sigma_{12}$ values represented by the blue broken lines in Fig.~\ref{fig:bif_dyn}. The last $60000$ iterates are plotted out of the $80000$ iterates run, to ensure the transients are discarded.}
\label{fig:pp2}
\end{figure}

In this section, we report the bifurcation structure of the action potentials to varying coupling strength, giving us an intuition on the emergence of chaotic and periodic attractors. This is achieved via both the forward continuation and backward continuation. The system is simulated for $40000$ iterations out of which the last $4500$ iterates are plotted for a specific value of the concerned parameter, see Fig.~\ref{fig:bif_dyn}. The model continuation is done within the parameter range $\sigma_{12} \in [0.09, 0.1]$ with forward continuation points marked in red and the backward continuation points in black in the same plot environment. We have set the rest of the coupling strengths as $\sigma_{21}=0.1$, $\sigma_{23}=0.05$, and $\sigma_{32}=0.06$. The local parameters for the oscillators are set as $a=0.6$, $b=0.6$, $c=0.89$, $k_0=-1$, $\alpha =5$, $\mu=0.0001$, and $\gamma=-0.5$. Plotting the points in the same environment allows us to report {\em coexistence} of the periodic and the chaotic attractors. At $\sigma_{12}=0.092$, we see that both forward and backward continuation indicate the existence of chaotic attractor, whereas, at $\sigma_{12}=0.094$, they indicate the existence of period-$4$ attractor. However, at $\sigma_{12}=0.096$, the forward continuation indicates the existence of a chaotic attractor, and the backward continuation indicates the existence of a period-$4$ solution. The forward and the backward continuation points not overlapping generates a hysteresis loop indicating the coexistence of the chaotic and periodic solutions. These are also supported by the phase portrait plots given in Fig.~\ref{fig:pp2} where out of $80000$ iterates the last $60000$ points are plotted to ensure that the transients are discarded.

\section{Codimension-$1$ and -$2$ bifurcation patterns}
\label{sec:bif2}
To investigate what type of complex dynamics our network is capable of, it requires to be studied in terms of sophisticated dynamical tools like codimension-$1$ and -$2$ bifurcation analysis. These give the readers an overall picture of the influence of relevant parameters on the behavior of the network. We have considered the coupling strengths $\sigma_{ij}$ as the main bifurcation parameters and have kept the local parameters of the oscillators fixed $a=0.6, b=0.6, c=0.89, k_{0}=-1, \alpha=5, \mu=0.0001$ and $\gamma=-0.5$. At first, we put forward three theorems corresponding to codimension-$1$ bifurcations that arise in our map when one of the eigenvalues of $\mathcal{J}$ at the fixed point $X^*$ has modulus equal to $1$. These are analytically easy to handle and are also supported by numerical results in the later half of this section. Given that the algebraic calculations for codimension-$2$ bifurcations are difficult and time-consuming, we only provide numerical results for those. It is \textsc{MatContM} that we use to report the numerical bifurcation patterns. Numerical bifurcation analysis using \textsc{MatContM} was also extensively employed to study the three-dimensional memristive Chialvo neuron map by Muni {\em et al.}~\cite{MuFa22}. The summary of codimension-one and codimension-two bifurcation types observed is presented in Table~\ref{Tab:bifpoints}.
\begin{table}[htbp]
\centering
\caption{Abbreviations of codimension-one and codimension-two bifurcations}
\label{Tab:bifpoints} 
\begin{tabular}{ |p{5.5cm}|p{1cm}||p{5cm}|p{1cm}|  }
 \hline
 \multicolumn{4}{|c|}{Codimension-1} \\
 \hline
 Fold (saddle node) bifurcation  & LP    & Neimerk-Sacker bifurcation&  NS\\
 \hline
Flip (period doubling) bifurcation&   PD  &   &\\
  \hline
 \multicolumn{4}{|c|}{Codimension-2} \\
 \hline
Double Neimark-Sacker& NSNS & Fold-Flip &LPPD\\
  Flip-Neimark-Sacker& PDNS  &  Fold-Neimark-Sacker  &LPNS\\
 1:1 resonance& R1 & 1:2 resonance&R2\\
 \hline
\end{tabular}
\end{table}\par

We first establish the codimension-$1$ bifurcation patterns (LP, PD, and NS) through Theorem~\ref{thm:saddleNode},~\ref{thm:periodDoubling}, and~\ref{thm:NeimarkSacker}.

\begin{theorem}
\label{thm:saddleNode}
Suppose
\begin{align}
\label{eq:saddleNodeEq}
\left[(1-a)(1+\sigma_{12}-D_1)+D_2b \right](1+J_0+J_1+J_2+J_3) + (L_0+L_1+L_2+L_3+L4) = 0.
\end{align}
Then model~\eqref{eq:model1}--\eqref{eq:model6} undergoes a saddle-node bifurcation.
\end{theorem}

\begin{proof}
Saddle-node bifurcation occurs when the Jacobian matrix $\mathcal{J}$ has an eigenvalue $1$ at $X^*$. Using the characteristic equation~\eqref{eq:poly}, we set $P_6(1)=0$. Solving this gives us
$$
a_0+a_1+a_2+a_3+a_4+a_5+a_6 = 0,
$$
which after some simple algebra reduces back to~\eqref{eq:saddleNodeEq}.
\end{proof}

\begin{theorem}
\label{thm:periodDoubling}
Suppose
\begin{align}
\label{eq:periodDoublingEq}
\left[(1+a)(1-\sigma_{12}+D_1)+D_2b \right](1+J_0-J_1+J_2-J_3) + (L_0-L_1+L_2-L_3+L4) = 0.
\end{align}
Then model~\eqref{eq:model1}--\eqref{eq:model6} undergoes a period-doubling bifurcation.
\end{theorem}

\begin{proof}
Period-doubling bifurcation occurs when the Jacobian matrix $\mathcal{J}$ has an eigenvalue $-1$ at $X^*$. Using the characteristic equation~\eqref{eq:poly}, we set $P_6(-1)=0$. Solving this gives us
$$
a_0-a_1+a_2-a_3+a_4-a_5+a_6 = 0,
$$
which after some simple algebra reduces back to~\eqref{eq:periodDoublingEq}.
\end{proof}

Another interesting bifurcation is the {\em Neimark-Sacker} bifurcation where $\mathcal{J}$ generates a complex eigenvalue having modulus $1$.

\begin{theorem}
\label{thm:NeimarkSacker}
Suppose
\begin{align}
\label{eq:NeimarkSackerEq}
&\mathcal{D}^{3/2}+J_3\mathcal{D}^{5/4}+J_2\mathcal{D} +J_1\mathcal{D}^{3/4}+J_0\sqrt{D} \nonumber\\
&+(\sigma_{12}-D_1-a)\left( \mathcal{D}^{5/4}+J_3\mathcal{D}+J_2\mathcal{D}^{3/4}+J_1\sqrt{D} + J_0\sqrt[4]{\mathcal{D}}\right) \nonumber \\
&+(D_1a-\sigma_{12}a+D_2b)\left(\mathcal{D} + J_3\mathcal{D}^{3/4}+J_2\sqrt{\mathcal{D}} + J_1\sqrt[4]{\mathcal{D}} + J_0 \right) \nonumber \\
&+\left(L_4\mathcal{D} + L_3\mathcal{D}^{3/4} + L_2\sqrt{\mathcal{D}}+L_1\sqrt[4]{\mathcal{D}}+L_0 \right) = 0.
\end{align}
Then model~\eqref{eq:model1}--\eqref{eq:model6} undergoes a Neimark-Sacker bifurcation.
\end{theorem}

\begin{proof}
Neimark-Sacker bifurcation occurs when an eigenvalue of $\mathcal{J}$ is complex with modulus $1$. This is possible for our model when $\lambda^4=\mathcal{D}$. We remind the reader that $\mathcal{D}$ is the determinant of the Jacobian $\mathcal{J}$. Here we utilise the fact from matrix algebra that the product of the eigenvalues will equal the determinant of $\mathcal{J}$, i.e., $\mathcal{D}=\lambda^6$. Thus Neimark-Sacker bifurcation occurs in our model~\eqref{eq:model1}--\eqref{eq:model6} if $P_6(\sqrt[4]{\mathcal{D}})=0$. This means
$$
a_0\mathcal{D}^{3/2}+a_1\mathcal{D}^{5/4}+a_2\mathcal{D}+a_3\mathcal{D}^{3/4}+a_4\sqrt{\mathcal{D}}+a_5\sqrt[4]{\mathcal{D}}+a_6=0,
$$
which after some algebraic manipulation reduces to~\eqref{eq:NeimarkSackerEq} (leveraging~\eqref{eq:a0}--\eqref{eq:a6}).
\end{proof}

\subsection{Numerical bifurcation analysis}
\label{sec:numbif}
Figure~\ref{fig:(sigma12,x1)} shows a codimension-1 bifurcation diagram of the map~\eqref{eq:model1}--\eqref{eq:model6} in the $(\sigma_{12},x_1)$-plane. For large values of $\sigma_{12}$, the system has a single fixed-point curve. As $\sigma_{12}$ decreases, a supercritical period-doubling bifurcation (PD) occurs with normal form $1.80e^{+02}$ at $(\sigma_{12}, x_1)=(-1.9032, -0.1102)$. A further decrease in $\sigma_{12}$ results in a fold bifurcation (LP1) at $(\sigma_{12}, x_1)=(-2.0530, -0.0477)$, producing two fixed-point curves. The upper branch then undergoes another fold bifurcation (LP2) at $(\sigma_{12}, x_1)=(-0.7598, 0.7134)$, generating a third branch of the fixed-point curve. This implies that between LP1 and LP2, the map~\eqref{eq:model1}--\eqref{eq:model6} has three fixed points. Additionally, a Neimark-Sacker (NS1) bifurcation was detected at $(\sigma_{12}, x_1)=(-2.0504, -0.0379)$ along the upper branch curve of the fixed-point curve close to LP1. Extending the numerical continuation along the curve of fixed points, we detected another Neimark-Sacker (NS2) and fold (LP3) bifurcations along the third branch of the fixed-point curves at $(\sigma_{12}, x_1)=(-0.9617, 1.3600)$ and $(\sigma_{12}, x_1)=(-1.1316, 2.7337)$, respectively.
\begin{figure}[h!]
    \centering
    \includegraphics[scale=0.5]{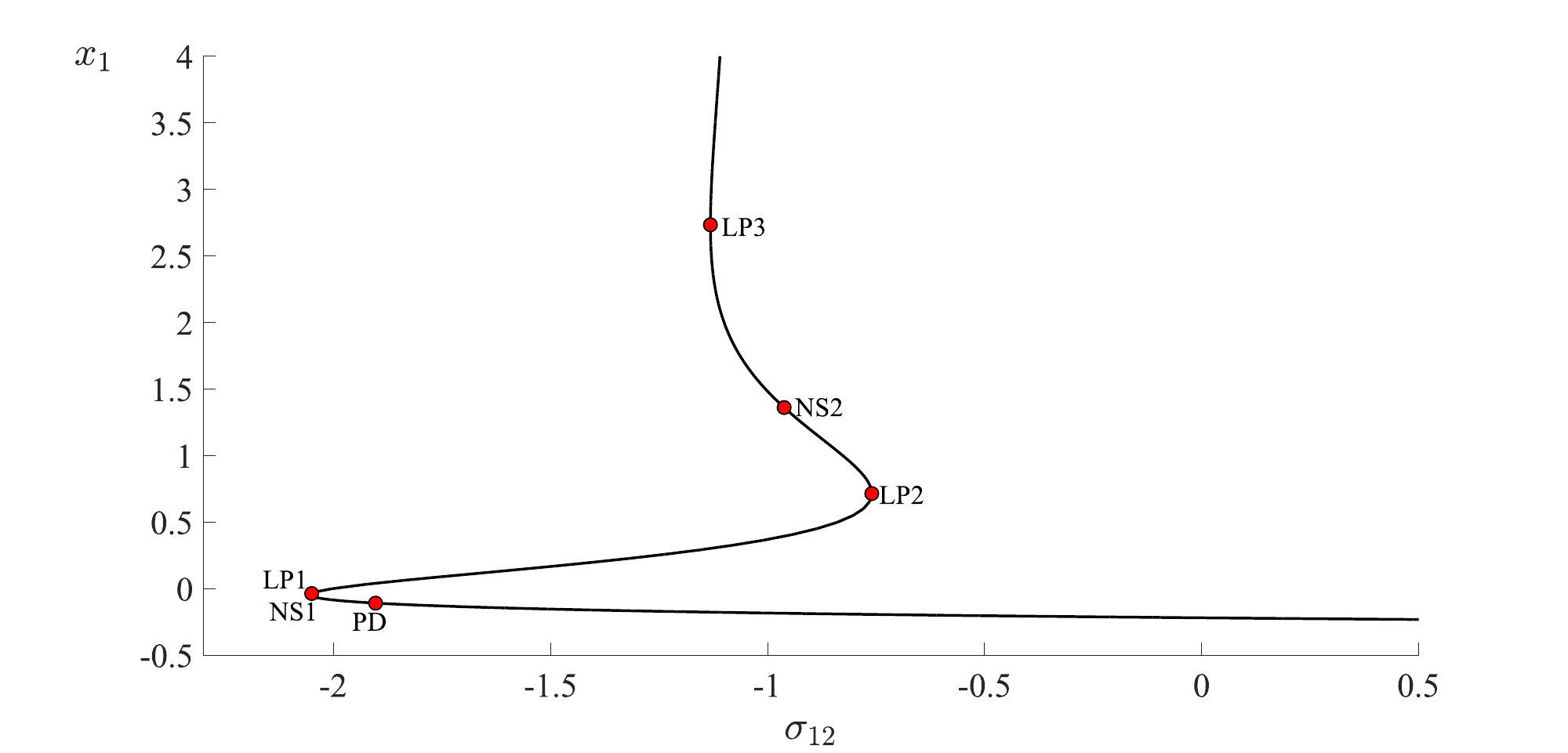}
    \caption{Codimension-1 bifurcation diagram of the map~\eqref{eq:model1}--\eqref{eq:model6} with $\sigma_{12}$ as the bifurcation parameter. Solid black curve correspond to fixed points of the map. The labels for the codimension-1 bifurcations are explained in Table~\ref{Tab:bifpoints}.}
    \label{fig:(sigma12,x1)}
\end{figure}

Figure~\ref{fig:(sigma12,sigma21)} depicts the codimension-1 bifurcation diagram of the map~\eqref{eq:model1}--\eqref{eq:model6} in the $(\sigma_{12},\sigma_{21})$-plane. The figure is composed of the curves of codimension-1 bifurcations detected in Fig.~\ref{fig:(sigma12,x1)}. The blue, green, and magenta curves represent the loci of the period-doubling bifurcation (PD), fold bifurcation (LP), and Neimark-Sacker bifurcation (NS), respectively. The continuation of the period-doubling bifurcation produces two codimension-2 points: the flip-Neimark-Sacker (PDNS) at $(\sigma_{12},\sigma_{21})=(-1.7226, 1.4820)$ and the 1:2 resonance (R2) at $(\sigma_{12},\sigma_{21})=(-1.5307, 2.2929)$. Tracing the LP1 curve, we detected a 1:1 resonance (R1) at $(\sigma_{12},\sigma_{21})=(-2.0530, -0.00004)$. This is a codimension-2 point from which the curve of Neimark-Sacker (NS1) emerges. Extending the continuation NS1 curve, we detected two codimension-2 points. First, we have the double Neimark-Sacker (NSNS) bifurcation at $(\sigma_{12},\sigma_{21})=(-1.9981, 0.8582)$, and the second is the flip-Neimark-Sacker (PDNS) bifurcation at $(\sigma_{12},\sigma_{21})=(-1.7226, 1.4820)$.

Next, the LP2 bifurcation is selected for continuation, and we detected another 1:1 resonance (R1) at $(\sigma_{12},\sigma_{21})=(-0.7598, 0.0003)$. Similarly, the curve of Neimark-Sacker (NS2) emanates from this codimension-2 point. Extending the continuation along the LP2 curve results in a fold-Neimark-Sacker (LPNS) bifurcation at $(\sigma_{12},\sigma_{21})=(-0.7598, 4.5233)$ and a fold-flip (LPPD) bifurcation at $(\sigma_{12},\sigma_{21})=(-0.7598, 5.4222)$. Lastly, continuation of the LP3 bifurcation produces the LP3 curve, and along the curve, we found a 1:1 resonance (R1) at $(\sigma_{12},\sigma_{21})=(-1.1316, 0.00006)$.
\begin{figure}[h!]
    \centering
    \includegraphics[scale=0.5]{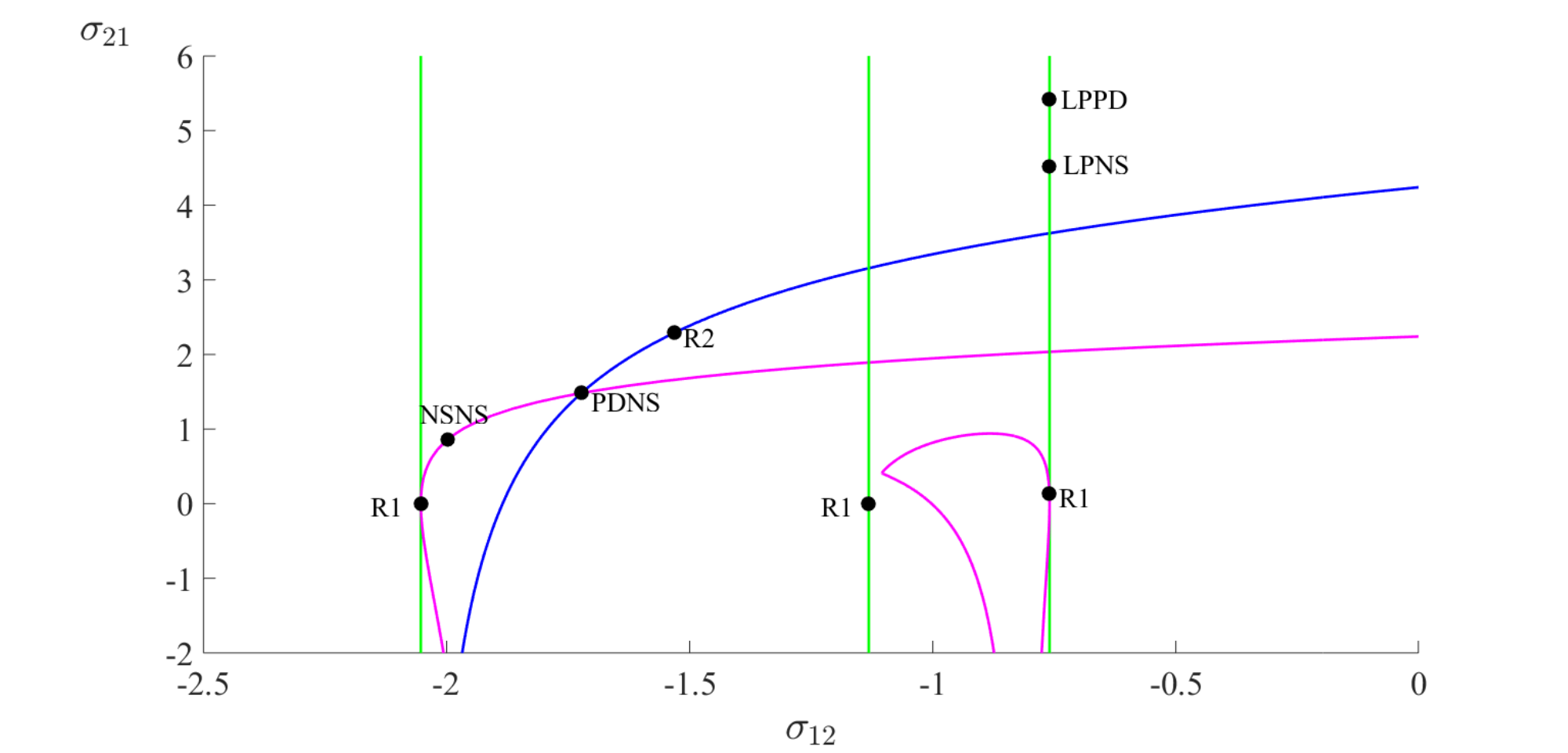}
    \caption{Codimension-2 bifurcation diagram in $(\sigma_{12},\sigma_{21})$-plane. The green, blue, and magenta curves are the loci of the LP, PD, and NS bifurcations. The labels for the codimension-2 bifurcations are explained in Table~\ref{Tab:bifpoints}.}
    \label{fig:(sigma12,sigma21)}
\end{figure}

\section{Synchronisation measures}
\label{sec:sync}
After analyzing our model in terms of dynamical tools, it is time to look into its collective behavior that arises from the dynamics of each of the oscillators involved in the network. This is usually achieved by studying the synchronization behavior of the whole network. To do so, we employ two commonly used quantitative metrics from the literature called the {\em cross-correlation coefficient}, and the {\em Kuramoto order parameter}.

\subsection{Cross-correlation coefficient}
Our model has three oscillators with the second oscillator connected to either the first and the third oscillator by a pair of links. This calls for formulating the cross-correlation coefficient between the first and the second oscillators and also between the second and the third oscillators before taking the average that will indicate the global synchronization pattern of the network as a whole. Let us first define the cross-correlation coefficient between the first and the second oscillator as
\begin{align}
\label{eq:G12}
\Gamma_{12} =\sum_{n=1}^T \frac{\langle \tilde{x}_1(n) \tilde{x}_2(n) \rangle}{\sqrt{\langle \tilde{x}_1(n)^2 \rangle \langle \tilde{x}_2(n)^2 \rangle}}.
\end{align}
Similarly, between the second and the third oscillator, it is given by
\begin{align}
\label{eq:G23}
\Gamma_{23} =\sum_{n=1}^T \frac{\langle \tilde{x}_2(n) \tilde{x}_3(n) \rangle}{\sqrt{\langle \tilde{x}_2(n)^2 \rangle \langle \tilde{x}_3(n)^2 \rangle}}.
\end{align}
Thus, the averaged cross-correlation coefficient is given by
\begin{align}
\label{eq:Gamma}
\Gamma = \frac{1}{2}\left(\Gamma_{12} + \Gamma_{23}\right).
\end{align}
The averages in~\eqref{eq:G12} and~\eqref{eq:G23} are calculated after the transient dynamics is discarded. The symbol $\tilde{x}_i(n)$ refers to the variance from the mean $\langle x_i(n) \rangle$, i.e, $\tilde{x}_i(n) = x_i(n) - \langle x_i(n) \rangle$, where $i=1,2,3$. The symbol $\langle \rangle$ denotes the average of the action potential over time. In our simulation for calculating the synchronization measures we take $80000$ iterates and discard the first $40000$ iterates to ensure no transients creep in. When $\Gamma = 1$, it means the network has reached complete in-phase synchrony, whereas $\Gamma=-1$ represents the network reaching a complete anti-phase synchrony. Any value $\Gamma \in (-1, 1)$ denotes partial synchronization to asynchronization.\par
\begin{figure}[h!]
\centering
\begin{tabular}{cccc}
  \includegraphics[scale=0.22]{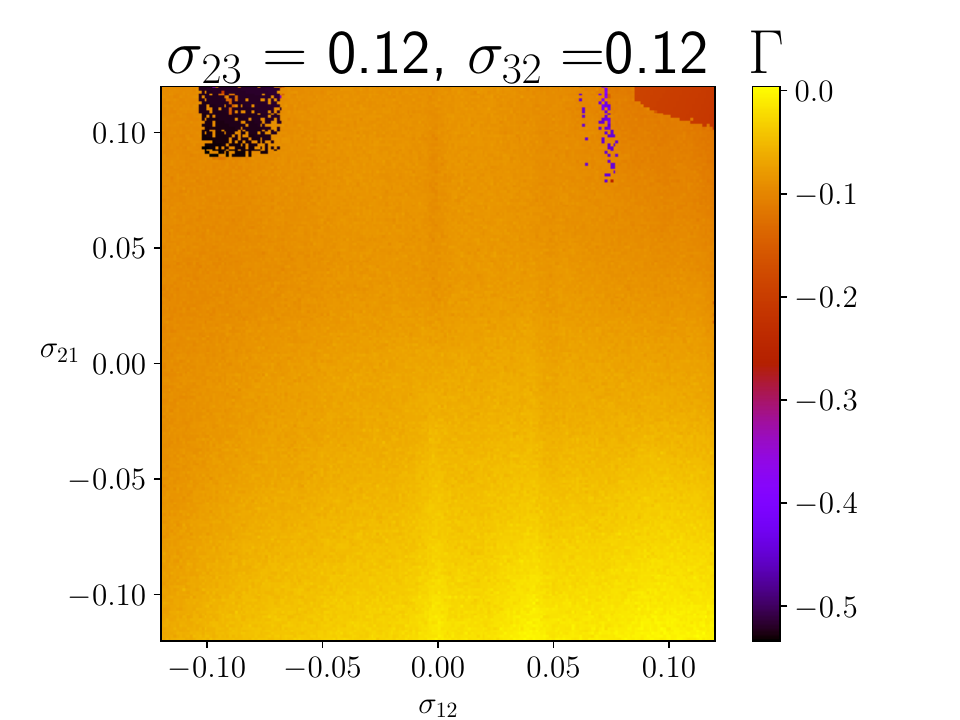} &   \includegraphics[scale=0.22]{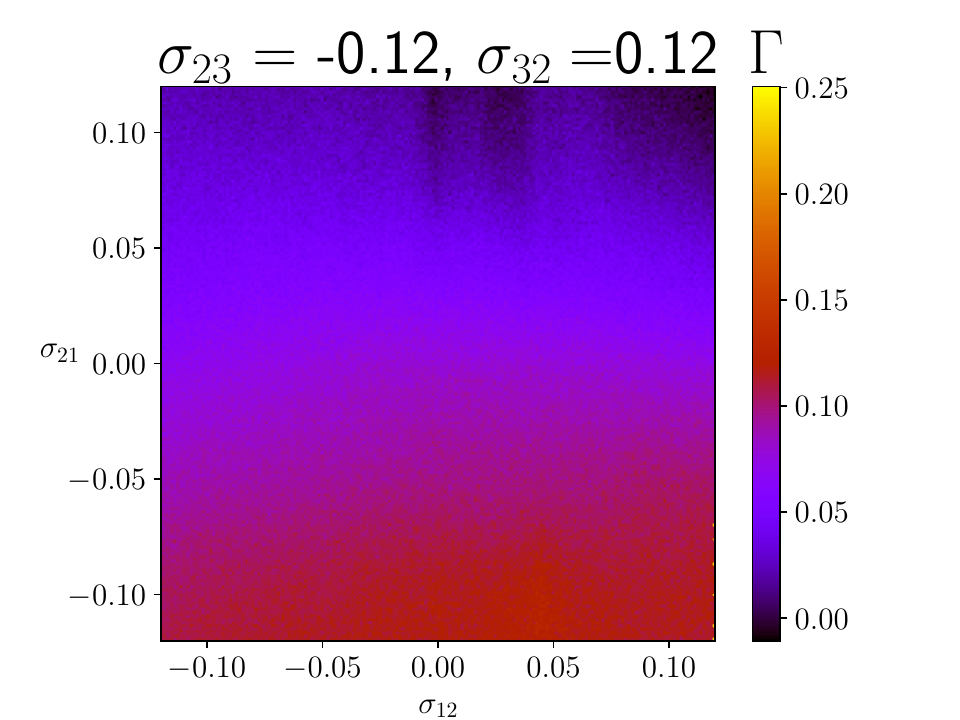} & \includegraphics[scale=0.22]{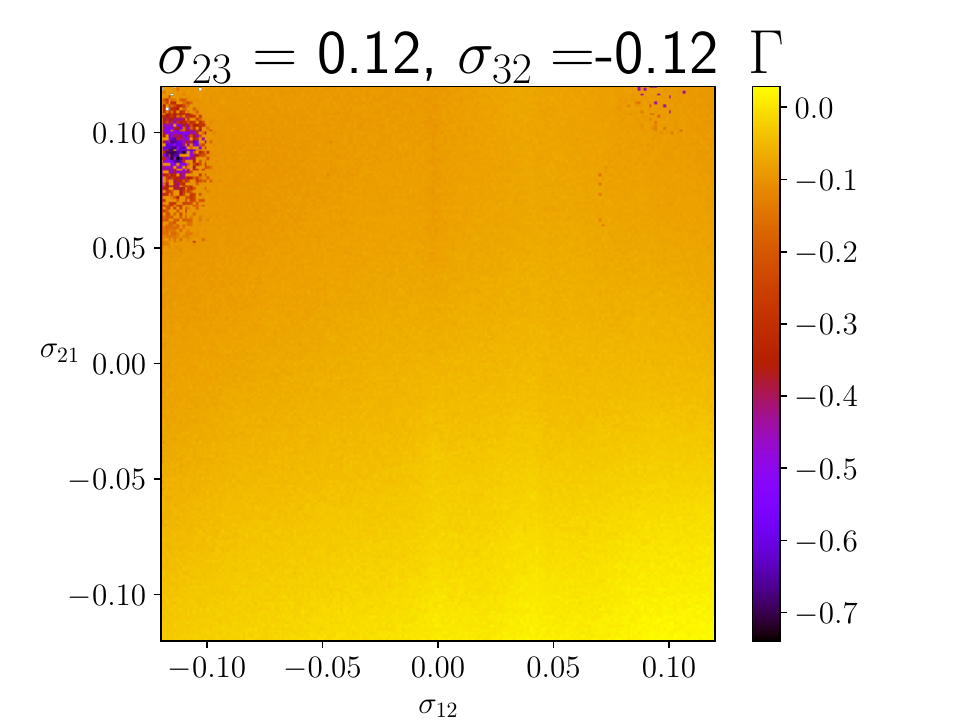} &   \includegraphics[scale=0.22]{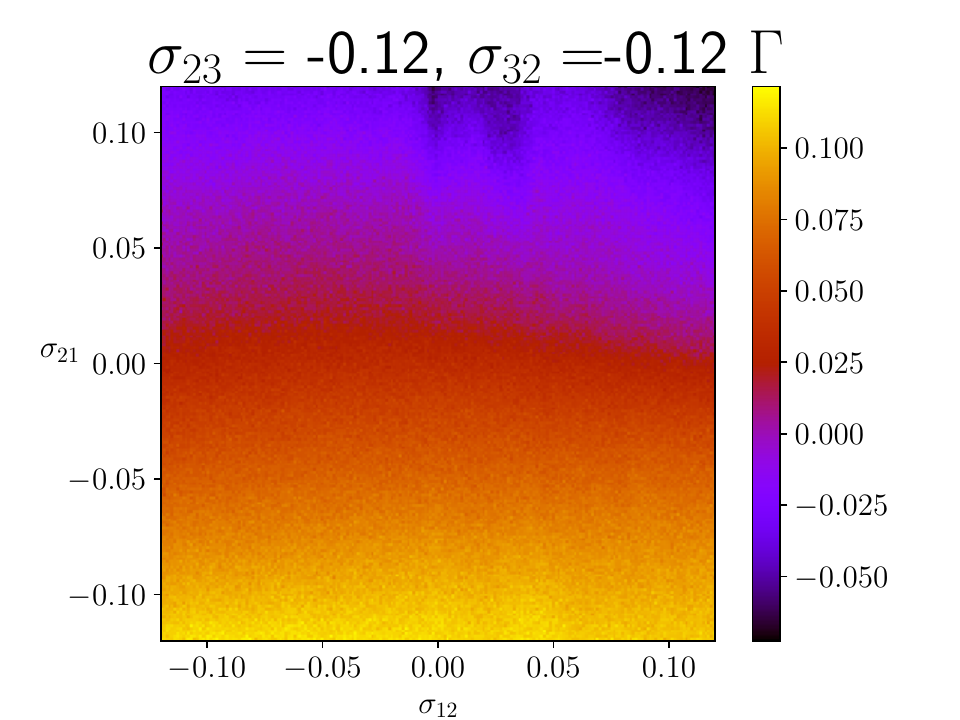}\\
(a) & (b) &(c) &(d)  \\[3pt]
  \includegraphics[scale=0.22]{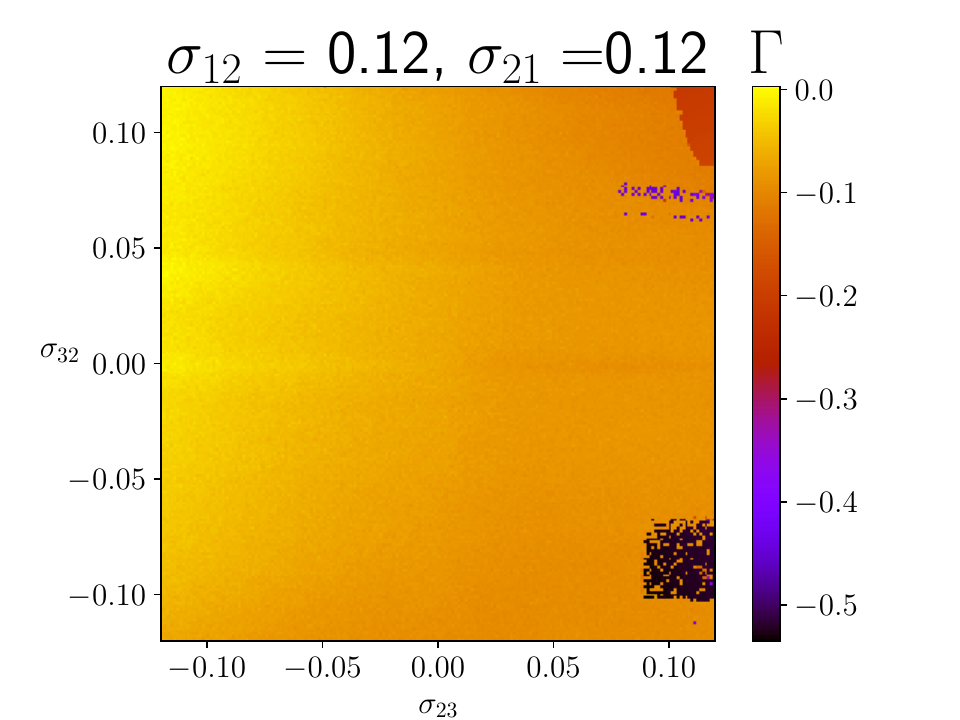} &   \includegraphics[scale=0.22]{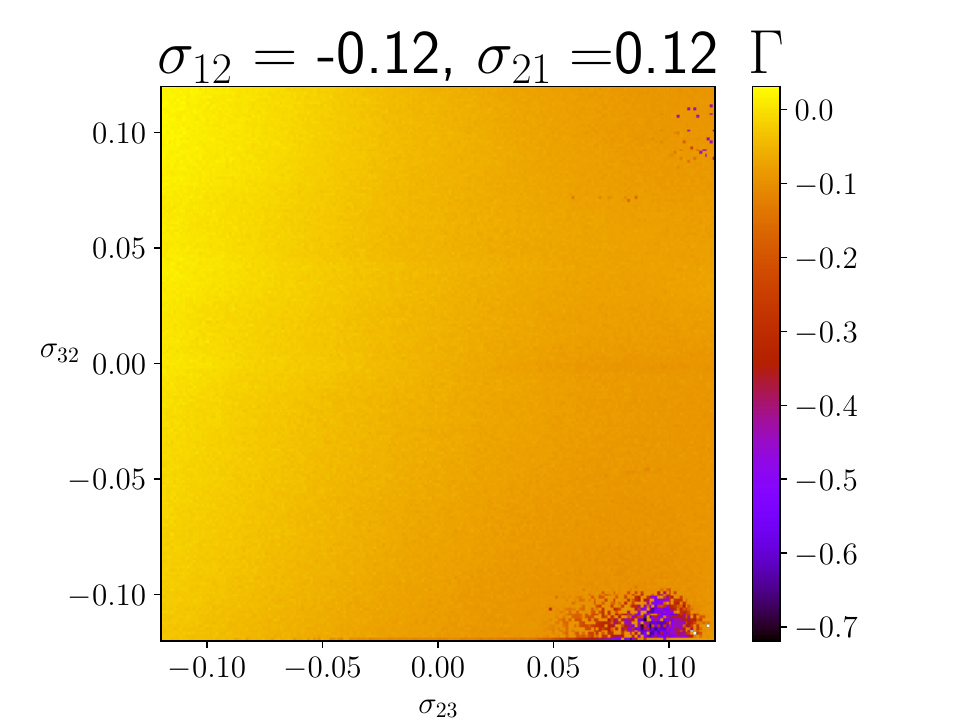} & \includegraphics[scale=0.22]{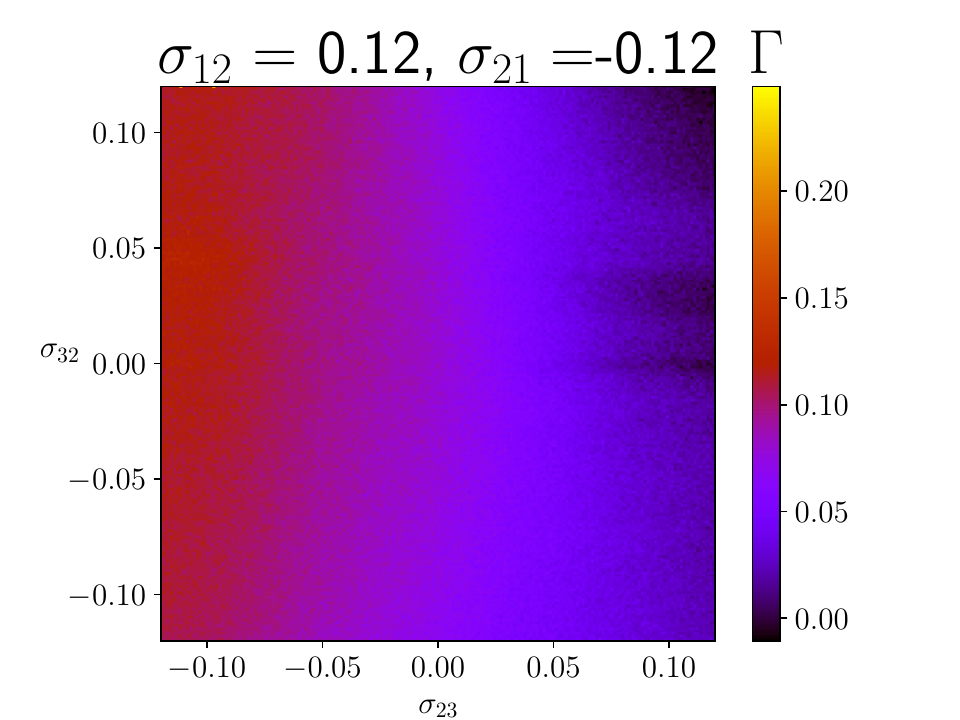} &   \includegraphics[scale=0.22]{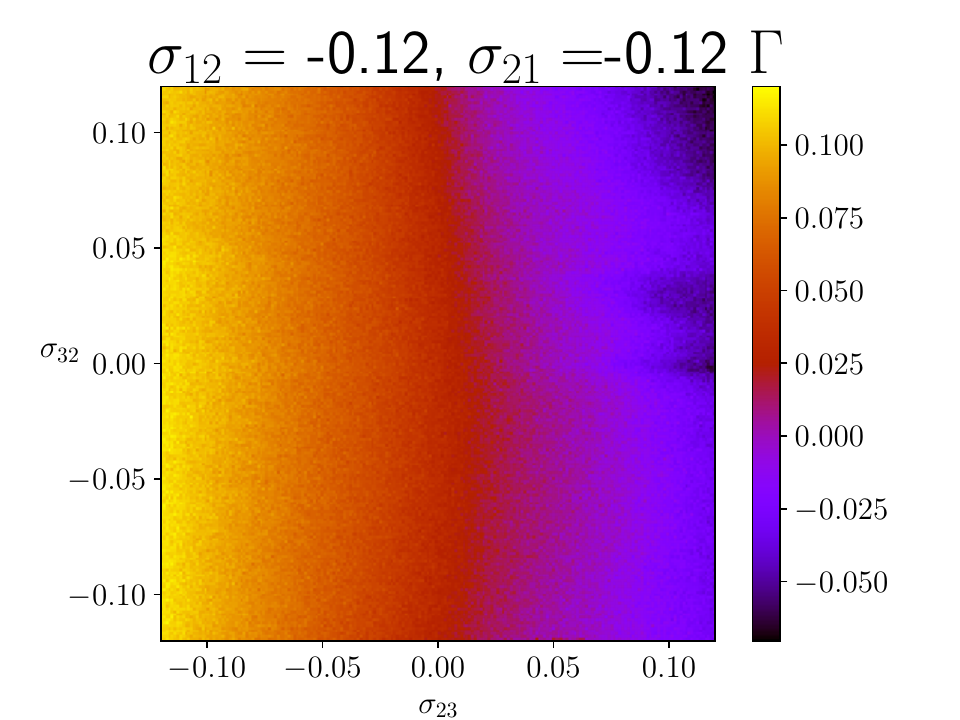}\\
(e) & (f) &(g) &(h) \\[3pt]
\end{tabular}
\caption{A collection of two-dimensional color-coded plots showing the cross-correlation coefficient $\Gamma$. The first row shows the $(\sigma_{12}, \sigma_{21})$ plane, whereas the second row shows $(\sigma_{23}, \sigma_{32})$ plane. The local parameter values are set to be $a=0.6$, $b=0.6$, $c=0.89$, $k_0 = -1$, $\alpha=5$, $\mu=0.0001$, and $\gamma=-0.5$. The other two coupling strengths are set as (a) $\sigma_{23}=\sigma_{32}=0.12$, (b) $\sigma_{23}=-\sigma_{32}=-0.12$, (c) $\sigma_{23}=-\sigma_{32}=0.12$, (d) $\sigma_{23}=\sigma_{32}=-0.12$, (e) $\sigma_{12}=\sigma_{21}=0.12$, (f) $\sigma_{12}=-\sigma_{21}=-0.12$, (g) $\sigma_{12}=-\sigma_{21}=0.12$, (h) $\sigma_{12}=\sigma_{21}=-0.12$. We mostly notice asynchrony and partial synchrony in the whole parameter domain.}
\label{fig:CrossCorrelationA}
\end{figure}

In Fig.~\ref{fig:CrossCorrelationA}, we visualize a collection of two-dimensional color-coded plots, where the grid pixels are colored according to the value of $\Gamma$ and the space is represented by the parameter combination given by either $(\sigma_{12},\sigma_{21})$ keeping fixed $\sigma_{12}$ and $\sigma_{21}$ (first row) or $(\sigma_{23}, \sigma_{32})$ keeping fixed $\sigma_{23}$ and $\sigma_{32}$ (second row). Both in the first and the second rows, the varying coupling strengths lie in $[-0.12, 0.12]$. The local parameter values are set to be $a=0.6$, $b=0.6$, $c=0.89$, $k_0 = -1$, $\alpha=5$, $\mu=0.0001$, and $\gamma=-0.5$. Panel (a) has $\sigma_{23}=\sigma_{32}=0.12$. We observe that $-0.533 \le \Gamma \le 0.004$. The cross-correlation coefficient has a maximum value of $\approx 0$. In the range $-0.1 \le \sigma_{12} \le -0.068$, $0.08885 \le \sigma_{21} \le 0.12$, the values of $\Gamma$ are the lowest $\approx -0.531$, illustrated by the black pixels. Furthermore, we see a reddish patch on the top right corner where $\Gamma \approx -0.2$. We also observe a purple patch near the upper boundary where $\Gamma \approx -0.44$. Otherwise, the whole parameter region is yellowish where $\Gamma \approx 0$. This tells us that overall the system mostly remains asynchronous because the model is a chain with no links between the first and the third oscillator contributing to freer oscillations as compared to a ring network. Panel (b) has $\sigma_{23}=-\sigma_{32}=-0.12$. We observe $-0.01 \le \Gamma \le 0.25$. The system as a whole remains asynchronous with the fact that it tends towards a more synchronous behavior as $\sigma_{21}$ decreases. Panel (c) has $\sigma_{23}=-\sigma_{32}=0.12$. At first, we report a handful of white pixels that appear in the upper left boundaries, corresponding to the diverging dynamics of the system. Other than that, the system has $-0.739 \le \Gamma \le 0.029$, meaning the system as a whole exhibits asynchronous behavior. In the domain $-0.12 \le \sigma_{12} \le -0.1$ and $0.05 \le \sigma_{21} \le 0.11$, there appears a pool of pixels in the purple to black range of color representing $-0.4 \le \Gamma -0.7$. This indicates that the system is approaching an anti-phase synchronization. In the rest of the domain $\Gamma \approx 0$. Panel (d) shows $\sigma_{23}=\sigma_{32}=-0.12$, with the system mostly oscillating in an asynchronous manner ($-0.07 \le \Gamma \le 0.121$). An interesting phenomenon is observed in the second row, where we fix $\sigma_{12}, \sigma_{21}$ and vary $\sigma_{23}, \sigma_{32}$. The qualitative behavior of the second row remains similar to the first row with some subtle differences. Panel (e) with $\sigma_{12}=\sigma_{21}=0.12$ becomes rotationally symmetric to panel (a), panel (f) with $\sigma_{12}=-\sigma_{21}=-0.12$ becomes rotationally symmetric to panel (c), panel (g) with $\sigma_{12}=-\sigma_{21}=0.12$ becomes rotationally symmetric to panel (b), and panel (h) with $\sigma_{12}=\sigma_{21}=-0.12$ becomes rotationally symmetric to panel (d).

\subsection{Kuramoto order parameter}
Another quantitative measure that has been proliferating in the synchronization literature is the {\em Kuramoto order parameter}, represented by the index $I$, first introduced to study the phase coherence behavior in Kuramoto oscillators.\par
\begin{figure}[h]
\centering
\begin{tabular}{cccc}
  \includegraphics[scale=0.22]{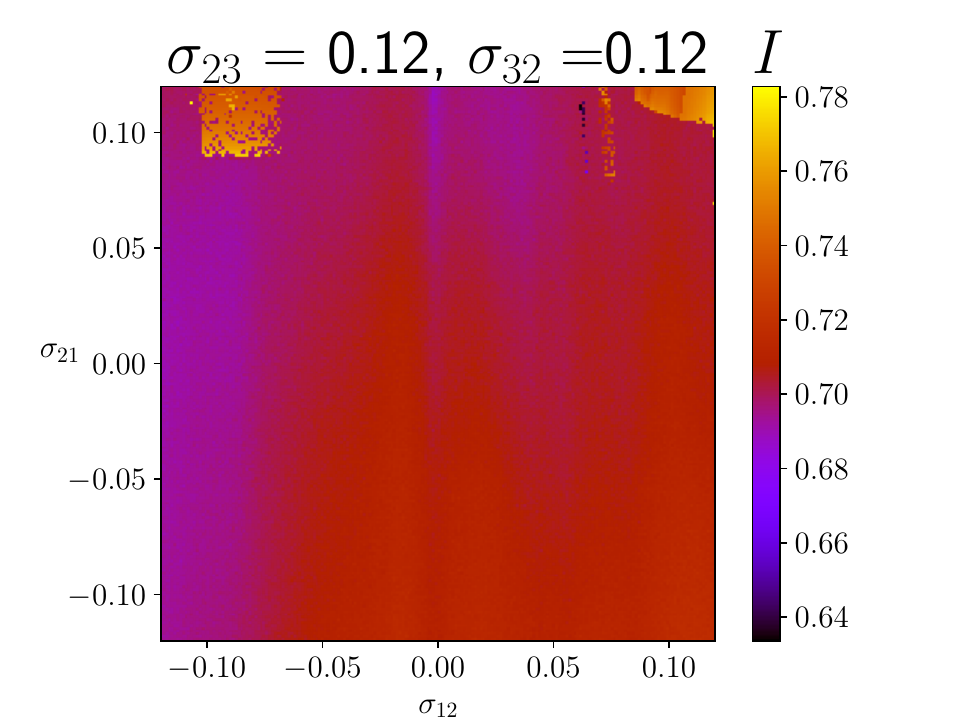} &   \includegraphics[scale=0.22]{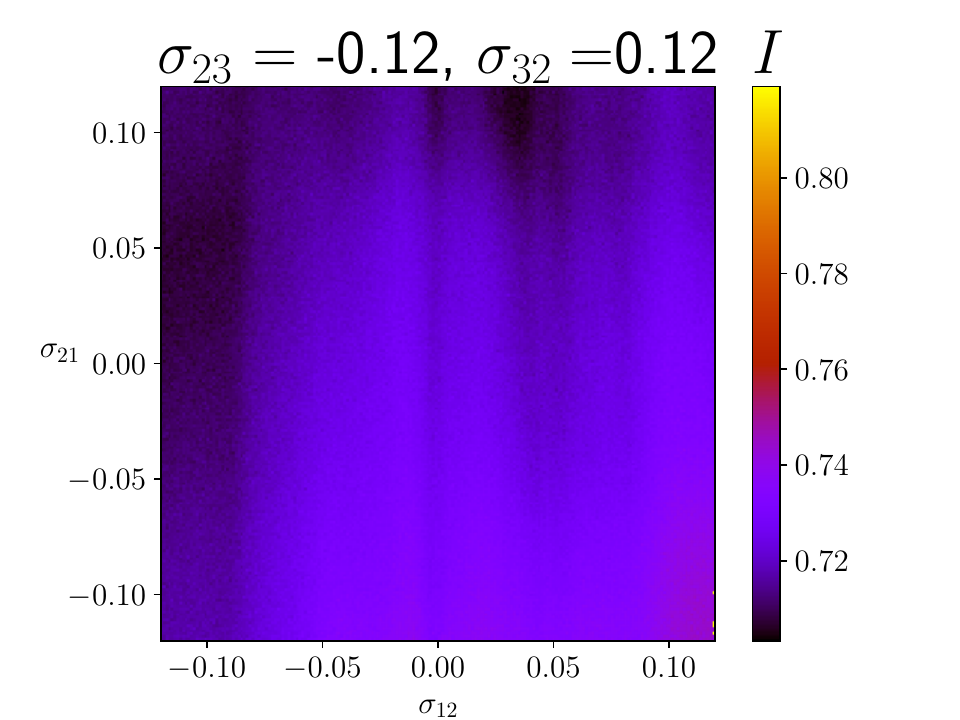} & \includegraphics[scale=0.22]{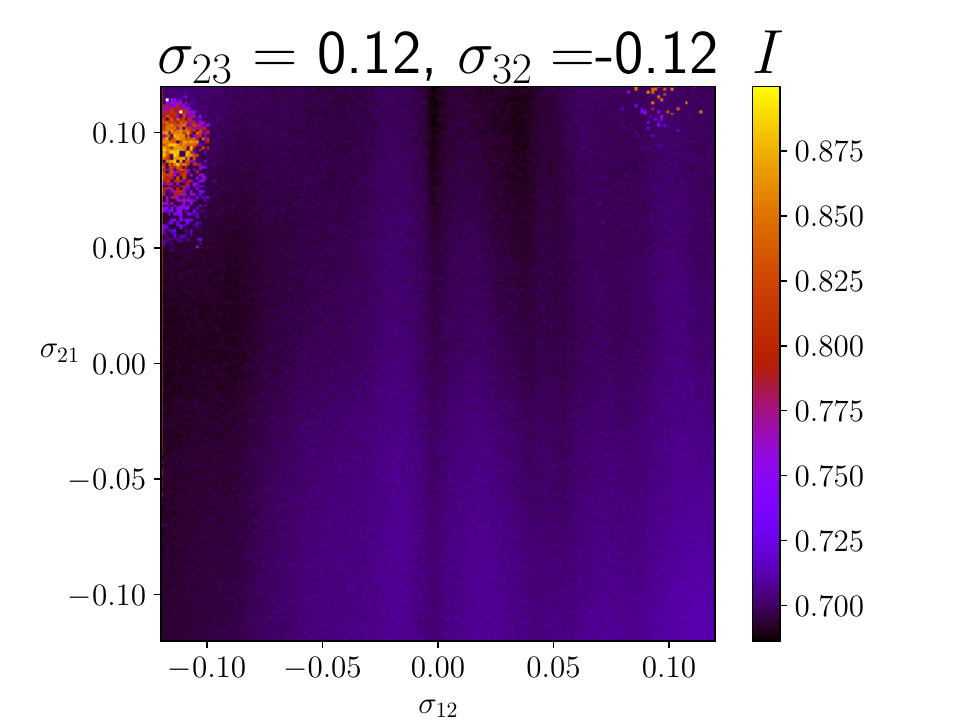} &   \includegraphics[scale=0.22]{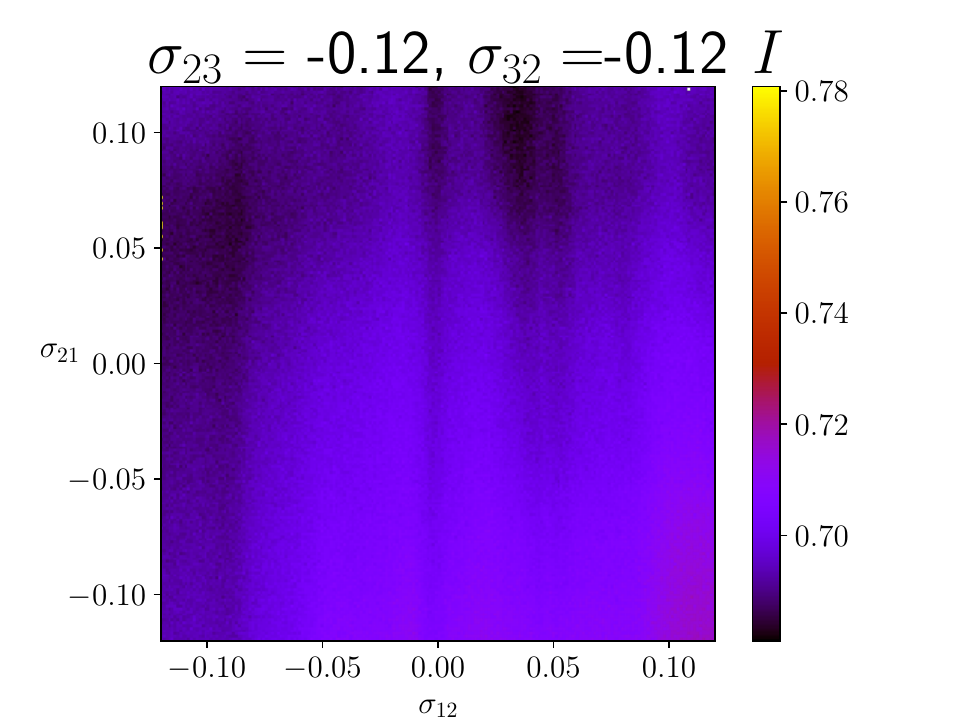}\\
(a) & (b) &(c) &(d)  \\[3pt]
  \includegraphics[scale=0.22]{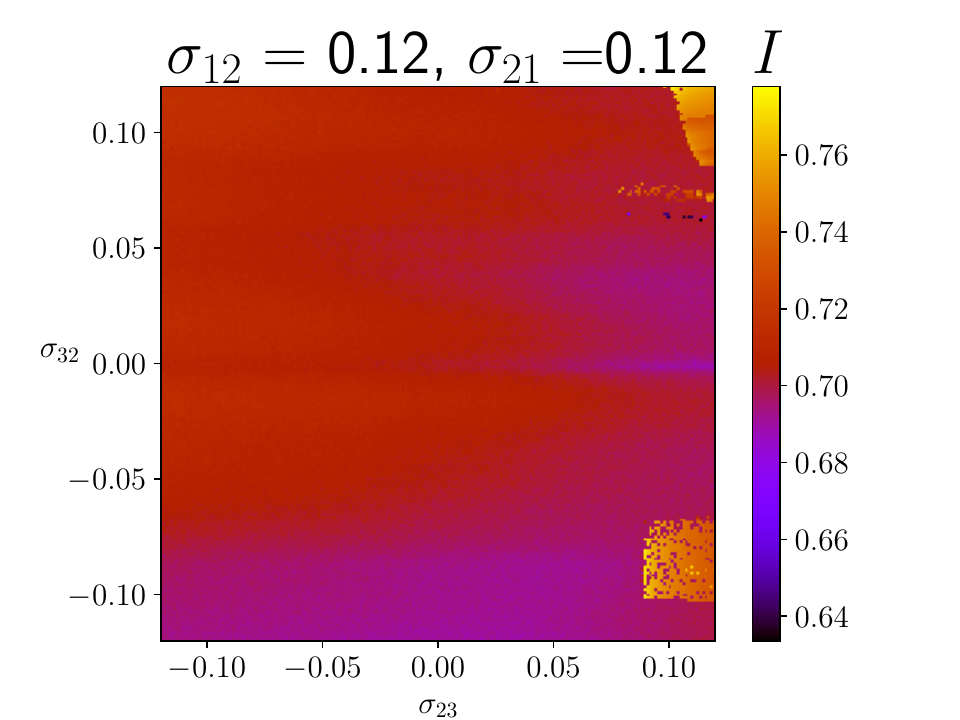} &   \includegraphics[scale=0.22]{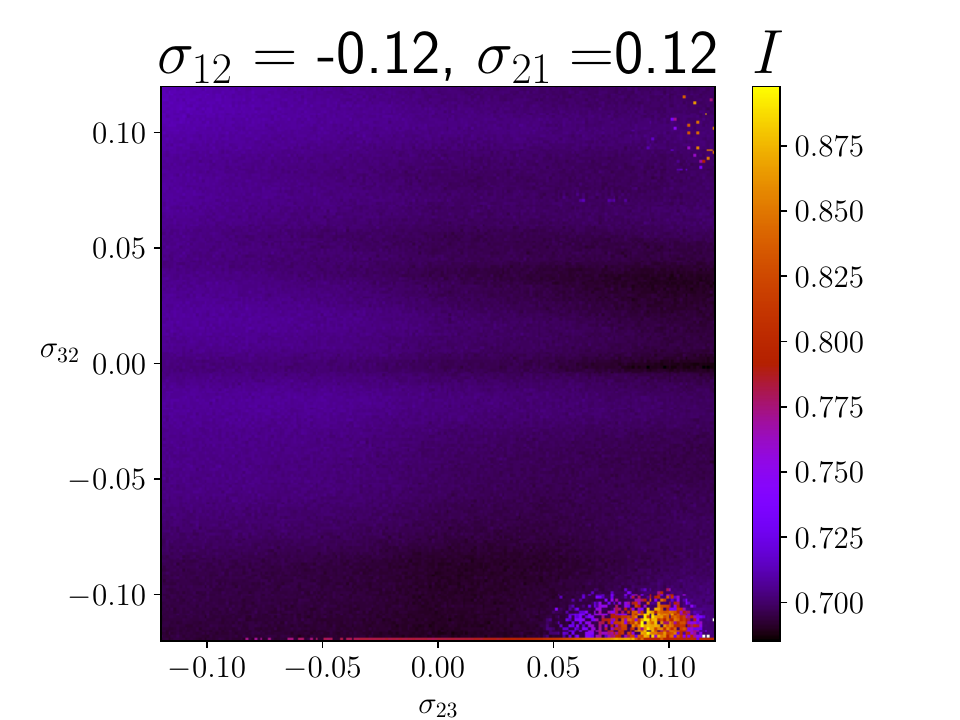} & \includegraphics[scale=0.22]{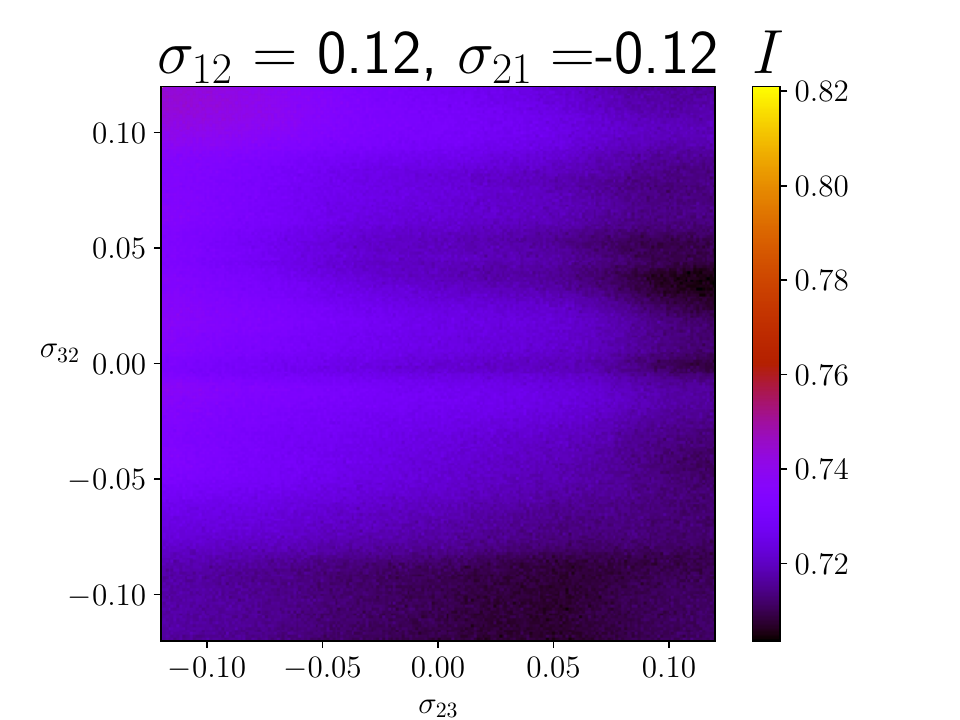} &   \includegraphics[scale=0.22]{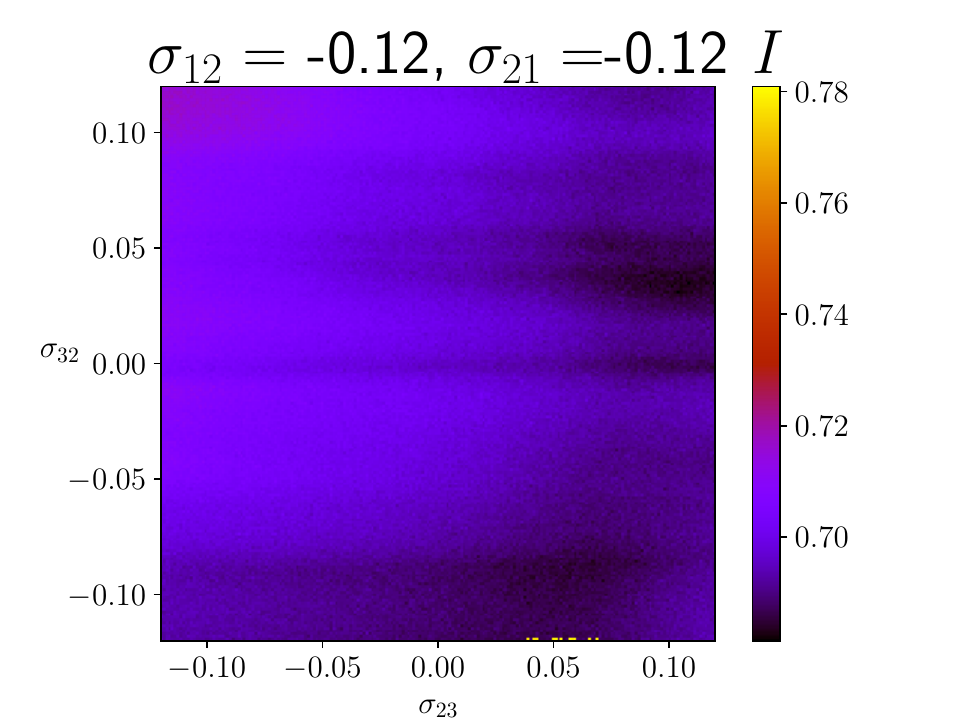}\\
(e) & (f) &(g) &(h) \\[3pt]
\end{tabular}
\caption{A collection of two-dimensional color-coded plots showing the Kuramoto order parameter $I$. The first row shows the $(\sigma_{12}, \sigma_{21})$ plane, whereas the second row shows $(\sigma_{23}, \sigma_{32})$ plane. The local parameter values are set to be $a=0.6$, $b=0.6$, $c=0.89$, $k_0 = -1$, $\alpha=5$, $\mu=0.0001$, and $\gamma=-0.5$. The other two coupling strengths are set as (a) $\sigma_{23}=\sigma_{32}=0.12$, (b) $\sigma_{23}=-\sigma_{32}=-0.12$, (c) $\sigma_{23}=-\sigma_{32}=0.12$, (d) $\sigma_{23}=\sigma_{32}=-0.12$, (e) $\sigma_{12}=\sigma_{21}=0.12$, (f) $\sigma_{12}=-\sigma_{21}=-0.12$, (g) $\sigma_{12}=-\sigma_{21}=0.12$, (h) $\sigma_{12}=\sigma_{21}=-0.12$. We mostly notice asynchrony and partial synchrony in the whole parameter domain.}
\label{fig:KuramA}
\end{figure}

In order to define $I$, we need to first wrap our heads around the instantaneous phase $\Theta_m$ of an oscillator $m$ at time step $n$, given by
\begin{align}
\label{eq:theta}
\Theta_m(n) = \tan^{-1}\left(\frac{y_m(n)}{x_m(n)}\right).
\end{align}
This is utilized to define the complex-valued index
\begin{align}
\label{eq:Imn}
I_m(n) = e^{i \Theta_m(n)},
\end{align}
where $i = \sqrt{-1}$. Furthermore, at time step $n$, the index $I(n)$ for our model~\eqref{eq:model1}--\eqref{eq:model6} is given by
\begin{align}
\label{eq:In}
I(n) = \left|\frac{1}{3}\sum_{m=1}^3 I_m(n) \right|,
\end{align}
where the notation inside the absolute value symbol represents the mean of all phases of the three oscillators inside the unit circle at iteration $n$. Finally, the index average over time is given by
\begin{align}
\label{eq:I}
I = \langle I(n) \rangle.
\end{align}
If $I \approx 0$, the system stabilizes in an asynchronous regime, whereas $I >0$ indicates partial synchrony and $I = 1$ indicates complete synchrony in the system. Like Fig.~\ref{fig:CrossCorrelationA}, we also visualize a collection of two-dimensional color-coded plots, where the grid pixels are colored according to the value of $I$ and the space is represented by the parameter combination given by either $(\sigma_{12},\sigma_{21})$ keeping fixed $\sigma_{12}$ and $\sigma_{21}$ (first row) or $(\sigma_{23}, \sigma_{32})$ keeping fixed $\sigma_{23}$ and $\sigma_{32}$ (second row), see Fig.~\ref{fig:KuramA}. From the first look of it, we see that there exists some kind of linear correspondence between both the synchronization measures $\Gamma$ and $I$. Panel (a) in Fig.~\ref{fig:KuramA}, has $0.633 \le I \le 0.7827$, indicating that the whole system remains in partial synchrony as was also reported in Fig.~\ref{fig:CrossCorrelationA}-(a). We also notice that in the range $-0.1 \le \sigma_{12} \le -0.07$, $0.08985 \le \sigma_{21} \le 0.12$, there exists a yellowish patch, like the black patch in Fig.~\ref{fig:CrossCorrelationA}-(a). The values of $I$ in this region are the highest $\approx [0.74, 0.78]$, indicating a behavior approaching synchrony. Similar behavior is also noticed in the top right corner depicting another pool of yellow pixels. A region corresponding to the purple patch in Fig.~\ref{fig:CrossCorrelationA}-(a) shows a value of $I \approx 0.7$ (in this case, the pixels are yellow). The rest of the figure has $I \in [0.633, 0.7]$ illustrating a partial synchronization. Panel (b) has $\sigma_{23}=-\sigma_{32}=-0.12$, with $0.703 \le I \le 0.82$ again depicting partial synchronisation. Near the right bottom boundary, the system tends to have $I \approx 0.82$, showing a high tendency towards synchronization. Panel (c) has $\sigma_{12}=-\sigma_{21}=0.12$ with $0.686 \le I \le 0.9$. Like Fig.~\ref{fig:CrossCorrelationA}-(c), we have a pool of pixels near the top left corner, where the system shows a high tendency towards synchronization with $I$ even reaching approximately $0.9$. The rest of the domain shows partial synchronization in the network. Panel (d) has $\sigma_{12}=\sigma_{21}=-0.12$ with $0.681 \le I \le 0.7808$. White pixels denote a diverging behavior in the dynamics of the network. Again the qualitative behavior of the second row remains similar to the first row with some subtle differences. Panel (e) with $\sigma_{12}=\sigma_{21}=0.12$ becomes rotationally symmetric to panel (a), panel (f) with $\sigma_{12}=-\sigma_{21}=-0.12$ becomes rotationally symmetric to panel (c), panel (g) with $\sigma_{12}=-\sigma_{21}=0.12$ becomes rotationally symmetric to panel (b), and panel (h) with $\sigma_{12}=\sigma_{21}=-0.12$ becomes rotationally symmetric to panel (d).

\section{Time series analysis via sample entropy}
\label{sec:sampleEntropy}
One important physical aspect of these network dynamical systems is the overall complexity. Thus the question arises, ``can we quantify the system complexity in terms of any entropy measure?". To answer this, we perform a statistical analysis of the network dynamics through the concept of {\em sample entropy}. We generate the time series data of the action potentials of all three oscillators and evaluate the sample entropy of each denoted by ${\rm SE}_{x_i}$ for $i=1, 2, 3$. Then we take the average of all three sample entropies to get the sample entropy of the whole network. The simulation is run for $40000$ iterates and the first $20000$ iterates are discarded to ensure the transients have died down. Then we compute ${\rm SE}_{x_i}$. Before we set up the formula of the sample entropy of the network, we introduce the definition of sample entropy for a time series data $\{x(n), n = 1, \ldots, \mathcal{N} \}$, following {\em Richmond et al.}~\cite{RiMo00}.\par
For a non-negative integer $p \le \mathcal{N}$ let the vectors $x_p(j)$ be defined as
\begin{align}
\label{eq:xpj}
x_p(j) = \left\{x(j+k) \mid 0 \le k \le p-1 \right\},\; 1 \le j \le \mathcal{N}-p+1,
\end{align}
where each of these $\mathcal{N}-p+1$ sets consists of $p$ data points, $x(j) \to x(j+p-1)$. From these, we can define the Euclidean distance as
\begin{align}
\Delta \left(x_p(j), x_p(n)\right) = \max_{0 \leq k \leq p-1} \left\{\lvert x(j+k) - x(n+k) \rvert \right\}. \nonumber
\end{align}
For a positive real threshold value $\epsilon$, $B_j^p(\epsilon)$ is defined as the ratio of the number of vectors $x_p(n)$ within $\epsilon$ of $x_p(j)$ (meaning $\Delta \left(x_p(j), x_p(n)\right) \leq \epsilon$) and the value $\mathcal{N} - p - 1$. Note that here $1 \le n \le \mathcal{N} - p$ with the constraint $n\ne j$. Thus, the term $B^p(\epsilon)$ is defined as
\begin{align}
B^p(\epsilon) = \frac{1}{\mathcal{N}-p} \sum_{j=1}^{\mathcal{N} - p} B_j^p(\epsilon). \nonumber
\end{align}
Similarly we can define $B^{p+1}(\epsilon)$. Thus the sample entropy measure of the given time series is defined as
\begin{align}
\label{eq:sampen}
{\rm SE} = \lim_{\mathcal{N} \to \infty}\left(-\ln \frac{B^{p+1}(\epsilon)}{B^p(\epsilon)} \right).
\end{align}
We apply this concept to the time series data of each of the action potentials before taking the average,
\begin{align}
    {\rm SE} = \frac{1}{3} \left({\rm SE}_{x_1} + {\rm SE}_{x_2} + {\rm SE}_{x_3}\right).
\end{align}
A high ${\rm SE}$ value is indicative of a higher unpredictability in the complex system, corresponding to a higher complexity. A similar approach was employed in computing the sample entropy of an ensemble of memristive Chialvo neurons arranged in a ring-star topology by Ghosh {\em et al.}~\cite{GhMu23}. Other relevant works considering sample entropy are~\cite{HaWe17, NeTe20, MoCa18, HeRa23}.\par
We utilize an open-source \texttt{Python} package \texttt{nolds}~\cite{Sc19} to compute the sample entropy of our time series via the \texttt{nolds.sampen()} function. This function is built following the algorithm by Richmond {\em et al.} Note that the default values of $p$ and $\epsilon$ in~\eqref{eq:sampen} are $2$ and $0.2\sigma$, where $\sigma$ is the standard deviation of the time series. This time we take $80000$ iterates (as noticed in the time series plots), out of which we discard the first $25000$ to get rid of the transients for computing the sample entropies. This makes $\mathcal{N}=55000$.\par
\begin{figure}[h]
\begin{subfigure}{.5\linewidth}
\centering
\includegraphics[scale=0.35]{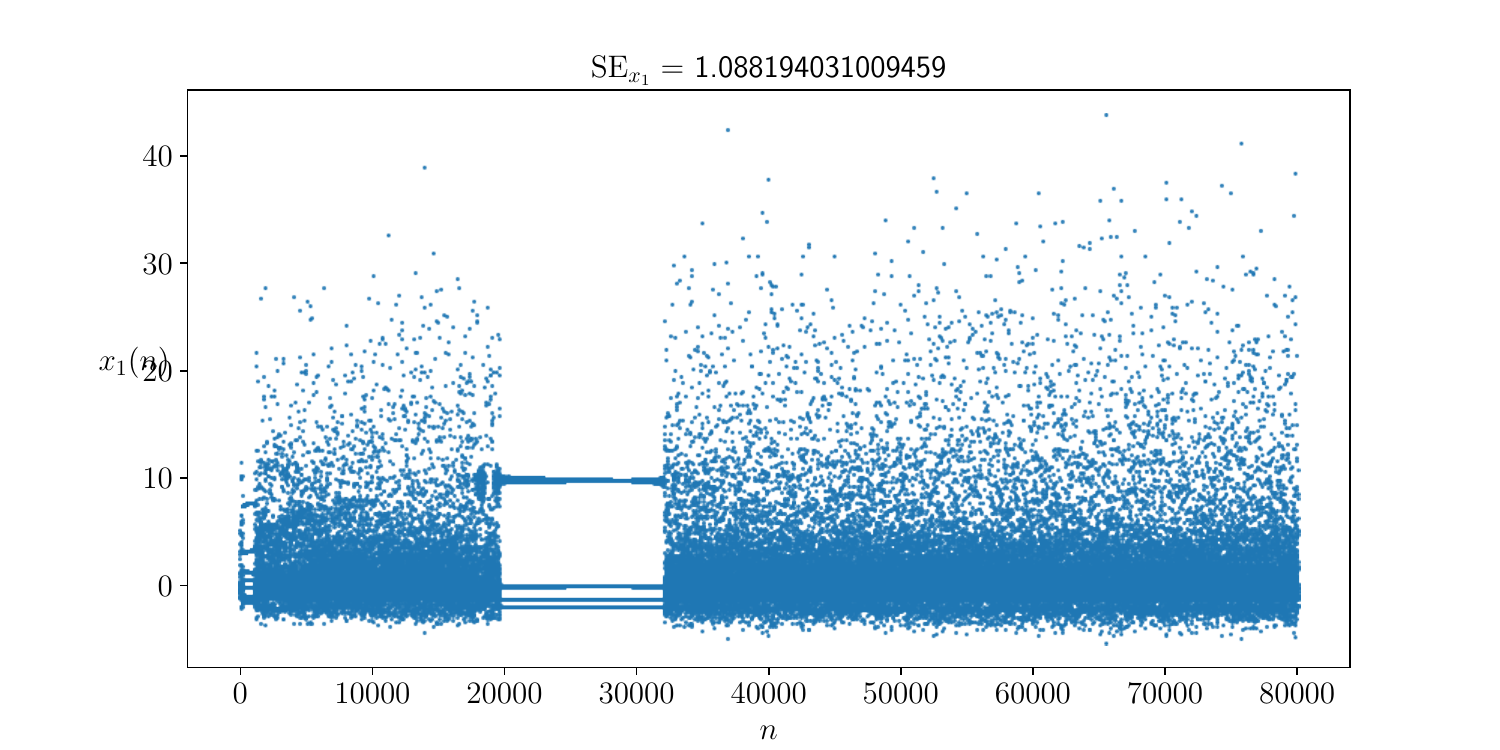}
\caption{$x_1(n)$}
\end{subfigure}%
\begin{subfigure}{.5\linewidth}
\centering
\includegraphics[scale=0.35]{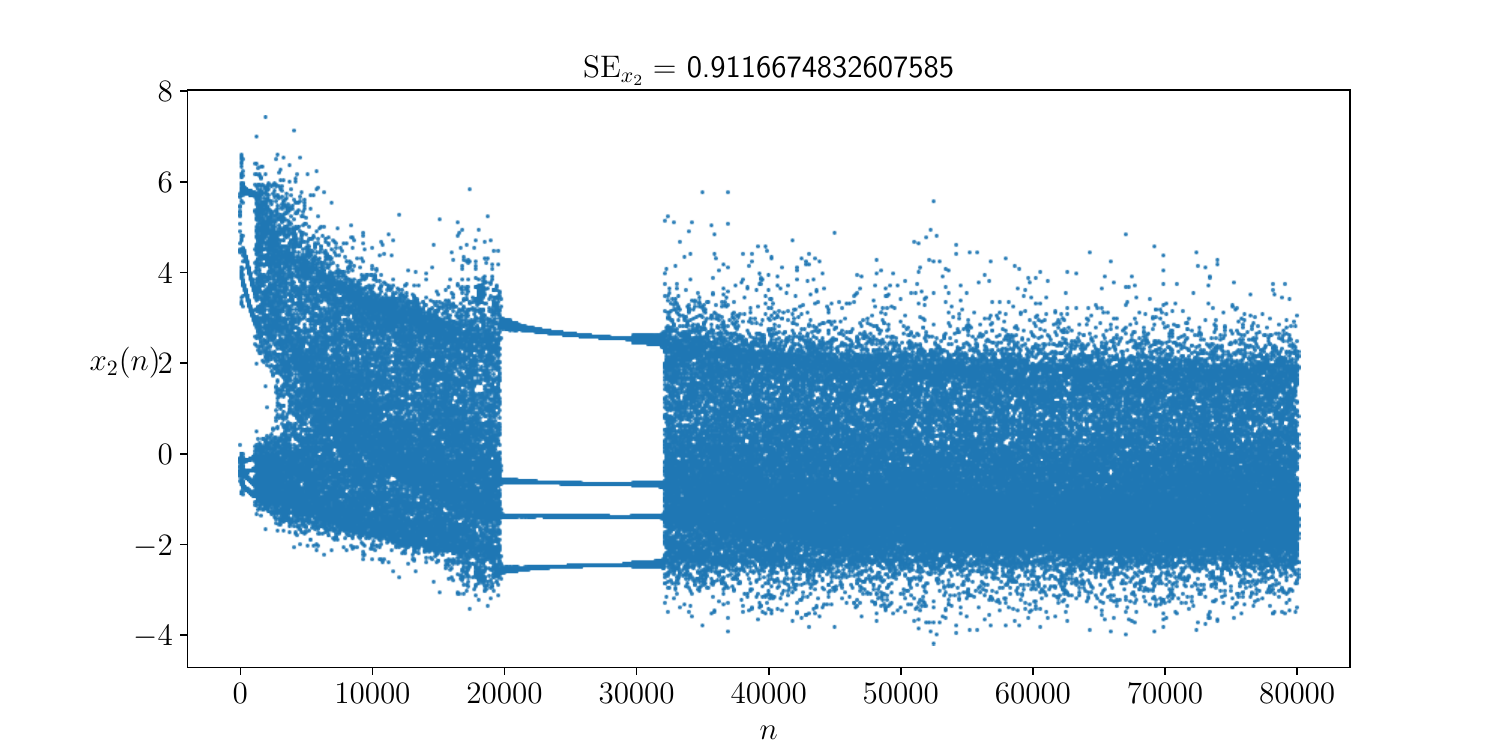}
\caption{$x_2(n)$}
\end{subfigure}\\[1ex]
\begin{subfigure}{1\linewidth}
\centering
\includegraphics[scale=0.35]{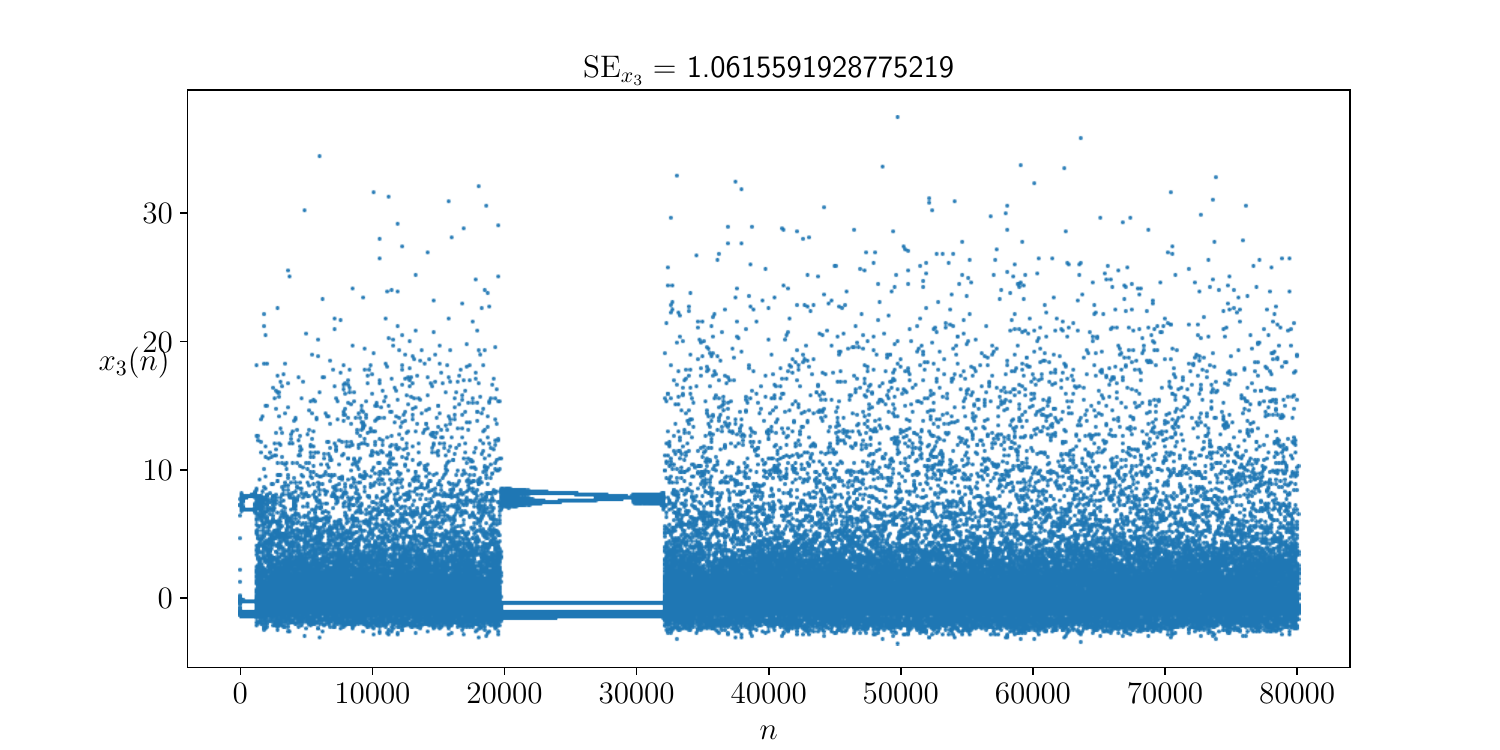}
\caption{$x_3(n)$}
\end{subfigure}
\caption{A collection of time series plots of the action potentials with their corresponding sample entropy values with $\sigma_{12}=0.092$, $\sigma_{21}=0.1$, $\sigma_{23}=0.05$, and $\sigma_{32}=0.06$. The local parameters are set as $a=0.6$, $b=0.6$, $c=0.89$, $k_0=-1$, $\alpha=5$, $\mu = 0.0001$, and $\gamma=-0.5$. We see a high complexity in the behavior corroborated by high sample entropy values. Qualitatively the time series also exhibits irregular chaotic bursts.}
\label{fig:TS}
\end{figure}
\begin{figure}[h]
\begin{subfigure}{.5\linewidth}
\centering
\includegraphics[scale=0.35]{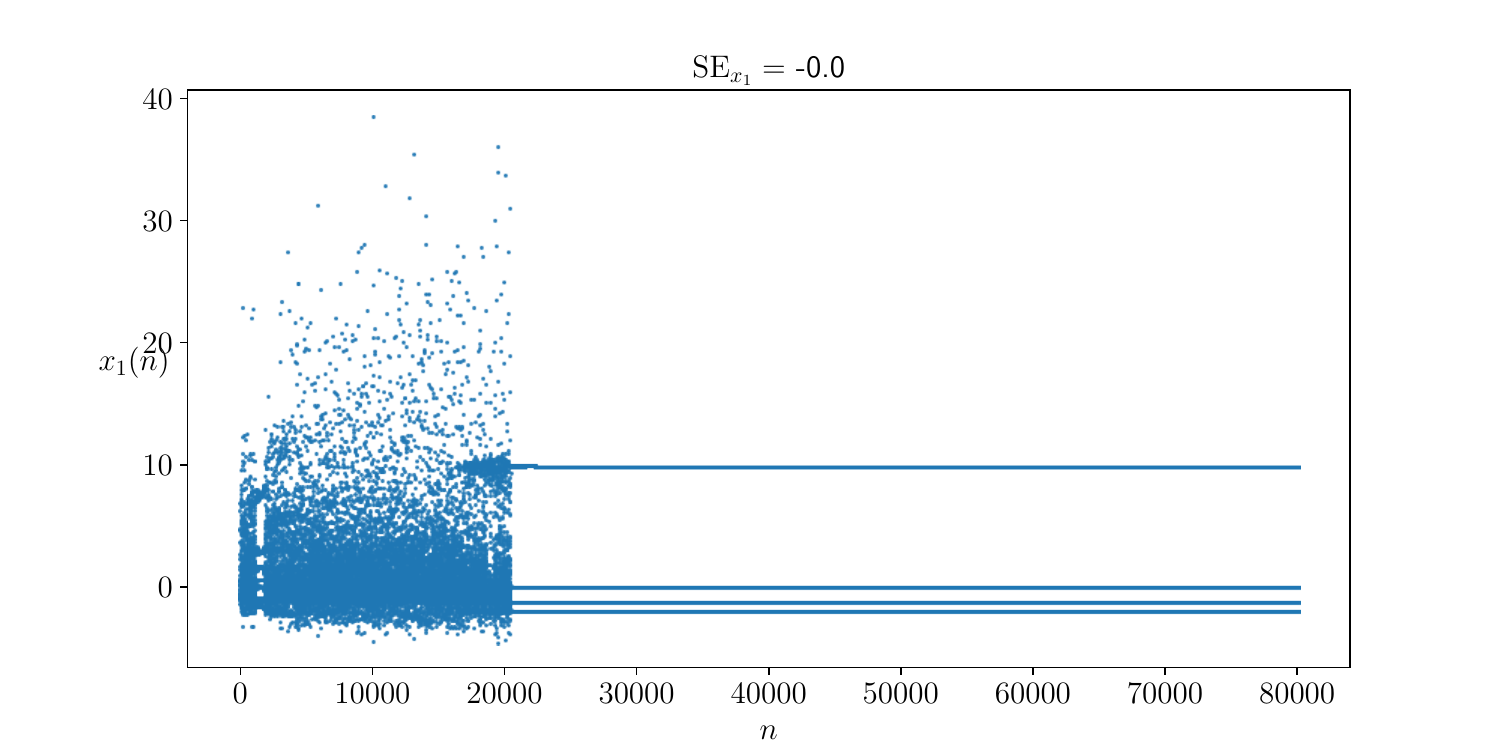}
\caption{$x_1(n)$}
\end{subfigure}%
\begin{subfigure}{.5\linewidth}
\centering
\includegraphics[scale=0.35]{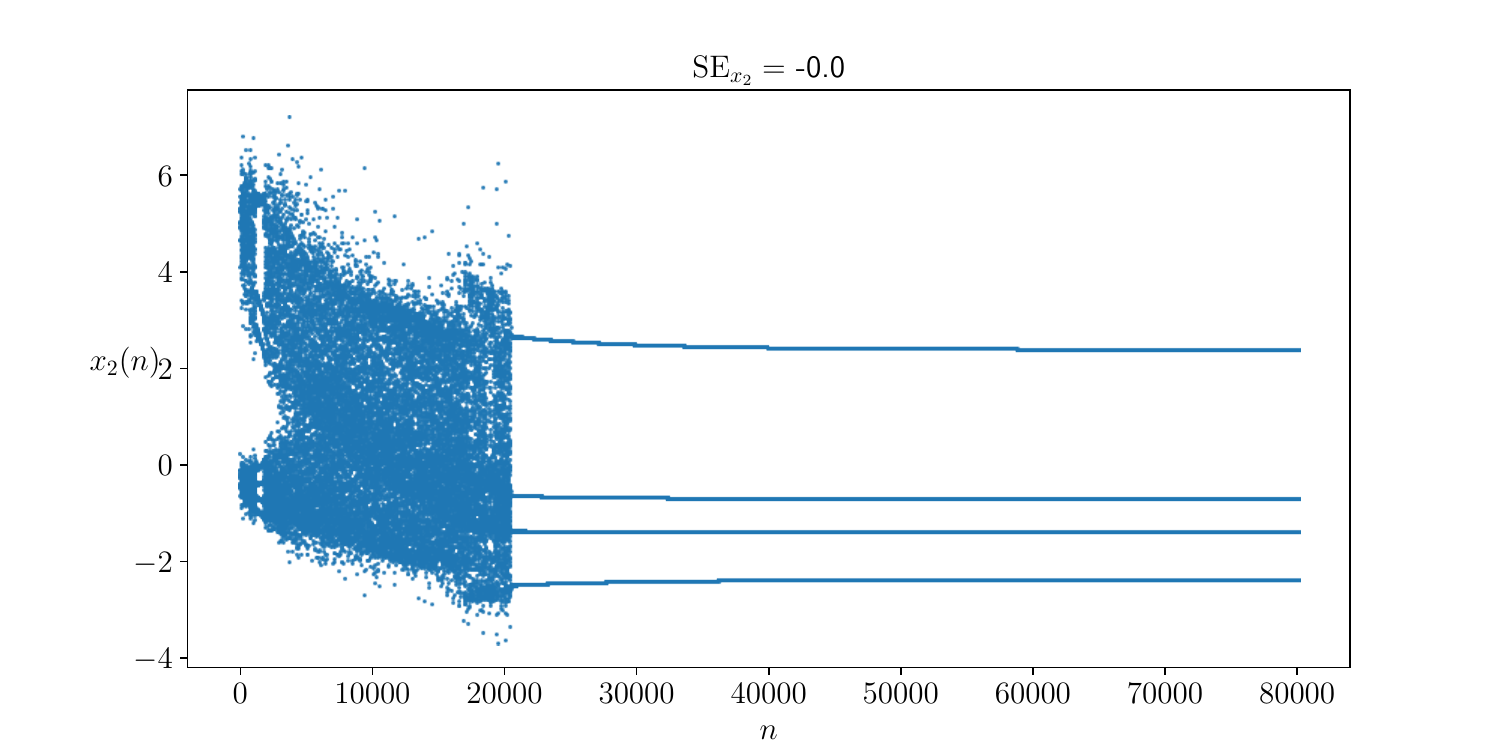}
\caption{$x_2(n)$}
\end{subfigure}\\[1ex]
\begin{subfigure}{1\linewidth}
\centering
\includegraphics[scale=0.35]{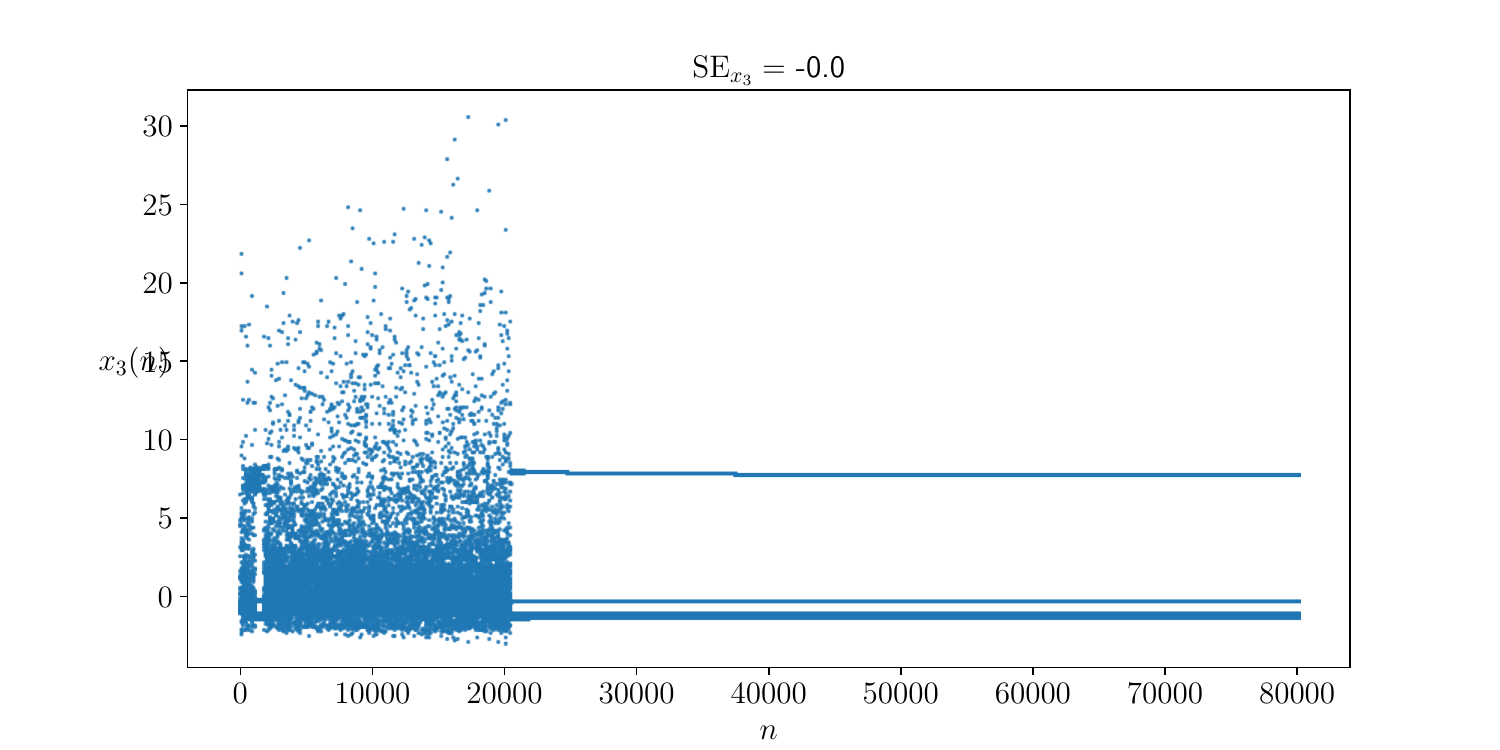}
\caption{$x_3(n)$}
\end{subfigure}
\caption{A collection of time series plots of the action potentials with their corresponding sample entropy values with $\sigma_{12}=0.096$, $\sigma_{21}=0.1$, $\sigma_{23}=0.05$, and $\sigma_{32}=0.06$. The local parameters are set as $a=0.6$, $b=0.6$, $c=0.89$, $k_0=-1$, $\alpha=5$, $\mu = 0.0001$, and $\gamma=-0.5$. We see a low complexity in the behavior corroborated by sample entropy value $\approx 0$. Qualitatively the time series also exhibits regular spikes.}
\label{fig:TS2}
\end{figure}

A collection of time series plots with their corresponding sample entropies are given in Fig.~\ref{fig:TS} and~\ref{fig:TS2}. The local parameters are set as $a=0.6$, $b=0.6$, $c=0.89$, $k_0=-1$, $\alpha=5$, $\mu = 0.0001$, and $\gamma=-0.5$. The coupling strengths are set as $\sigma_{21}=0.1$, $\sigma_{23}=0.05$, and $\sigma_{32}=0.06$. In Fig.~\ref{fig:TS}, we have $\sigma_{12}=0.092$. We have previously seen that for the above parameter combination, there exists a chaotic attractor, see Fig.~\ref{fig:pp2}-(a). The time series behavior for all three action potentials corroborates this, with points exhibiting a dense distribution over the time frame. In the range $n=20000 \to 32000$ approximately, the orbit oscillates periodically, whereas everywhere else it shows irregular bursting corresponding to chaos. We have ${\rm SE}_{x_1} \approx 1.08819$, ${\rm SE}_{x_2} \approx 0.91167$, and ${\rm SE}_{x_3} \approx 1.06156$. In Fig.~\ref{fig:TS2} we have $\sigma_{12}=0.096$. Previous numerics have shown us that for this parameter combination, there exists a periodic attractor with a period $4$. We see that after approximately $n =20000$, the orbit oscillates periodically with all the sample entropies equal to $0$. These results exhibit that chaotic behavior indicates a higher complexity in the system corresponding to a higher ${\rm SE}$. \par
An important note to the reader is to highlight that instead of line plots, the time series in this paper has been represented by discrete points because the model is discrete in time. This gives the illusion that the time series plots look like bifurcation diagrams. It is to inform the reader that the plots manifesting a behavior similar to the dynamics reaching a period-$4$ attractor is a regular oscillatory behavior in the time series (had it been plotted with lines instead of dots).  

\begin{figure}[h]
\centering
\begin{tabular}{cc}
  \includegraphics[scale=0.3]{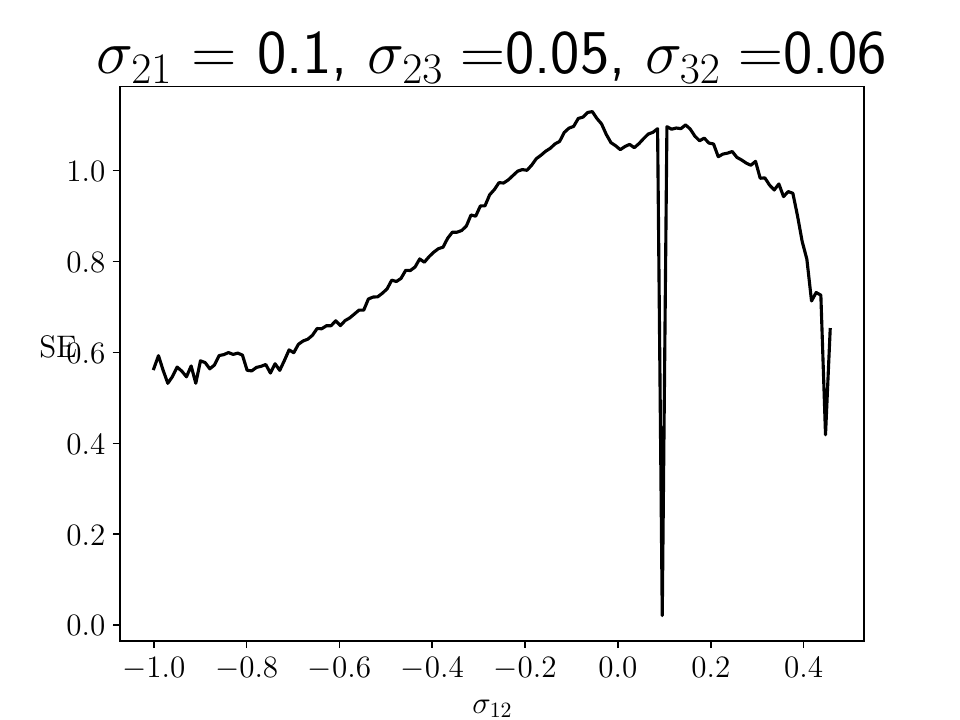} &   \includegraphics[scale=0.3]{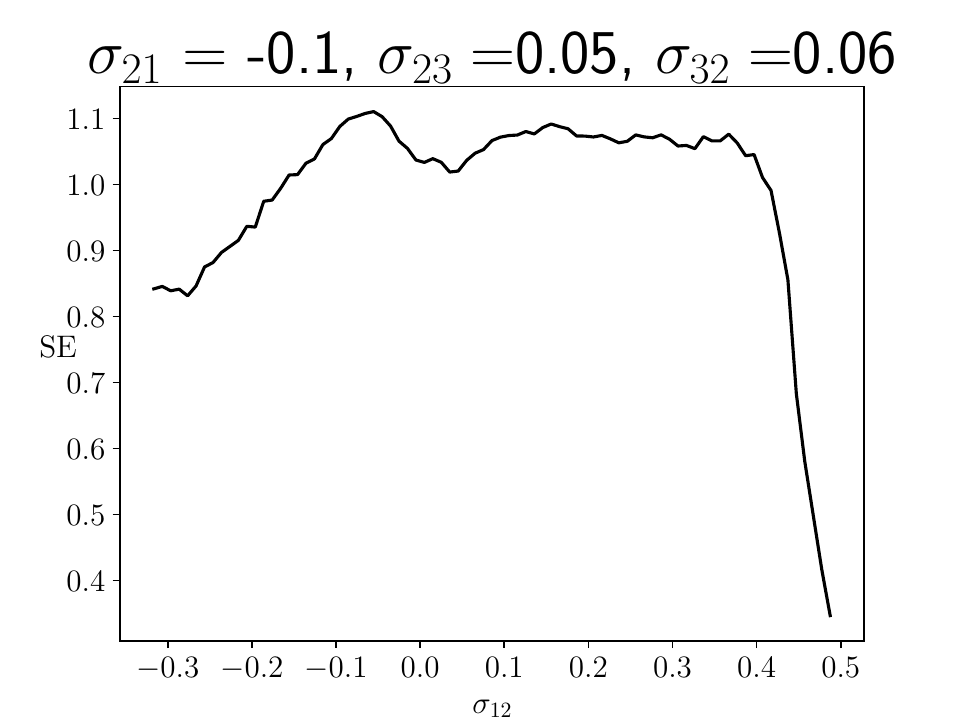}\\
  (a) & (b) \\
  \includegraphics[scale=0.3]{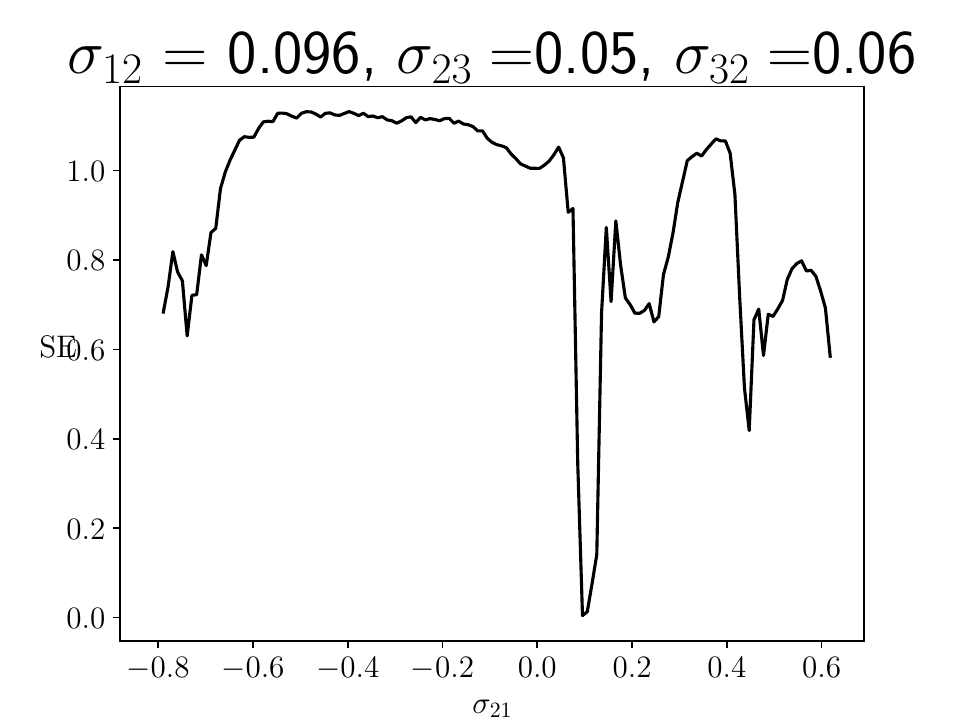} &   \includegraphics[scale=0.3]{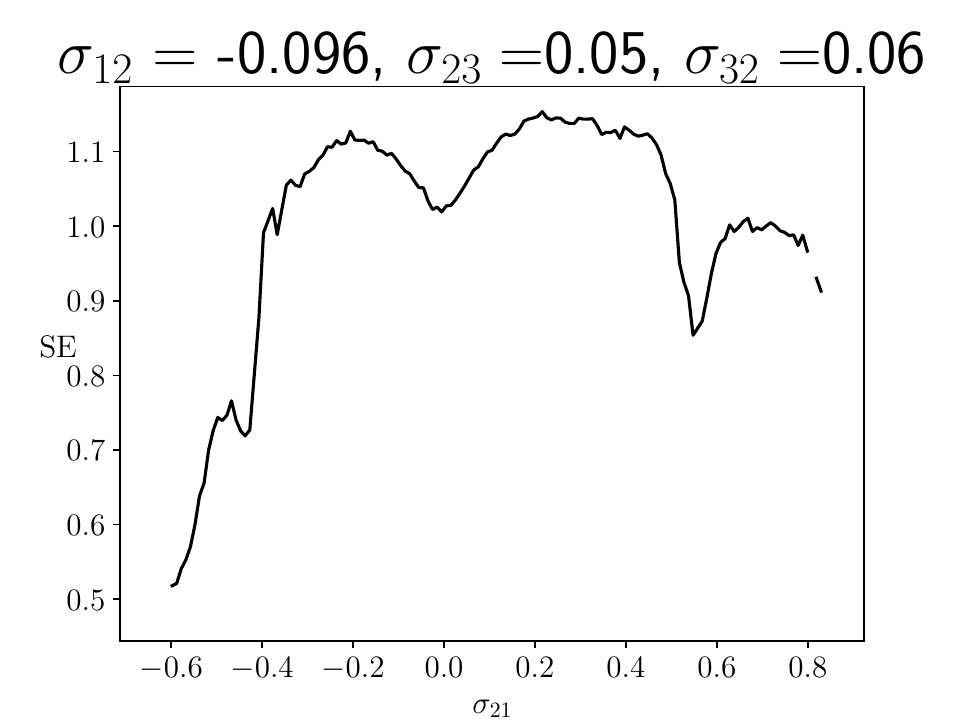}\\
 (c) &(d)  \\[3pt]
  \includegraphics[scale=0.3]{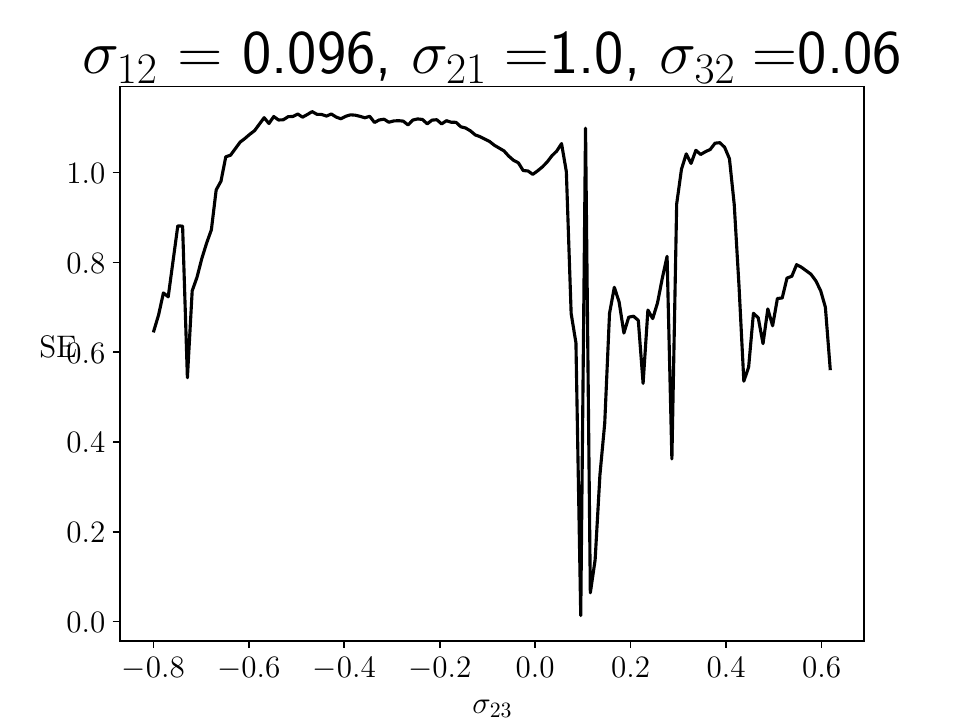} &   \includegraphics[scale=0.3]{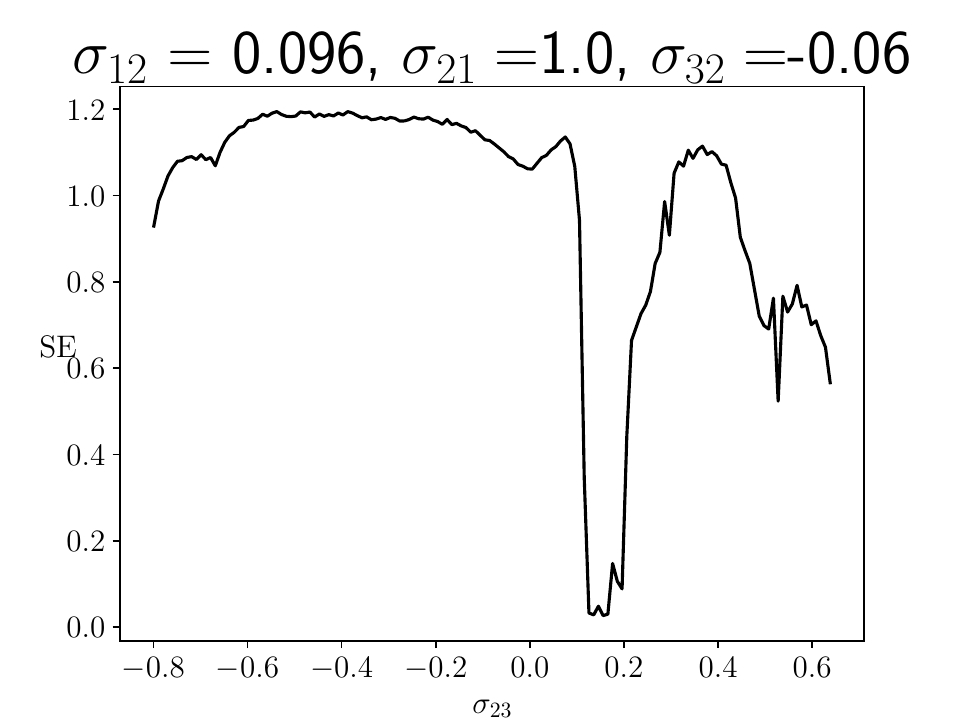}\\
  (e) & (f) \\
  \includegraphics[scale=0.3]{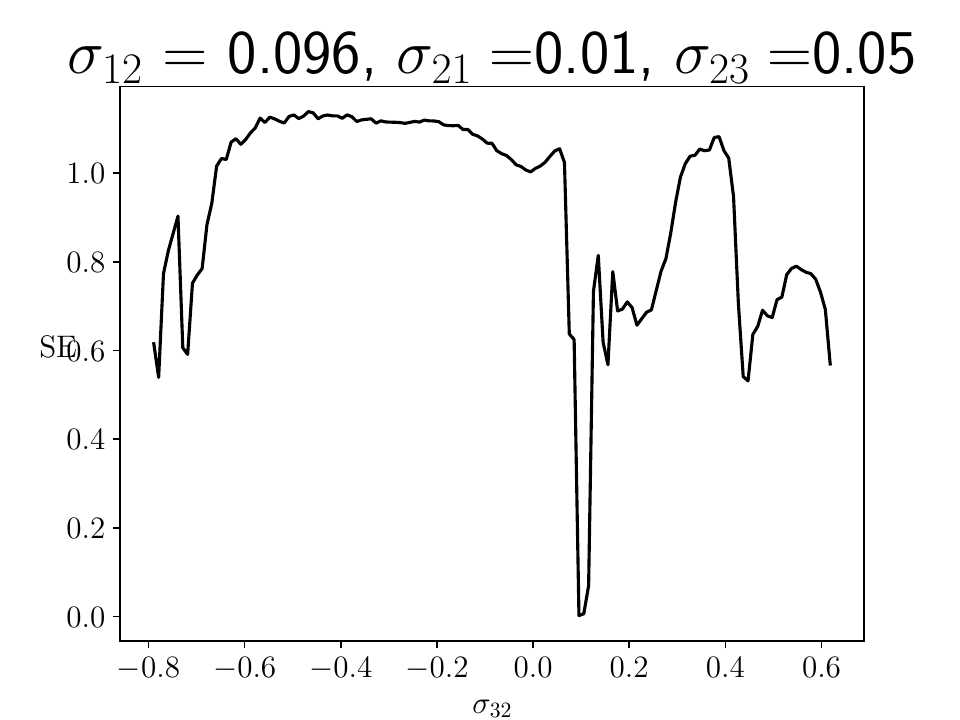} &   \includegraphics[scale=0.3]{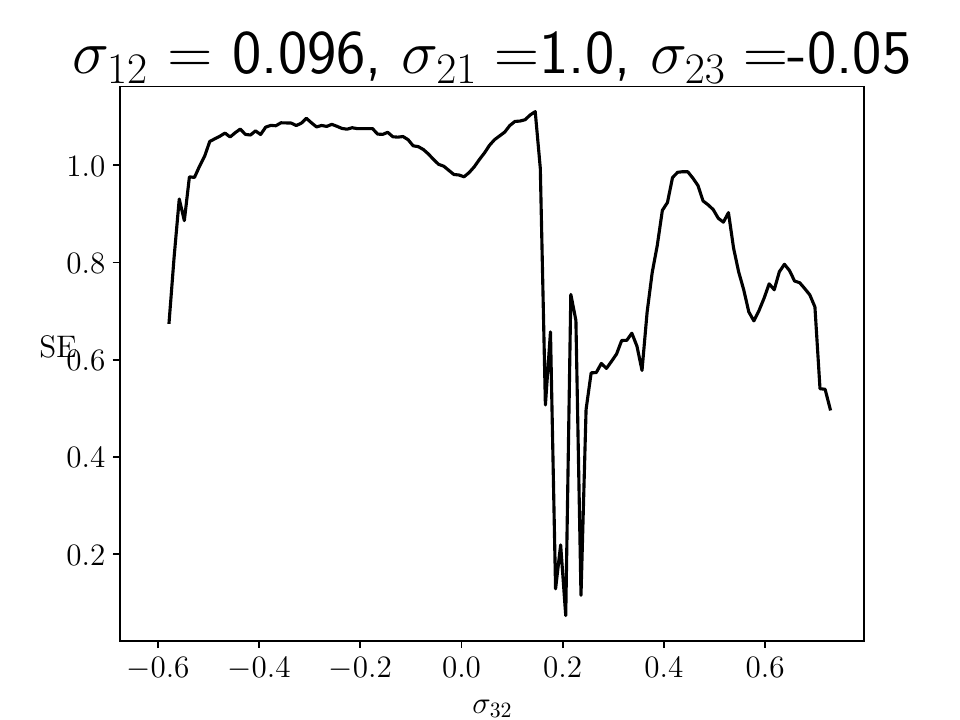}\\
 (g) &(h)  \\[3pt]
\end{tabular}
\caption{A collection of one-parameter bifurcation plots of the sample entropy of the network with varying $\sigma_{ij}$. The local parameters have been set to $a=0.6$, $b=0.6$, $c=0.89$, $k_0=-1$, $\alpha=5$, $\mu=0.0001$, and $\gamma=-0.5$. }
\label{fig:SampleEntropy}
\end{figure}

Next, we plot one-parameter bifurcation diagrams for the sample entropy of the network with the varying coupling strengths $\sigma_{ij}$, see Fig.~\ref{fig:SampleEntropy}. Again the local parameters have been set to $a=0.6$, $b=0.6$, $c=0.89$, $k_0=-1$, $\alpha=5$, $\mu=0.0001$, and $\gamma=-0.5$. We run $40000$ iterates and remove the first $20000$ to ensure no residual transients. In the first row, we vary $\sigma_{12}$ and fix $\sigma_{23}=0.05, \sigma_{32}=0.06$. Panel (a) has $\sigma_{21}=0.1$. We see a rise in the complexity of the system reaching a maximum of ${\rm SE} \approx 1.1298$ at $\sigma_{12}\approx -0.056$ as $\sigma_{12}$ increases from $-1$ before it meets a sharp dip of ${\rm SE} \approx 0.021$ at $\sigma_{12}\approx 0.093$. Following this, the complexity gradually goes down as $0<\sigma_{12}<0.4$. beyond this range, the dynamics of the network diverge. Panel (b) has $\sigma_{21}=-0.1$. In this case, the network remains in a relatively high complexity state ${\rm SE}>0.8$ in the domain $-0.3<\sigma_{12}<0.367$ approximately, after which it sees a sharp fall in the complexity till $\sigma_{12} \approx 0.486$. beyond this range of $\sigma_{12}$ the network diverges. The maximum of ${\rm SE} \approx 1.11068$ is reached at $\sigma_{12} \approx 1.112$. In the second row, $\sigma_{21}$ is the primary bifurcation parameter, fixing $\sigma_{23}=0.05$, and $\sigma_{32}=0.06$. Panel (c) has $\sigma_{12}=0.096$. Again, the complexity of the dynamics of the system is moderately high in the range $0.6<{\rm SE}<1.132$ approx, except for a few dips which happen when $\sigma_{21}>0$. One of the dips happens at $\sigma_{21} \approx 0.1$ where ${\rm SE} \approx 0$. The dynamics diverges beyond the range $-0.79<\sigma_{21}<0.62$ approximately. Panel (d) has $\sigma_{12}=-0.096$. The complexity of the tri-oscillator model, in this case, exhibits similar behavior where $0.5<{\rm SE}<1.153$ is moderately to substantially high. Note that there is a break from $0.797<\sigma_{21}<0.814$ indicating a diverging behavior. Also, the dynamics are divergent beyond the range $-0.6<\sigma_{21}<0.83$. In the third row, we vary $\sigma_{23}$ as the primary bifurcation parameter and $\sigma_{12}=0.096$, $\sigma_{21}=0.01$. Panel (e) has $\sigma_{32}=0.06$. The complexity is moderately to substantially high ($0.55<{\rm SE}<1.136$) when $\sigma_{23}\le 0$. Otherwise, the sample entropy value fluctuates between a high and a low value with the lowest $\approx 0$. The dynamics diverges beyond $-0.8<\sigma_{23}<0.62$ approximately. Panel (f) has $\sigma_{32}=-0.06$ and exhibits a similar complexity as in panel (e). Lastly the fourth row sees $\sigma_{32}$n as the primary bifurcation parameter with $\sigma_{12}=0.096$, $\sigma_{21}=0.01$. Panel (g) has $\sigma_{23}=0.05$. For $\sigma_{32}<0$, the system exhibits a moderate to substantial complexity ($0.54<{\rm SE}<1.138$), and when $\sigma_{32}>0$ the complexity fluctuates between a low and a high value with the lowest being ${\rm SE} \approx 0$ at $\sigma_{32}\approx 0.1$. Beyond $-0.792<\sigma_{32}<0.616$, the system diverges. Panel (h) has $\sigma_{23}=-0.05$ and shows a similar qualitative behavior as panel (g).




\section{Conclusions and future directions}
\label{sec:conclusions}
In this paper, we have investigated a heterogeneous chain network to model the dynamics of neuron ensembles in the nervous system. This model can be realized as a unit that could repeat itself to generate more complicated neuron aggregates replicating the real-world functionalities of the nervous system. The model is built on two popular neuron maps: the Chialvo map (peripheral nodes) and the Rulkov map (central node) with bidirectional linear couplings between two neurons. The motivation behind this was to build a heterogeneous model that mimics the functionalities of three kinds of neurons present in the nervous system and the synaptic connections for information transfer among them. Heterogeneity is incorporated in two ways: first, the central node oscillates following the dynamics of the Rulkov map, whereas the end nodes oscillate following the Chialvo map, and second the coupling between two nodes $i$ and $j$ is bidirectional with $\sigma_{ij} \ne \sigma_{ji}$. One future direction in the study of the dynamics of this small network is to incorporate noise modulation via additive or multiplicative noises and characterize a battery of spatiotemporal patterns.

After we put forward our model which is a nonlinear system of six coupled equations, we look deep into the dynamical properties of the model. The first step is figuring out the fixed point and performing a stability analysis of the same. To do so, we need to first build the $6 \times 6$ Jacobian matrix and look into the eigenvalues of the matrix at the fixed point. We also set up the noninvertibility criterion of the model, i.e., where the determinant of the Jacobian matrix goes to $0$. The next step toward unfolding the dynamics of the model was to look into various bifurcation patterns. At first, we plot the last $500$ points of the action potentials from every simulation with a varying primary bifurcation parameter. We notice that the dynamics fluctuate between a chaotic attractor and a period-$4$ attractor. We also observe a coexistence of the two, supported by the existence of hysteresis loops. The concept of coexistence is also verified from the phase portraits where the same parameter set generates these two different attractors on two separate simulation runs. Then we perform a codimension-$1$ and -$2$ pattern analysis using \textsc{MatContM} as a tool and discover the existence of saddle-node, period-doubling, and Neimark-Sacker bifurcation patterns. These three are codimension-$1$ bifurcations and are also supported by analytical proofs. Furthermore, \textsc{MatContM} allows us to observe rich codimension-$2$ patterns like double Neimark-Sacker, flip-Neimark-Sacker, $1:1$ resonance, fold-flip, fold-Neimark-Sacker, and $1:2$ resonance. These show that our heterogeneous neuron model is a repository of a wide array of engrossing dynamical properties. Thus this model can be in future utilized in designing a ring network (infinite chains), a star network (repetition of the chain with one central node and an infinite number of peripheral nodes), and a combination of the two to further study the rich spatiotemporal behaviors that might arise due to the complexity induced. Another interesting candidate is a multiplex network made up of our model as the building block.

We have taken a step forward to also study the synchronization behavior of this small network model via the cross-correlation coefficient and the Kuramoto order parameter. Both these measures indicate that the model mostly remains in an excitatory state exhibiting asynchrony and partial synchrony (in-phase and anti-phase). These are illustrated using two-dimensional color-coded plots in this paper. These color-coded plots correspond to two-dimensional bifurcation diagrams revealing parameter regions where the system turns from partial synchrony to complete asynchrony and vice versa, indicating a global behavior of the system. Increasing the number of nodes to determine whether there are solitary nodes, chimera patterns, cluster states, and wave structures as spatiotemporal patterns using these measures is an interesting avenue to investigate. Also, another future aspect is to build a metric to look into whether there exists a ``weak chimera'' in the tri-oscillator model. Another important step in studying a dynamical system is to look into its time series and perform a complexity analysis using an entropy metric. In this paper, we see time series with both chaotic and regulatory behavior. To quantify this we utilised the concept of sample entropy. We see that for an irregular and chaotic time series, the sample entropy value is high whereas when the time series is regular, the sample entropy value is close to zero. Using this metric on a noise-modulated tri-oscillator model and also a model with an infinite number of nodes in the thermodynamic limit is an important aspect to look into. An analytical relationship is also required to be set up to check how all these measures relate to each other. Note that throughout the paper, we have kept the local parameters of each of the three oscillators fixed and varied the coupling strengths between the oscillators as primary bifurcation parameters. 

In the future, we want to look at the dynamical properties of this model using coupling strengths which change over every iteration number, making the network temporally heterogeneous. One question that also arises is what kind of conservative properties these kind of neuron maps have, for example, the conservation of Hamiltonian energy in the continuous time systems. Can we come up with an equivalent quantity that is being conserved in discrete-time neuron maps? Furthermore, as discrete-time systems are more computationally efficient, we suppose it would be an interesting problem to look into heterogeneous models of discretized versions of continuous-time neuron models, via a small-network topology. Small networks are reduced order models which are undoubtedly the best candidates to study before we consider networks in the thermodynamic limit.

As with any other model, our model is not perfect. But of course, we can work on making our model come closer to a real-world scenario. One step towards that is to fit our model from medically available EEG data from reliable sources. This would by itself be a captivating field to persuade. One challenge the authors have faced is to come up with a Lyapunov exponent study of the network itself, where the nodes are coupled. One approach could be motivated by Caligiuri {\em et al.}~\cite{CaEg23}. This remains an open question for a static network like our model. Another way to make the model closer to reality is to perturb every node or the coupling strengths with external forces. 

Our approach in this paper has been an amalgamation of both analysis and numerics which we believe will aid mathematical modelers, engineers, quantitative biologists, and neuroscientists the same. This model lays a step towards understanding the intricate dynamics of more topologically complicated ensembles of neurons involved in signal processing in the nervous system.

\clearpage

\bibliographystyle{plain}
\bibliography{main}

\end{document}